\documentclass[letterpaper,twocolumn,10pt]{article}
\usepackage{usenix2019_v3}

\usepackage{tikz}
\usepackage{amsmath,amssymb,amsfonts,amsthm,mathrsfs}

\usepackage{graphicx}
\usepackage{subcaption}
\usepackage{caption}

\usepackage{filecontents}

\usepackage{pifont}
\usepackage{booktabs,siunitx}
\usepackage{longtable}
\usepackage{supertabular}
\usepackage{url}
\usepackage{amsmath,amssymb}
\usepackage[square,sort,comma,numbers]{natbib}
\usepackage{setspace}
\usepackage{diagbox}

\newcommand{\quotebox}[1]{
\begin{quote}
\normalsize	
\textit{``#1''}
\end{quote}
}

\usepackage{hyperref}\hypersetup{
    colorlinks = true,
    linkcolor = blue,
    anchorcolor = blue,
    citecolor = blue,
    filecolor = blue,
    urlcolor = black
}
\usepackage{breakurl}

\usepackage[normalem]{ulem}

\usepackage{listings}
\lstdefinelanguage{JavaScript}{
  keywords={GET, ns-1168.awsdns-18.org},
  keywordstyle=\color{blue!70}\bfseries,
  ndkeywords={ns-1168.awsdns-18.org, export, boolean, throw, implements, import, The, Shoestring, Gardener},
  ndkeywordstyle=\color{red}\bfseries,
  identifierstyle=\color{black},
  sensitive=false,
  comment=[l]{//},
  morecomment=[s]{/*}{*/},
  commentstyle=\color{purple}\ttfamily,
  stringstyle=\color{black}\ttfamily,
  morestring=[b]',
  morestring=[b]"
}

\usepackage{xcolor}
\definecolor{codegreen}{rgb}{0,0.6,0}
\definecolor{codegray}{rgb}{0.5,0.5,0.5}
\definecolor{codepurple}{rgb}{0.58,0,0.82}
\definecolor{backcolour}{rgb}{0.95,0.95,0.92}
\definecolor{ao}{rgb}{0.0, 0.5, 0.0}
 \definecolor{antiquefuchsia}{rgb}{0.57, 0.36, 0.51}
 
\definecolor{lx-red}{rgb}{1.0, 0.13, 0.32}
\definecolor{bittersweet}{rgb}{1.0, 0.44, 0.37}
\definecolor{blue-violet}{rgb}{0.54, 0.17, 0.89}
\newcommand{\LX}[1] {{\color{blue-violet}#1}}
\newcommand{\LXcomment}[1] {{\color{lx-red}LX: #1}}
\newcommand{\XY}[1] {{\color{antiquefuchsia}XY: #1}}

\usepackage{color, colortbl}

\usepackage{makecell}

\definecolor{cadmiumorange}{rgb}{0.93, 0.53, 0.18}
\definecolor{electriccrimson}{rgb}{1.0, 0.0, 0.25}

\newcommand{\xiaojing}[1] {{\color{electriccrimson}{#1}}}

\newcommand{\ignore}[1]{}

\newcommand\qy[1]{{\color[RGB]{34,125,81}{\textbf{\{qy: {\em#1}\}}}}}

\newcommand{\staticFramework}{\textit{SAF{}}}  
\newcommand{\dynamicFramework}{\textit{DAF{}}}

\newcommand{\comparer}{\textit{PLCC{}}}

\newcommand{\toolname}{\textit{Lalaine{}}} 
\newcommand{\appNumber}{\textit{485,024}} 
\newcommand{\ldataItem}{\textit{l-data item}}

\newcolumntype{C}[1]{>{\centering\let\newline\\\arraybackslash\hspace{0pt}}m{#1}}

\makeatletter
\newcommand{\defeq}{\mathrel{\aban@defeq}}
\newcommand{\aban@defeq}{%
  \vbox{\offinterlineskip\check@mathfonts
    \ialign{\hfil##\hfil\cr
      \fontsize{\ssf@size}{\z@}\normalfont def\cr
      \noalign{\kern1\p@}
      $\m@th=$\cr
      \noalign{\kern-.5\fontdimen22\textfont2}
    }%
  }%
}
\makeatother

\usepackage[title]{appendix}

\usepackage{amsmath,amssymb,amsfonts,amsthm,mathrsfs}
\usepackage{algorithmic}

\theoremstyle{definition}
\newtheorem{inconsistency}{Inconsistency}[]

\theoremstyle{definition}
\newtheorem{definition}{Definition}[]
\begin{document}
%-------------------------------------------------------------------------------

%don't want date printed
\date{}

\title{Lalaine: Measuring and Characterizing Non-Compliance of Apple Privacy Labels at Scale}

%for single author (just remove % characters)
%\author{
%{\rm Your N.\ Here}\\
%Your Institution
%\and
%{\rm Second Name}\\
%Second Institution
% copy the following lines to add more authors
% \and
% {\rm Name}\\
%Name Institution
%} % end author

\author{
\normalsize Yue Xiao$^{1}$, Zhengyi Li$^{1}$, Yue Qin$^{1}$, Xiaolong Bai$^{2}$, Jiale Guan$^{1}$, \normalsize  Xiaojing Liao$^{1}$, 
\normalsize Luyi Xing$^{1}$
\\
\small $^1$Indiana University Bloomington, 
\small $^2$Orion Security Lab, Alibaba Group\\
} % end author

\pagestyle{empty}  % no page number for the second and the later pages
\thispagestyle{empty} % no page number for the first page

\maketitle
\begin{abstract}

As a key supplement to privacy policies that are known to be lengthy and difficult to read, Apple has launched app privacy labels, which purportedly help users more easily understand an app's privacy practices. However, false and misleading privacy labels can dupe privacy-conscious consumers into downloading data-intensive apps, ultimately eroding the credibility and integrity of the labels. 
Although Apple releases requirements and guidelines for app developers to create privacy labels, little is known about whether and to what extent the privacy labels in the wild are correct and compliant, reflecting the actual data practices of iOS apps. 

This paper presents the first systematic study, based on our new methodology named \toolname{}, to evaluate data-flow to privacy-label (\textit{flow-to-label}) consistency. \toolname{} analyzed the privacy labels and binaries of 5,102 iOS apps, shedding light on the prevalence and seriousness of privacy-label non-compliance. 
We provide detailed case studies and analyze root causes for privacy label non-compliance that complements prior understandings. This has led to new insights for improving privacy-label design and compliance requirements, so app developers, platform stakeholders, and policy-makers can better achieve their privacy and accountability goals. \toolname{} is thoroughly evaluated for its high effectiveness and efficiency. We are responsibly reporting the results to stakeholders. \looseness=-1

\end{abstract}
\section{Introduction}

% Congress letter:

% https://energycommerce.house.gov/sites/democrats.energycommerce.house.gov/files/documents/Apple%20Letter%20re%20App%20Privacy%20Label%202-2021.pdf

In December 2020, Apple launched its app privacy labels, which purportedly help users better understand an app's privacy practices before they download the app on any Apple platform. Without meaningful, accurate information, Apple's tool of
illumination and transparency may become a source of consumer confusion and harm --- a concern raised in the United States congress~\cite{energycommerce}. False and misleading privacy labels can dupe privacy-conscious consumers into downloading data-intensive apps, ultimately eroding the credibility and integrity of the labels. A privacy label without credibility and integrity also may dull the competitive forces encouraging app developers to improve their data practices. %The Congress of the United States urges that:``A privacy label is no protection if it is false. We urge Apple to improve the validity of its App Privacy labels to ensure consumers are provided meaningful information about their apps’ data practices and that consumers are not harmed by these potentially deceptive practices ''~\cite{congress-letter}. 

Apple defines a set of compliance requirements and provides guidelines for app developers to create privacy labels. In particular, the Apple privacy label is not a direct abstraction of the app's privacy policy: Apple defines a four-layer structure for privacy label (i.e., with four data usages --- first layer, six purposes --- second layer, 14 data types --- third layer, and 32 data items --- fourth layer, see \S~\ref{background:privacyLabel}). The recent work~\cite{li2022understanding}, through a user study with 12 iOS developers, showed that developers had knowledge gaps about privacy concepts needed in Apple privacy labels or suffered from extra overhead for creating the labels, posing serious practical challenges for creating correct privacy labels.
%
%or even have never heard about privacy label requirements.
%and thus might not be able create correct privacy labels. 
%
Despite the complexity and serious implications, Apple officially states \ignore{(in the beginning of privacy labels)}that it has not verified the information on privacy labels provided by app developers. 
%Y. Li et al~\cite{} performs a longitude study to understand how promptly developers create and update privacy labels in response to this emerging privacy-compliance expectation.
%
Hence, it is imperative to understand whether and to what extent the privacy labels in the wild are correct and compliant, reflecting the actual data practices of iOS apps.
The recent work~\cite{li2022understandinglongitutde} performs a longitude study to understand how promptly developers create and update privacy labels in response to this emerging privacy call.
However, to the best of our knowledge, there was no systematic study to evaluate data-flow to privacy-label (\textit{flow-to-label}) consistency, which we aim to address in this paper.

\vspace{2pt}\noindent\textbf{Methodology for privacy-label compliance check.} %In this paper, we report the first study on 
We present a new, automatic methodology called \toolname{} to check the compliance of privacy labels for iOS apps at scale. Specifically, \toolname{} checks the consistency between disclosure statements in privacy labels and actual data flows or practices of iOS apps \ignore{each abstracted as a tuple $(date~item, ~usage~purpose)$adapted for Apple privacy label's structure and semantics} (see the modeling in \S~\ref{sec:model}). Essential \ignore{for a systematic analysis }is to formally define the inconsistency model for privacy labels, which \ignore{has not been done and}cannot directly adopt prior inconsistency models for privacy policies~\cite{andow2020actions,bui2021consistency} %\ignore{(e.g., \textit{Omitted Disclosure}, \textit{Incorrect Disclosure}~\cite{})} 
due to privacy labels' different structure and semantics (e.g., with four layers, emphasizing data-usage purposes\ignore{framed by Apple} versus the party/entity that receives the data, see \S~\ref{s:preConsisModel}). In particular, one of the key prior inconsistency types for privacy policies, namely \textit{Omitted Disclosure}, is too coarse-grained to characterize non-compliance issues with privacy labels: for example, a privacy label that mistakenly discloses its data-usage purpose (e.g., \textit{Third-party Advertising} compared to \textit{Developer's Advertising or Marketing}, based on the six usage purposes defined by Apple~\cite{iosprivacylabel}), a privacy label that honestly discloses a usage purpose while missing to disclose all purposes for that data, and a privacy label that simply discloses nothing, would have all been simply considered as \textit{Omitted Disclosure} based on the prior inconsistency models (see the comparison in \S~\ref{s:preConsisModel}). To characterize key issues in privacy label compliance and guide the design of useful tools which can help developers achieve the emerging compliance goal, we formally define a new inconsistency model adapted for privacy labels (\S~\ref{sec:model}).

%missed to disclosure all usage purposes (e.g., app functionality, third-party advertising, developer's advertising, out of six fixed purposes defined by Apple~\cite{}) 

Based on our new inconsistency model, we design and implement \toolname{} as an automated, end-to-end system by adapting and synthesizing a set of innovative techniques including automatic iOS app UI execution, natural-language processing (NLP), and dynamic and static binary analysis (Figure~\ref{f:overview}).
Notably, based on Apple, a data item is deemed ``collected''\ignore{ by a vendor } only if it is transmitted to the Internet~\cite{iosprivacylabel}. That is, Apple's accountability criteria is more strict than many prior works on Android~\cite{gordon2015information,zimmeck2019maps,arzt2014flowdroid} which commonly assumed the data as leaked out once an API returning the data was invoked. Hence, for a precise compliance check, \toolname{} includes dynamic end-to-end app execution to find out the actual data ``collection''. Due to the known, limited scalability of dynamic analysis, we could not fully run all 366,685 iOS apps we collected. To address the challenge, \toolname{} proposes a novel, optimized strategy by filtering apps of most interest: (1) apps that involve sensitive iOS APIs whose return values fall under Apple's framing of the 32 date items for privacy labels (\S~\ref{sec:saa}) and (2) apps showing inconsistency between their privacy labels and privacy policies. \toolname{} further intersected the two app sets to yield three sub-sets with different (non-)compliance characteristics (e.g., apps being flow-to-label inconsistent while flow-to-policy consistent, flow-to-label consistent while flow-to-policy inconsistent, see \S~\ref{sec:dap}), and sampled a total of 6,332 apps for full dynamic execution, maintaining both scalability and generality of our study. Another key challenge is to infer the vendors' actual purposes for the collection of each data item, \ignore{which is essential for Apple privacy-label compliance check} which we tackled by adapting a modeling and learning-based approach (\S~\ref{sec:plcc}). Our thorough evaluation shows that \toolname{} can detect privacy label non-compliance effectively and efficiently (\S~\ref{s:overallEval}). \looseness=-1

%For our compliance check, \toolname{} first infers the \ldataItem{} that is actually collected based on both network traffic and evidence found in dynamic instrumentation. Another key challenge is to infer the vendors' actual purposes for the collection of each \ldataItem{}s (e.g., \textit{App functionality}, \textit{Third party advertising}, falling under categories defined by Apple), which we tackled by adapting a modeling and learning-based approach. 

\ignore{
    Based on our new inconsistency model, we design and implement \toolname{} as an automated, end-to-end system by adapting and synthesizing a set of innovative techniques including automatic iOS app UI execution, natural-language processing (NLP), and dynamic and static binary analysis (Figure~\ref{f:overview}).
    Specifically, \toolname{} first collects a large-scale of iOS apps from Apple app store (U.S.) with their corresponding privacy labels. Meanwhile, \toolname{} collects the apps' privacy policies, which \toolname{} compares with privacy labels based on NLP and semantic analysis to better understand ``label-to-policy'' inconsistencies and the new challenges (\S~\ref{sec:plpcc}). Apps that involve sensitive iOS APIs whose return values fall under Apple's framing of the 32 date items for privacy labels (found by \toolname{}'s light-weight static analysis on app binaries) and those showing inconsistent between privacy labels and privacy policies, are filtered and sampled, and then go through an end-to-end dynamic analysis pipeline with fully automatic UI execution, dynamic instrumentation and network monitoring. 
    Based on context information that features the data practices (e.g., API caller, stack traces, server endpoints, Web requests and responses) collected by the dynamic analysis pipeline, \toolname{} infers the actual data collected and purposes of collection, and aligns them, abstracted as tuples $(data~item, ~usage~purpose)$, to the 32 data items and 6 purposes defined by Apple. \toolname{} finally reports privacy label non-compliance by comparing those tuples with disclosures in the vendors' privacy labels. Our thorough evaluation shows that \toolname{} can detect privacy label non-compliance effectively and efficiently (\S~\ref{s:overallEval}). 
}

% (and privacy polices for additionally comparing inconsistencies between)

%an iOS app's actual data practice, abstracted as $(date item, usage purpose)$
%\toolname{} relies on the extraction of actual data practices

%data-purposes pair from dynamic code behavior and analysis of inconsistency with its corresponding privacy label.

\vspace{3pt}\noindent\textbf{Measurement and findings.}
Looking into the apps with privacy label non-compliance reported by \toolname{}, we are surprised to find the pervasiveness of privacy label non-compliance in iOS apps, with a big impact on credible and transparent disclosure of app privacy practices.
More specifically, among 5,102 iOS apps being fully tested, we discover 3,423 apps detected with privacy label non-compliance:  3,281 of them fail to disclose data and purposes, 1,628 contrarily specify purposes, and 677 apps inadequately disclose purposes.
We observe \textit{User ID}, \textit{Device ID} and \textit{Location} data are prone to be omitted by developers. Also, game apps and education apps are commonly observed with privacy label non-compliance, especially omitting the disclosure of Device ID, User ID, and Precise Location, while those apps  are usually youth using.

Also important are the root causes of privacy label non-compliance, as discovered in our study, which include opaque data collection from diverse third-party partners, and misleading privacy label disclosure guidance, etc.
In particular, while app developers are responsible for the disclosure of data collection by third-party partners~\cite{iosprivacylabel}, the app developer is typically unaware of the data collection and usage practices employed by third-party partners, and even misled by the privacy label disclosure guidance provided by the third-party partners.
When examining 18 guidances provided by high-profile third-party partners (e.g., Flurry, On Signal, and Appsflyer), we found that 14 of them are either inadequately (N=4) or incorrectly (N=9) disclosing data collection practices, or using vague statements (N=1). As an instance, we observe that \textit{Flurry Analytics SDK} collected precise location data from 76 apps. However, in its privacy-label disclosure guidance, it didn't suggest app developers fill out the privacy label to declare the collection of user location data~\cite{FlurrySDKguidance} (but only ``Device ID'' , ``Product Interaction'', and ``Other Usage Data'').
These findings shed light on new challenges in reality for creating precise and complete privacy labels that were less understood by prior work~\cite{li2022understanding,li2022understandinglongitutde} (\S~\ref{sec:root}). 

\vspace{3pt}\noindent\textbf{Responsible disclosure}. We are in the process of reporting all the findings (i.e., apps with non-compliant privacy labels and the inconsistencies between app privacy labels and privacy policies) to Apple. Part of the results have been reported.

\vspace{3pt}\noindent\textbf{Contributions}. We summarize the key contributions below.

\vspace{2pt}\noindent$\bullet$ We performed a systematic study on Apple privacy label compliance, in particular focusing on inconsistency between privacy labels and app data flows/practices for the first time. Our results shed lights on the prevalence, seriousness of privacy-label non-compliance and dure challenges for developers to achieve the emerging compliance goal. \looseness=-1

\vspace{2pt}\noindent$\bullet$ We introduce the first methodology with end-to-end implementation called \toolname{} that can automatically assess the ``flow-to-label'' inconsistency. \toolname{} is based on our new, formally defined inconsistency model for privacy labels (by adapting prior inconsistency models for privacy policies but cannot directly use them). Our tool can be used by app developers for achieving compliance goals and by app stores for vetting purpose. We will release the tool to the public. \looseness=-1

\vspace{2pt}\noindent$\bullet$ We provide detailed case studies and analyze root causes for privacy label non-compliance that complements recent understandings on privacy labels. This has led to new insights for improving privacy-label design and compliance requirements, so app developers, platform stakeholders, and policy-makers can better achieve their privacy and accountability goals. \looseness=-1

\ignore{
Static analysis, as widely used in prior work to identify privacy and security problems~\cite{egele2011pios,gibler2012androidleaks, livshits2013automatic,enck2011study}, can scale to a large number of apps. However, examining app code without executing it, static analysis is prone to false positives as it can only deliver potential privacy-violation behaviours. Besides, some highly-stealthily malicious code may leverage runtime contexts (e.g., download additional
code from cloud/server~\cite{grace2012unsafe}, dynamically load endpoints~\cite{wang2021understanding}) to hide data exfiltration channels, which is hard to be captured by static analysis and further cause false negatives.

Dynamic analysis, one the other hand, can execute apps and examine their runtime properties. Previous dynamic analysis tools (e.g, TaintDroid~\cite{tio32taintdroid} and DroidBox~\cite{droidbox}) used crafted-DVM that supports taint tracking (i.e., observing modified data by traversing through system). Other work~\cite{rastogi2013appsplayground,hao2014puma,reyes2018won} used the UI automation techniques that can automatically launch and interact with an app and simultaneously collecting dynamic state information during app execution. 

Hybrid analysis, combining static and dynamic analysis, can increase the coverage, scalability, or visibility of the analyses~\cite{reardon201950}. Our approach is to first perform a simple static analysis to identify sensitive API from binary code, and then instrument it by dynamic analysis to confirm the violations.

%However, performing dynamic analysis at scale is challenging due to the need to actually run apps with user input.

\vspace{2pt}\noindent\textbf{Flow-to-privacy label consistency},
Sensitive information leakage is one of the most prominent security threats to mobile ecosystem. Literature has proposed a variety of analysis tools (apply static analysis~\cite{grace2012unsafe,gordon2015information,zimmeck2016automated,gibler2012androidleaks} and/or dynamic analysis~\cite{reyes2018won,hao2014puma,lindorfer2014andrubis,ren2018bug,song2015privacyguard,yang2013appintent}) that are designed to analyze or explore the information flows inside the mobile apps. Further, whether such information flows are considered as a privacy leakage are determined by whether the corresponding data object are disclosed in its privacy privacy. However, those prior work does not differentiate who received the privacy-sensitve data (e.g., first-party vs. third-party). To solve this weakness, Andow et al.~\cite{andow2020actions} proposed an entity-sensitive flow-to-policy consistency model which consider both the data type and the entity receiving the data and sentiment of the statement from the privacy policy, which increased the precision of compliance check (information flow vs privacy policy).

However, none of the previous work considers whether the purpose of such data collection is compliance or not. Defined by the GDPR~\cite{gdpr}, \textit{Personal data shall be collected for specified, explicit and legitimate \textbf{purposes} and not further processed in a manner that is incompatible with those purposes.} Besides, CCPA~\cite{ccpa} requires businesses \textit{list the categories of personal information businesses collect about consumers and the \textbf{purposes} for which they use the categories of information.} In addtion, Apple also require developer that  \textit{(2) Keep in mind that even if you collect the data for \textbf{purpose} other than analytics or advertising, it still needs to be declared. (1) if you collect data solely for the purpose of app functionality, declare the data on your label and indicate that it is only being used for that \textbf{purpose}. } Hence, any missing, incorrect and inadequate purpose declaration in privacy label is an violation with Apple policy and an infringement to regulations (e.g., GDPR~\cite{gdpr}, CCPA~\cite{ccpa}).

Here, we propose a purpose-sensitive flow-to-privacy label consistency model to determine if an application’s 
privacy label discloses relevant data flows.

\vspace{2pt}\noindent\textbf{Challenge in iOS Dynamic Analysis}
To begin with, the emulator provided by Xcode~\cite{xcode} can only execute application with source code. There is no existing open-source emulator~\cite{szydlowski2011challenges} that can be extended to perform additional analysis tasks for iOS application binary which can be dumped from Apple App Store. 
What's more, It is hard to explore a wide range of user interaction. Simply launching the application can only achieve 16\% code coverage~\cite{szydlowski2011challenges}.
}
\section{Background}
\label{sec:background}

\begin{figure}[t]
    \centering
    \setlength{\abovecaptionskip}{0.cm}
    \includegraphics[width=0.9\columnwidth]{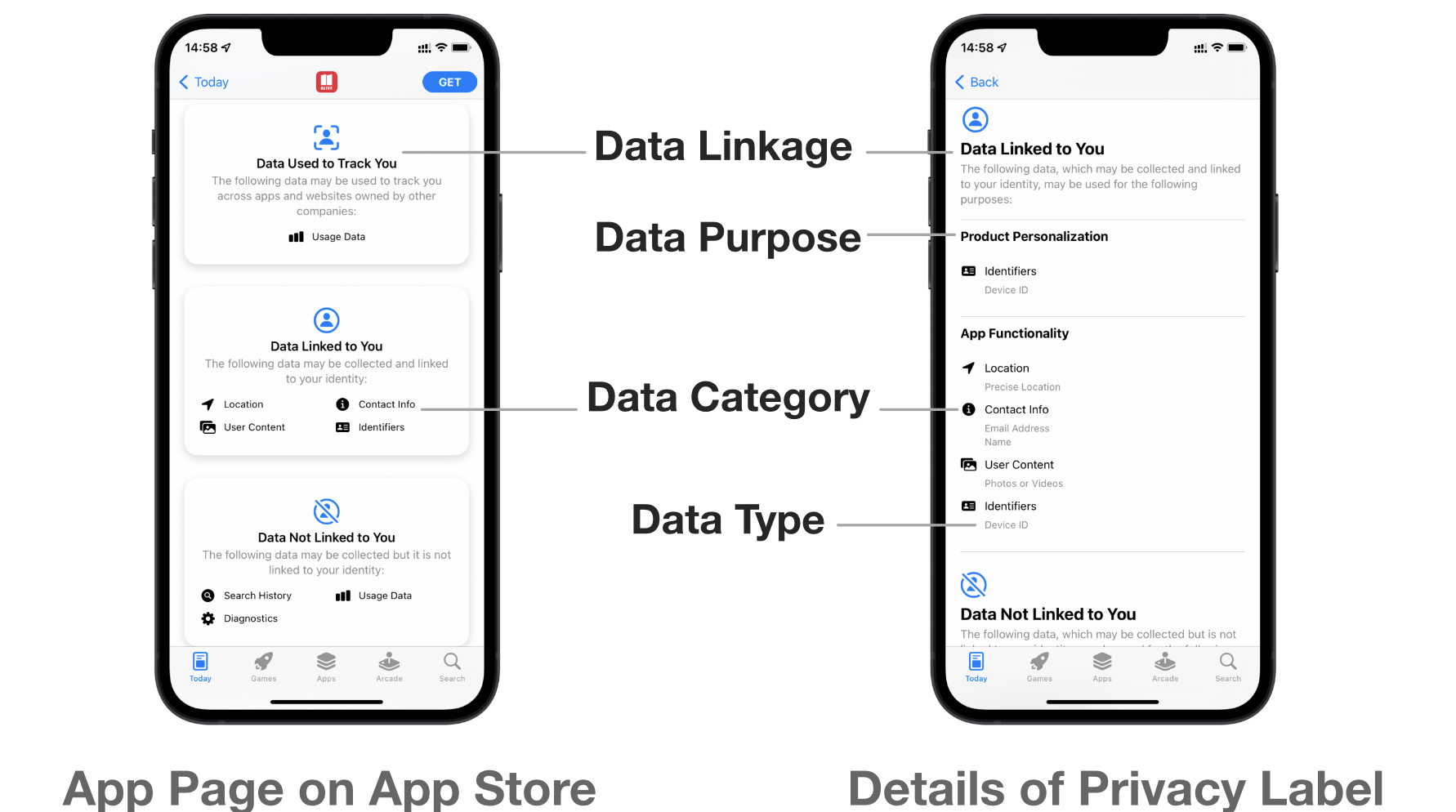}
    \vspace{10pt} 
    \caption{Privacy label views on App Store. The left is the privacy label card views from the app details page. On the right is the details of privacy label, which will appear after clicking any of the card views on the left page.}

    \label{fig:privacyLabel}
    \vspace{-15pt}
\end{figure}

\vspace{-5pt}
\subsection{Privacy Labels of iOS Apps}
\label{background:privacyLabel}
As is illustrated in Figure~\ref{fig:privacyLabel}, privacy label of an iOS app is a ``nutrition label"-like privacy disclosure which lists what data is collected from the iOS app and how it is used. 
Privacy label has a specific focus on data collection, to help app users better understand how apps handle privacy-sensitive data. 
As indicated in~\cite{iosprivacylabel},  ``collect" refers to transmitting data off the device in a way that allows app developers and/or their third-party partners to access it for an extended period. 
In addition, Apple asked developers to provide privacy labels following certain key requirements~\cite{iosprivacylabel}: (1) The app developer are responsible for the disclosure for data collection of the whole app, including those collected by third-party partners; the app developer are required to disclose \textit{all} purposes for collecting a data. (2) The app developers need to keep privacy labels accurate and up to date.

% than what is necessary to service the transmitted request in real time.
%

%
%Since December 8, 2020, the App Store announced a requirement for iOS App developers to fill out privacy labels when publishing iOS apps~\cite{privacynutritionlabels}. Note that the ecosystem of the privacy label is built on the honor system. The privacy label has not been verified by Apple~\cite{}.

\vspace{2pt}\noindent\textbf{Taxonomy of privacy label.} The privacy disclosure of the privacy label follows a four-layer taxonomy~\cite{iosprivacylabel}: data usage, purpose, data type and data item.
At the highest layer, these labels fall into four categories of data usage: \textit{Data Used to Track You}, \textit{Data Linked to You}, \textit{Data Not Linked to You}, or \textit{Data Not Collected}. Under each category, Apple defines six purposes except for the category \textit{Data Not Collected}: \textit{Third-Party Advertising}, \textit{Developer's Advertising or Marketing},  \textit{Analytics}, \textit{Product Personalization}, \textit{App Functionality}, and \textit{Other Purposes}. For each purpose, the privacy label lists what data types (e.g., \textit{Contact Info}) and specific data items (e.g., \textit{Email Address, Phone Number, Physical Address, etc.}) are being collected. 
The taxonomy of privacy label defines 14 data types and 32 data items~\cite{iosprivacylabel} covering various privacy-sensitive data ranging from personally identifiable information (PII) to health information.
%
%Those required privacy labels are intended to improve user awareness and make it tougher for dishonest apps developers to track, steal or sell information wrongfully.
%
If the app developer does not fill out the privacy label, the privacy disclosure will be \textit{``No Details Provided"}. 
%In our study, we focus on the whether the disclosure of data items and  purposes in privacy label is consistent with its privacy policy and code behavior. 

%The disclosure of privacy label cover a wide range of practices, including client-side and server-side practices. In this paper, we are only concerned about client-side practices affecting iOS applications.

\ignore{
    In our study, we investigated whether the data items (the fourth layer) and purposes (second-layer) have been disclosed \textit{correctly} and \textit{completely}. 
    For data types (the third layer), each data type covers several \ignore{fixed}data items (e.g., \textit{Identifiers} covers two data items: User ID and Device ID) defined by Apple. 
    We did not scrutinize the first layer. This is because that for some data (e.g., vendor-specific data items like Apple device ID), linking data to the user's identity (e.g., via their account) are usually taken place at the server-side, which may not be precisely identified from client-side behavior. For example, although both Apple apps \textit{Contacts} and \textit{Apple news} collect a device identifier, 
    \LX{based on their privacy labels, the \textit{Contacts} app does not link it to user identity, while \textit{Apple news} news does.}
    %The Contacts app does not link data to your identity, while Apple news does link it to user. 
    %
    However, it can be hard to differentiate those two from client side.  \looseness=-1
    %Explained by Apple, \textit{``whether the device ID is linked to you or not linked to you depends on the nature of the ID and whether Apple will link it to user identity~\cite{privacylabelsupport}''.} 
}

%\vspace{2pt}\noindent\textbf{Privacy-label requirements.} 
%In our study, we defined the inconsistency model (Sectiosn XXX) considering XXXX.
%
 %In our study, we investigated the timeliness of the privacy label and found that XXXX (see Section XX). 

%Further, Apple defined certain criteria guiding optional disclosure~\cite{}. First, data that is processed only on device is not considered ``collected'' and does not need to be disclosed in the privacy label. Second, optional disclosure is allowed if the data satisfies \texit{all} of the following criteria: (1) not used for tracking, (2) not used for third-party advertising, (3) collected infrequently, (4) data provided by users with apparent awareness of collection. \XY{(5) Data collected via web traffic must be declared, unless redirect to browser}\XY{add policy of webview?}
%
%\LX{\sout{ In our study, we found that XXX piece of data collection behaviors in XXX apps only meet part of those criteria, but developers consider them as optional disclosure and choose not disclose them.}}

%Data types that meet all of the following criteria are optional to disclose:

\ignore{
    $\bullet${ Data that is processed only on device is not considered ``collected'' and does not need to be disclosed in the privacy label; Optional disclosure are required to meet all of the following  criteria: (1) not used for tracking purposes, (2) not used for Third-Party Advertising, (3) collect infrequently, (4) require user's awareness. }  In our study, we found that  \XY{XXX piece of data collection behaviors in XXX apps only meet part of those criteria, but developers consider them as optional disclosure and choose not disclose them.} 
}

%In short, today’s online privacy policies are failing consumers because finding information in them is difficult, consumers do not understand that there are differences between privacy policies, and policies take too long to read. We set out to design a clear, uniform, single-page summary of a company’s privacy policy that would help to remedy each of these three concerns.

%It has further been established through numerous studies that people do not read privacy policies and make mistaken assumptions based upon seeing that a site has a link to a privacy policy \cite{turow2005open}.

%A recent study estimated that if consumers were somehow convinced to read the policies of all the companies they interact with, it would cost an estimated 365 billion dollars per year in lost productivity \cite{mcdonald2008cost}

%privacy policies are: (1) inconspicuously posted; (2) poorly written; and (3) rarely read by website visitors

\ignore{
\subsection{iOS Security}
The iOS uses several mechanisms to secure the platform which includes Code signing, encryption and sandboxing. %
Code signing: If Apple approves of the application, it adds its own signature to the application and ensures only executables reviewed and distributed by Apple are allowed to run. 
Encryption ensures apps can’t be reverse-engineered, which means only decrypted app can be used to perform static analysis. %protecting the app developer’s investment, and that only the purchaser can launch the app.
The sandboxing mechanism ensures that individual apps cannot access others’ data or other parts of the file system by using a mandatory access control policy to limit the abilities of exploited or malicious code
}

\vspace{-8pt}
\subsection{Consistency Model}
\label{s:preConsisModel}
\ignore{
\xiaojing{definition of consistency model}
The consistency model is defined as a contract between rule-followers and rule-regulators. This model guarantees that the system works correctly if the rule-followers obey specific regulations. In the privacy domain, the rule-follower is usually a mobile app, and the rule-regulators are privacy policy. The consistency model defined a set of rules between them. If the actual data collection behavior in the mobile app follows those rules, it is considered consistent with its privacy policy. }

% regulation vs. practice
A consistency model is defined as a contract between regulators and rule-followers. In the privacy domain, a consistency model measures to what extent privacy practices (e.g., data collection behaviors in mobile apps) execute the rules (e.g., a privacy statement indicating which party collects what data) specified by the associated privacy disclosures (e.g., privacy policies)~\cite{andow2020actions,bui2021consistency}.
More specifically, a disclosure can correctly and completely indicate critical information in a privacy practice if the privacy practice follows all rules in the disclosure. Otherwise, an inappropriate disclosure occurs, indicating that certain rules are violated by the privacy practice %\LX{or even the rules are omitted in the disclosure (insufficient disclosure)}
.
The consistency model summarizes such inappropriate disclosures into several categories, according to the type of the rules obeyed by the privacy practice.
%
% 对于给定的规则，practice有没有遵守； 或对于给定的pracrtice， regulation有没有尽到告知义务
%This model enables the identification of the regulation obeyed by privacy practices to avoid unexpected data leakage or can be used to verify whether the informing obligation is fulfilled by the regulator. 

\begin{table*}[ht]

\renewcommand\arraystretch{1.5}
\centering
  \setlength{\abovecaptionskip}{0.cm}
\scriptsize

\setlength{\tabcolsep}{2pt}{% 调整列间距
\caption{Comparison with prior consistency models. }
 \label{tb:comparison}
\begin{tabular}{l|c|c|c|c}

\toprule
\multicolumn{2}{c|}{\diagbox{}{\textbf{Method}}}& \textbf{PoliCheck~\cite{andow2020actions}} &  \textbf{PurPliance~\cite{bui2021consistency}}  & \textbf{Our work} \\ %\hline 
%\hline
%{{Characteristics}}  & entity($e$)-sensitive & entity($e$)-and-purpose($q$)-sensitive & entity-sensitive-purpose($q$) \\ 
%\hline
%\makecell{{Differentiator} \\{of inconsistency}} & predicate sentiment: $c$ & predicate sentiment: $c, k$ &purpose portfolio: $\mathcal{P}$ \\ 
% \hline
\hline
\multicolumn{2}{l|}{Disclosure} & Privacy Policy &  Privacy Policy& Privacy Label \\ 
\hline
% \multicolumn{2}{l|}{Representation of}&\multirow{2}{*}{$\{s|s:(e,c,d)\}$} & \multirow{2}{*}{$\{s|s:((e, c, d),(d,k,q))\}$} &\multirow{2}{*}{$\{s|s:(d, q)\}$}\\
% \multicolumn{2}{l|}{disclosure}&&&\\
\multicolumn{2}{l|}{Disclosure Representation}&{$\{s|s:(e,c,d)\}$} &{$\{s|s:((e, c, d),(d,k,q))\}$} &{$\{s|s:(d, q)\}$}\\
\hline
\multicolumn{2}{l|}{Data Flow Representation}&{$\{f|f:(e, d)\}$} &{$\{f|f:(e,d,q)\}$} &{$\{f|f:(d, q)\}$}\\
\hline
% \multicolumn{2}{l|}{Representation of}&\multirow{2}{*}{$\{f|f:(e, d)\}$} & \multirow{2}{*}{$\{f|f:(e,d,q)\}$} &\multirow{2}{*}{$\{f|f:(d, q)\}$}\\
% \multicolumn{2}{l|}{data flows}&&&\\
% \hline
\multicolumn{2}{l|}{Definition of flow-$f$-}&
\multirow{2}{*}{\begin{tabular}{@{}c@{}}{$\mathbb{S}_f = \{s:(e_s,c_s, d_s)$}\\{$|d_f\sqsubseteq d_s \wedge  e_f\sqsubseteq{e_s}\}$}\end{tabular}} 
 &\multirow{2}{*}{\begin{tabular}{@{}c@{}}{$\mathbb{S}_f =\{s:((e_s, c_s, d_s),(d_s,k_s,q_s))$}\\{$|e_f\sqsubseteq e_s \wedge d_f\sqsubseteq d_s  \wedge q_f \sqsubseteq q_s\} $}\end{tabular}} &\multirow{2}{*}{$\mathbb{S}_f =  \{s:(d_s,q_{s})| d_{f}\sqsubseteq d_{s} \vee d_{f}\sqsupset d_{s}\}$}\\
\multicolumn{2}{l|}{related disclosure}&&&\\
% \hline
% \multicolumn{2}{c|}\makecell{{Representation of}\\ {data flows}} & $\{f|f:(e, d)\}$ & $\{f|f:(e,d,q)\}$ &$\{f|f:(d, q)\}$ \\ 
% \hline
% \multicolumn{2}{c|}\makecell{{Definition of flow-}\\{related disclosure}\\ } & \makecell{ $\mathbb{S}_f = \{s:(e_s,c_s, d_s)|d_f\sqsubseteq d_s \wedge  e_f\sqsubseteq{e_s}\}$}
% & \makecell{$\mathbb{S}_f =\{s:((e_s, c_s, d_s),(d_s,k_s,q_s))$\\$|e_f\sqsubseteq e_s \wedge d_f\sqsubseteq d_s  \wedge q_f \sqsubseteq q_s\} $} &$\mathbb{S}_f =  \{s:(d_s,q_{s})| d_{f}\sqsubseteq d_{s} \vee d_{f}\sqsupset d_{s}\}$\\
\hline

%\makecell{{Definition of Flow-}\\{Regulation consistency}\\ } & \makecell{ $\exists s \in \mathbb{S}_{f}$  s.t. $c_s=\operatorname{collect}  \wedge$\\$ \nexists s^  {\prime} \in$ $\mathbb{S}_{f}$  s.t.  $c_{s^{\prime}} =$ not\_collect}  
%& \makecell{$\exists s \in \mathbb{S}_{f}$ s.t. $c_s=\operatorname{collect},   k_s =\operatorname{for} $ \\ $\wedge$
%        $ \nexists s^{\prime} \in$ $\mathbb{S}_{f}$ s.t. \\ $c_{s^{\prime}}=$ not\_collect $\vee$ $k_{s^{\prime}}=$ not\_for} 
%& \makecell{ $\mathbb{S}_f\neq\emptyset \wedge  \{q_{s,i}|s\in \mathbb{S}_f\}= \{q_{f,j}\}$, \\ $P_{\mathbb{S}_f}=\{p |p \in \mathcal{P}_s, s\in \mathbb{S}_f\}$}\\ 
% %\hline
% \makecell{{Inconsistent}\\{ Disclosure Type}}   &\makecell{$\bullet$ Omitted Disclosure\\$\mathbb{S}_f=\emptyset$\\ \\$\bullet$ Incorrect Disclosure\\$\exists s \in \mathbb{S}_{f}$  s.t. $c_s=\operatorname{not\_collect} $}

% &\makecell{$\bullet$ Omitted Disclosure\\$\mathbb{S}_f=\emptyset$\\ \\$\bullet$ Incorrect Disclosure\\$\exists s \in \mathbb{S}_{f}$  s.t. $c_s=\operatorname{not\_collect} \vee k_s=\operatorname{not\_for} $}
% &\makecell{$\bullet$ Omitted Disclosure\\$\mathbb{S}_f=\emptyset$\\ \\ $\bullet$ Incorrect
% Disclosure\\$\mathbb{S}_f\neq\emptyset\wedge \{q_{s,i}|s\in \mathbb{S}_f\} \nsubseteq \{q_{f,j}\}\wedge  \{q_{f,j}\}\nsubseteq \{q_{s,i}|s\in \mathbb{S}_f\}$\\ \\$\bullet$ Inadequate Disclosure\\$\mathbb{S}_f\neq\emptyset\wedge \{q_{s,i}|s\in \mathbb{S}_f\} \subset \{q_{f,j}\}$}\\
% \hline

\multirow{2}{*}{\begin{tabular}[c]{@{}l@{}}
{Inconsistent}\\{Disclosure }\\{Type}\end{tabular}} &\makecell{Omitted \\Disclosure}    &\makecell{$\mathbb{S}_f=\emptyset$}&\makecell{  $\mathbb{S}_f=\emptyset$ }&\makecell{$\mathbb{S}_f=\emptyset$ (neglect disclosure);\\ $\mathbb{S}_f\neq\emptyset\wedge \{q_{s}|s\in \mathbb{S}_f\} \nsubseteq \{q_{f}\}\wedge  \{q_{f}\}\nsubseteq \{q_{s}|s\in \mathbb{S}_f\}$ (contrary disclosure);\\ $\mathbb{S}_f\neq\emptyset\wedge \{q_{s}|s\in \mathbb{S}_f\} \subset \{q_{f}\}$ (inadequate disclosure)}\\ 
\cline{2-5}
&\makecell{Incorrect \\Disclosure}   &\makecell{$\exists s \in \mathbb{S}_{f}$ \\  s.t. $c_s=\operatorname{not\_collect} $}&\makecell{$\exists s \in \mathbb{S}_{f}$ \\ s.t. $c_s=\operatorname{not\_collect} \vee k_s=\operatorname{not\_for} $}& N/A\\

\bottomrule

% \label{tb:comparison}
\end{tabular}}
\vspace{-10pt}
\end{table*}

%因为不emphasize sentiment contradiction, 所以在define flow-related disclosure时可以不match purposem, 所以可以对omitted disclosure 有更详细的归因

\ignore{

\begin{table*}[ht]
\renewcommand\arraystretch{1.5}
\centering
\footnotesize
\setlength{\tabcolsep}{2pt}{% 调整列间距
\caption{Comparisons with previous consistency models.}
\begin{tabular}{c|c|c|c}
\toprule
\diagbox{}{\textbf{Method}}& \textbf{PoliCheck} &  \textbf{PurPliance}  & \textbf{TagCheck} \\ \hline 
\hline
{{Characteristics}}  & entity($e$)-sensitive & entity($e$)-and-purpose($q$)-sensitive & entity-sensitive-purpose($q$) \\ 
\hline
\makecell{{Differentiator} \\{of inconsistency}} & predicate sentiment: $c$ & predicate sentiment: $c, k$ &purpose portfolio: $\mathcal{P}$ \\ 
\hline
\makecell{{Regulation}} & Privacy Policy &  Privacy Policy& Privacy Label \\ 
\hline
\makecell{{Representation of} \\{Regulation Rules}} & $\{s|s:(e,c,d)\}$ & $\{s|s:((e, c, d),(d,k,q))\}$ &$\{s|s:(d, \mathcal{P}),  \mathcal{P}=\{q\}\}$ \\ 
\hline
\makecell{{Representation of}\\ {Data flows}} & $\{f|f:(e, d)\}$ & $\{f|f:(e,d,q)\}$ &$\{f|f:(d, \mathcal{P}),  \mathcal{P}=\{q\}\}$ \\ 
\hline
\makecell{{Definition of Flow-}\\{related Regulation}\\ } & \makecell{ $\mathbb{S}_f = \{s:(e_s,c_s, d_s)|d_f\sqsubseteq d_s \wedge  e_f\sqsubseteq{e_s}\}$}
& \makecell{$\mathbb{S}_f =\{s:((e_s, c_s, d_s),(d_s,k_s,q_s))$\\$|e_f\sqsubseteq e_s \wedge d_f\sqsubseteq d_s  \wedge q_f \sqsubseteq q_s\} $} &$\mathbb{S}_f =  \{s:(d_s,\mathcal{P}_s)|d_s: d_{s}\sqsubseteq d_{f} \vee d_{s}\sqsupseteq d_{f}\}$\\
\hline

\makecell{{Definition of Flow-}\\{Regulation consistency}\\ } & \makecell{ $\exists s \in \mathbb{S}_{f}$  s.t. $c_s=\operatorname{collect}  \wedge$\\$ \nexists s^  {\prime} \in$ $\mathbb{S}_{f}$  s.t.  $c_{s^{\prime}} =$ not\_collect}  
& \makecell{$\exists s \in \mathbb{S}_{f}$ s.t. $c_s=\operatorname{collect},   k_s =\operatorname{for} $ \\ $\wedge$
        $ \nexists s^{\prime} \in$ $\mathbb{S}_{f}$ s.t. \\ $c_{s^{\prime}}=$ not\_collect $\vee$ $k_{s^{\prime}}=$ not\_for} 
& \makecell{ $\mathbb{S}_f\neq\emptyset \wedge  \{q_{s,i}|s\in \mathbb{S}_f\}= \{q_{f,j}\}$, \\ $P_{\mathbb{S}_f}=\{p |p \in \mathcal{P}_s, s\in \mathbb{S}_f\}$}\\ 
% \hline

% \makecell{{Flow-Regulation}\\ {Inconsistency Type} }  & \makecell{Narrowing Definitions\\
% $\mathbb{S}_{N_f}=\{s|s:(e_f\sqsupset e_s \wedge d_f\equiv d_s) $\\$\vee ((e_f\sqsubseteq e_s \vee e_f \sqsupset e_s)\wedge d_f \sqsupset d_s)\}$\\
% Logical Contradictions\\$\mathbb{S}_{C_f}=\{s|s:(e_f\sqsubseteq e_s \wedge (d_f\sqsubseteq d_s \vee d_f\approx d_s))$\\$\vee  (e_f\sqsupset e_s \wedge (d_f\sqsubset d_s \vee d_f \approx d_s))$\\
% $\vee (e_f\approx e_s \wedge(d_f\sqsubseteq d_s \vee d_f\sqsupset d_s \vee d_f \approx d_s))\}$ \\
% Omitted Disclosure\\$\mathbb{S}_f=\emptyset$\\ \\ Incorrect Disclosure\\$\exists s \in \mathbb{S}_{f}$  s.t. $c_s=\operatorname{not_collect} \vee $ \\ $ \mathbb{S}_{f}\neq\emptyset \wedge \mathbb{S}_{f} \cap\mathbb{S}_{N_f}\neq \emptyset \wedge \mathbb{S}_{f} \cap\mathbb{S}_{C_f}= \emptyset$ \\Ambiguous Disclosure\\$ \mathbb{S}_{f}\neq\emptyset \wedge \mathbb{S}_{f} \cap\mathbb{S}_{C_f}= \emptyset$\\
% }
% & \makecell{Narrowing Definitions\\
% $\mathbb{S}_{N_f}=\{s|s:(d_f\sqsupset d_s \wedge p_f\equiv p_s) $\\$\vee ((d_f\sqsubseteq d_s \vee d_f \sqsupset d_s)\wedge p_f \sqsupset p_s)\}$\\
% Logical Contradictions\\$\mathbb{S}_{C_f}=\{s|s:(d_f\sqsubseteq d_s \wedge (p_f\sqsubseteq p_s \vee p_f\approx p_s))$\\$\vee  (d_f\sqsupset d_s \wedge (p_f\sqsubset p_s \vee p_f \approx p_s))$\\
% $\vee (d_f\approx d_s \wedge(p_f\sqsubseteq p_s \vee p_f\sqsupset p_s \vee p_f \approx p_s))\}$} & ---\\
\hline
\makecell{{Inappropriate}\\{ Disclosure Type}}   &\makecell{$\bullet$ Omitted Disclosure\\$\mathbb{S}_f=\emptyset$\\ \\$\bullet$ Incorrect Disclosure\\$\exists s \in \mathbb{S}_{f}$  s.t. $c_s=\operatorname{not\_collect} $}
%\vee $ \\ $ \mathbb{S}_{f}\neq\emptyset \wedge \mathbb{S}_{f} \cap\mathbb{S}_{N_f}\neq \emptyset \wedge \mathbb{S}_{f} \cap\mathbb{S}_{C_f}= \emptyset$ \\$\bullet$ Ambiguous Disclosure\\$ \mathbb{S}_{f}\neq\emptyset \wedge \mathbb{S}_{f} \cap\mathbb{S}_{C_f}= \emptyset$\\}
&\makecell{$\bullet$ Omitted Disclosure\\$\mathbb{S}_f=\emptyset$\\ \\$\bullet$ Incorrect Disclosure\\$\exists s \in \mathbb{S}_{f}$  s.t. $c_s=\operatorname{not\_collect} \vee k_s=\operatorname{not\_for} $}
&\makecell{$\bullet$ Omitted Disclosure\\$\mathbb{S}_f=\emptyset$\\ \\ $\bullet$ Incorrect
Disclosure\\$\mathbb{S}_f\neq\emptyset\wedge \{q_{s,i}|s\in \mathbb{S}_f\} \nsubseteq \{q_{f,j}\}\wedge  \{q_{f,j}\}\nsubseteq \{q_{s,i}|s\in \mathbb{S}_f\}$\\ \\$\bullet$ Inadequate Disclosure\\$\mathbb{S}_f\neq\emptyset\wedge \{q_{s,i}|s\in \mathbb{S}_f\} \subset \{q_{f,j}\}$\\
\{q_{s,i}|s\in \mathbb{S}_f\}=$\{q_s|s\in \mathbb{S}_f\}$}, \mathcal{P}_{f}=\{q_{f,i}\}\\
%  \multirowcell{3}{Flow-Regulation\\ Inconsistency Type}   & \makecell{Omitted Disclosure\\$\mathbb{S}_f=\emptyset$\\ } & \multirowcell{1.5}{Logical Contradictions\\$asdfasdfasdfasdf$\\ } \\%\cline{2-2}\cline{3-3}
%  & \makecell{ Incorrect Disclosure\\$asdfasdfasd$\\ } &   \multirowcell{1.5}{Narrowing Definitions\\$asdfasdfasdfasdf$\\}&   \\%\cline{2-2}
%  & \makecell{Ambiguous Disclosure} & & \\
\bottomrule
% \hline
\label{tb:comparison}
\end{tabular}}

\end{table*}}
\ignore{
\begin{table*}[ht]
\centering
\footnotesize
\caption{consistency model comparison (TO DO)}
\begin{tabular}{c|c|c|c}
\hline
\hline

\textbf{-} & \textbf{PoliCheck} &  \textbf{PurPliance}  & \textbf{TagCheck} \\ \hline \hline

\makecell{Characteristics} & entity-sensitive & purpose-sensitive & purpose-sensitive \\ \hline

\makecell{Contradictions\\ rules example} & \makecell{$e_{i} \equiv \varepsilon e_{j} \wedge d_{i} \sqsupset_{\delta} d_{j}$ \\(Flurry, collect, Dev Info)\\(Flurry, not\_collect, IMEI) }  & \makecell{$d_{i} \equiv \delta d_{j} \wedge p_{i} \sqsubset_{\pi} p_{j}$\\(Device ID, for, Advertising)\\(Device ID, not\_for, Marketing)} & $\ominus$\\ \hline

\makecell{Flow consistency\\ definition} & \makecell{ $\exists s \in S_{f}$  \\ such that $s . c=\operatorname{collect} \wedge \nexists s^{\prime} \in$ $S_{f}$  \\such that $s^{\prime} . c=$ not\_collect.}  & \makecell{$\exists s \in S_{f}$ \\such that $s . c=\operatorname{collect}, s . k=\operatorname{for}, \wedge \nexists s^{\prime} \in$ $S_{f}$ \\such that $s^{\prime} . c=$ not\_collect $V$ $s^{\prime} . k=$ not\_for.} & \\ \hline

 \multirowcell{3}{Flow inconsistency \\ type}   & \makecell{omitted disclosure} & \multirowcell{3}{$\ominus$} \\\cline{2-2}
 & \makecell{ incorrect disclosure} & &  \\\cline{2-2}
 
 & \makecell{ambiguous disclosure} & &\\\hline

\end{tabular}
\label{table:ConsistencyModel}
\end{table*}
}

\noindent\textbf{Commonly-used inconsistency detection logic}.
An inconsistency detection logic captures the difference between a data flow in a privacy practice and its associated rules in the privacy disclosure.
%
%This formally specifies whether the privacy practice obeys the regulation rule, and how the rule is obeyed.
%
A \ignore{regulation }rule or a data flow can typically be formalized as a triplet in the form of $(e, c, d)$, where $e$ is subject, $c$ is predicate and $d$ is object.
Here, the predicate can be along with positive or negative sentiments.
%
%Previous works~\cite{andow2020actions,bui2021consistency} study the inconsistency logic between two such triplets where the sentiment (i.e., positive or negative) of the predicates are opposite.
%
For example, as shown in \autoref{tb:comparison}, PoliCheck \cite{andow2020actions} defines the subject $e$ as \textit{platform entities} (e.g., ﬁrst-party or third-party), the object $d$ as \textit{data objects} (e.g., email address), and the predicate $c$ as \textit{data collection action} with positive (i.e., collect) and negative (i.e., not collect) sentiments.
In addition, PurPliance~\cite{bui2021consistency} considered the data usage (i.e., purpose) in the consistency model, and extended the representation of a rule in privacy disclosures to be $((e,c,d), (d,k,q))$ where $k$ is the predicate associated with data usage (\ignore{\textit{data usage action}, }e.g., ``used for" and ``not used for"), $q$ is the purpose of data usage (e.g., advertising for third party). \looseness=-1

%defines the subject as \textit{data objects} (e.g., email address), the object as \textit{entity-sensitive purpose} (e.g., advertising for third party), and the predicate as \textit{data usage action} with positive (i.e., used for) and negative (i.e., not used for) sentiments.
%
Previous works study the inconsistency between privacy disclosure and data flows whose predicate sentiments $c, k$ are opposite.
More specifically, the prior inconsistency detection logic~\cite{andow2020actions,bui2021consistency} generally are defined between two representations with \textit{different} predicate sentiments $c$ and $k$, \textit{correlated} entities $e$, data objects $d$ and data usage $q$, i.e., with the definitions of flow-$f$-related disclosure $\mathbb{S}_f = \{s:(e_s,c_s, d_s)|d_f\sqsubseteq d_s \wedge  e_f\sqsubseteq{e_s}\}$~\cite{andow2020actions} or $\mathbb{S}_f =\{s:((e_s, c_s, d_s),(d_s,k_s,q_s))|e_f\sqsubseteq e_s \wedge d_f\sqsubseteq d_s  \wedge q_f \sqsubseteq q_s\} $~\cite{bui2021consistency}.
The correlation between $e, d, q$ of two representations is modeled using four semantic relations: synonym ($d \equiv d'$), approximation ($d \approx d'$), hyponym ($d \sqsubset d'$), and hypernym ($d \sqsupset d'$).
%and subsumption (  or $s' \sqsubset s$).
%
%Also, the objects $o, o'$ of two triplets have similar correspondence according to the four semantic relations.
%
%Based on the definitions of flow-related disclosure, PoliCheck and PurPliance propose 16 inconsistency logic, respectively, such as
%XXX\xiaojing{formal presentation} ([(Flurry, collect, DevInfo) vs.(Advertiser, not collect, DevInfo) ]) and XXX ([(DevID, for, Advertising) vs.(DevInfo, not for, Advertising) ]). 

It is worth noting that in privacy label, the data usage purposes of a privacy-sensitive data item are presented only with the positive sentiment.
In our study, we analyze inconsistency disclosure with specific focuses on data usage purpose (see Section~\ref{sec:model}) instead of predicate sentiment.

%\qy{privacy label: no contradiction in sentiment. --> to discuss: }

%compatible: inconsistency results from opposite sentiment are not excluded from our work.
% Some previous studies apply similar inconsistency logic

\noindent\textbf{Types of Inconsistency Disclosure.}
% ame predicate sentiment
%Consistency model categorizes different inconsistency logic into several types.
%, in order to reveal whether the regulation fulfills the informing obligation of a privacy practice.
% a privacy practice exceeds the regulation, 
Previous studies~\cite{andow2020actions,bui2021consistency} summarize several inconsistency types to inspect to what extent data flows can be disclosed by a privacy disclosure (see Table~\ref{tb:comparison}).
The typical inconsistency types include omitted disclosure and incorrect disclosure~\cite{andow2020actions,bui2021consistency}. 
\textit{Omitted disclosure} indicates a data flow which is not discussed by any policy statements (i.e., $\mathbb{S}_f=\emptyset$).
And \textit{incorrect disclosure} recognizes a data flow if a policy statement indicates that the flow will not occur (i.e., a negative sentiment collection statement) and there is not a contradicting positive sentiment statement (i.e., $\nexists s \in \mathbb{S}_{f}$  s.t. $c_s=\operatorname{collect} \vee k_s=\operatorname{for} $). 
In this paper, we did not include \textit{logical contradictions}, which is defined on a pair of policy statements in the disclosure that have contradictions.
This is because that compared with privacy policy, privacy label is a better-structured privacy disclosure where the logic contradicted privacy statement does not exist.
Since logical contradiction is out of scope of this work, the prior type of ambiguous disclosure~\cite{andow2020actions,bui2021consistency} will not be discussed in this paper.
%
%In contrast to previous work~\cite{andow2020actions,bui2021consistency}, as mentioned earlier, we are able to rationalize the privacy label compliance issue with a focus on inconsistent data usage purpose. 
%
In contrast to the previous works on ``flow-to-policy'' consistency, to analyze flow-to-label consistency, we need to focus on inconsistency of data usage purpose. 
Accordingly, we formally define three types of inconsistency for privacy labels : \textit{neglect disclosure, contrary disclosure  and inadequate disclosure} (Section~\ref{sec:model}) to fine-grain the type of \textit{Omitted disclosure} in PurPliance~\cite{bui2021consistency}.

\ignore{

\xiaojing{commonly-used consistency rules}

\vspace{2pt}\noindent\textbf{Contradictions rules in policy statements}. 
PoliCheck \cite{andow2020actions} proposed 16 types of conflicting policy statements in a privacy policy by taking into account the entities (third-party vs first-party) of the data collection. Each policy statement can be represented as $s$ is a tuple, $s=(e, c, d)$, where data type $d \in D$ is either collected or not collected, $c \in\{$ collect, not\_collect $\}$, by an entity $e \in E$. In each conflicting pair $(s_{1}=(e_{i},collect, d_{i}), s_{2}=(e_{j}, not\_collect, d_{j})$, the key conflict point is the action, one discusses collecting, while the other claims not collecting. The 16 conflicted rules comes from the combination of 4 semantic relationships between entities ($e_{i} \equiv_{\varepsilon} e_{j}$, $e_{i}\sqsubset_{\varepsilon} e_{j}$, $e_{i}\sqsupset_{\varepsilon}e_{j}$, $e_{i} \approx e_{j}$) and 4 semantic relationship between data ($d_{i} \equiv_{\delta} d_{j}$, $d_{i}\sqsubset_{\delta} d_{j}$, $d_{i}\sqsupset_{\delta}d_{j}$, $d_{i} \approx d_{j}$), where $\varepsilon$ is entity ontology, $\delta$ is data ontology,  $\equiv$ means semantic equivalence, $\sqsupset$ or $\sqsubset$ represents subsumptive relationship, $\approx $ means semantic approximation.
Different from PoliCheck, PurPliance \cite{bui2021consistency} defined 16 types of conflicting policy statements by incorporating data usage purposes. Each data usage can be represented as a turple $du=(d,f,p)$, where data type $d \in D$ is either for or not\_for, $f \in\{$ for or not\_for$\}$ an purpose $p \in P$. Similar with PoliCheck, a collision in the conflicting pair occurs at the ``action'', one claimed data collection for one purpose, while the other claimed not for that purpose. The 16 conflicted rules comes from the combination of 4 semantic relationships between data and purpose. 
A contradiction rule example is shown in table \ref{table:ConsistencyModel}.

\vspace{2pt}\noindent\textbf{Flow Consistency Model}. PoliCheck defined A data flow $f$ is consistent with an application’s privacy policy $S$ if and only
if and only if $\exists s \in S_{f}$ such that $s . c=\operatorname{collect} \wedge \nexists s^{\prime} \in$ $S_{f}$ such that $s^{\prime} . c=$ not\_collect. Further, PoliCheck defined three types of flow-to-policy inconsistency by considering contradictions rules in policy statements. The omitted disclosure means that a data flow $f$ can not find any policy statement at any semantic granularity for the flow’s data type and entity. The incorrect disclosure means there exists a policy statement that claimed it does not collect or share the data type that same in data flow $f$. The ambiguous disclosure of a data flow $f$ means there exists conflicting statements in privacy policy regarding this data flow. PurPliance has the  the similar definition of flow Consistency but add purpose alignment: A data flow $f$ is consistent with an application’s privacy policy $S$ if and only
if and only if $\exists s \in S_{f}$ such that $s . c=\operatorname{collect}, s . k=\operatorname{for}, \wedge \nexists s^{\prime} \in$ $S_{f}$ such that $s^{\prime} . c=$ not\_collect $V$ $s^{\prime} . k=$ not\_for. However, PurPliance didn't define the flow-to-policy inconsistency model. 

\xiaojing{the same and differences for our study}

Different from Policheck and PurPliance which define the consistency model on a single network flow, our consistency model compare flow to policy consistency at an data-aggregated level (purpose portfolio). We group flows sent the same data and a data item can have multiple purposes, represented as $(data, [purpose 1, purpose 2, purpose 3])$. As required by Apple, the app developer need to disclose all the purposes regarding to a single data item. Any missing, incorrect or inadequate purpose disclosure is considered as a violation to Apple policy. Hence, we defined three new inconsistency types based on aggregated purpose portfolio of data. 

\quotebox{Keep in mind that even if you collect the data for reasons other than analytics or advertising, it still needs to be declared}

Besides, different from contraction rules in the internal policy statement, Policy Label explicitly demonstrate which data is collected for which purpose and it does not have the negative action (``not collect'', ``not for''). Hence, we didn't consider the contractions rules in our consistency model. 
}
\subsection{Scope of Problem}
Our study focuses on the inconsistency between privacy label and data flows in iOS apps to check the non-compliance. %of privacy labels
Between the privacy policy and privacy label, we perform a new measurement study to understand their divergence.
We did not directly compare data flows with privacy policies since such a task has been conducted extensively in the literature~\cite{han2013comparing,razaghpanah2018apps,zimmeck2019maps,andow2020actions}. \looseness=-1
%We did not perform the comparison between data flow and privacy policy as it has been studied extensively in the past literature work~\cite{}. 

Regarding the four layers of privacy labels (\S~\ref{background:privacyLabel}), our study investigated whether the data items (the fourth layer) and purposes (the second-layer) have been disclosed \textit{correctly} and \textit{completely}.

For data types (the third layer), each data type covers several \ignore{fixed}data items (e.g., \textit{Identifiers} covers two data items: User ID and Device ID) defined by Apple. 
We did not scrutinize the first layer. This is because that for some data (e.g., vendor-specific data items like Apple device ID), linking data to the user's identity (e.g., via their account) are usually taken place at the server-side, which may not be precisely identified from client-side behavior. For example, although both Apple apps \textit{Contacts} and \textit{Apple news} collect a device identifier, based on their privacy labels, the \textit{Contacts} app does not link it to user identity, while \textit{Apple news} does.
%The Contacts app does not link data to your identity, while Apple news does link it to user. 
%
However, it can be hard to differentiate those two from client side.  \looseness=-1

\noindent\textbf{Scope of data items}. Apple defined 32 privacy-sensitive data items (the fourth layer in privacy labels), which if collected, should be disclosed in privacy labels~\cite{iosprivacylabel} (e.g., Device ID, Precise Location, referred to as \ldataItem{}s in our study). \ldataItem{}s \ignore{typically}originate from three sources:
(1) Apple system-level API (e.g., Device ID, Location, Contacts);
(2) app-level code including third-party libraries (e.g., User ID, Advertising Data, Crash Data);
(3) user input through graphical interface (e.g., Sensitive Info including sexual orientation, religious or philosophical beliefs, etc.).  In \S~\ref{sec:method}, we propose complementary static and dynamic analysis techniques, called \toolname{}, aiming at capturing data items from all three sources for compliance check of privacy labels. \looseness=-1

%a dynamic analysis on iOS apps aiming at capturing data items from all three sources (see Section XXX).

\ignore{
    In our study, we conduct a dynamic analysis on iOS apps aiming at capturing data items from all three sources (see Section XXX).
    \xiaojing{method for mapping here}.
    \XY{Specifically, the data items that originate from iOS system-level APIs are well-defined and traceable by monitoring invocation of system-level APIs built in Apple framework. Hence, similar to previous work~\cite{le2015antmonitor,song2015privacyguard,reyes2018won}, we inspect the description of API documentation and manually built a mapping between the return value and data items. If the return value can be matched to network traffic, we can pinpoint the collected data items. }
    \XY{In addition, for data items originated from either app-level code and user inputs, we infer the data items through the naming of key in network traffic. For example, the \textit{blood\_type} as a key can be
    semantically aligned to the data item \textit{Health}.}
    In this way, we recognize XX data items associated with XX data types as shown in Table XX.
    It is worth noting that the mapping between the data objects in the network traffic and privacy label's taxonomy has a limited coverage due to the vantage point of our study. \LXcomment{Just a reminder: Luyi plans to make some edits here.}
}

\ignore{

1. we didn't consider internal inconsistency in privacy policy

2. we compare flow to policy consistency at an data-aggregated level (purpose portfolio), a data item can have multiple purposes. We aggregate purposes based on each data item: $(data, [purpose 1, purpose 2, purpose 3])$ 

3. we defined three inconsistent type based on data-aggregated level.

In recent years, consistency rules are evolving from simple but coarse-grained and imprecise ones to complex but fine-grained and precise ones. The most simple rule only considers the data-level consistency, which checks whether the return value of sensitive API is literally described in the policy \cite{slavin2016toward,zimmeck2016automated}. However, without considering the semantic context around data (e.g., collect or not collect), the consistency check is error-prone. Hence, PPChecker \cite{yu2016can} integrated ``action'' (e.g., collect or not collect) into the consistency model and introduced an inconsistent case  where the privacy policy declares that the app will not collect user information, but the app does. 
Recent work (e.g., PoliCheck \cite{andow2020actions}) take into account the entities (third-party vs first-party) of the personal data and proposed an entity-sensitive consistency model. They introduce a new set of entity-sensitive conflicting rules. For example,  such a case is considered inconsistent if the data receiver is a third-party from a mobile app but declared in policy to be first-party.
PurPliance \cite{bui2021consistency} propose a more comprehensive and precise consistency model by extending the PoliCheck  \cite{andow2020actions} through incorporating data usage purposes. They define that A flow is consistent with a privacy policy if there exists a privacy policy statement where the data, entity and purpose are semantically subsumptive or equivalent  to those extracted from the flow.

Different from the traditional consistency model, which involves two parties (mobile app vs. privacy policy), our consistency model incorporates a new rule-regulator/follower (privacy label). The privacy label, as a rule-regulator, governs the code behavior in the iOS app. As a rule-follower, the privacy label should also be consistent with the privacy policy because it is considered as a rundown of the privacy policy. The best privacy practice in our consistency model is that app code behavior is consistent with the privacy label, and the privacy label is accordant with the privacy policy. For the consistency check, we reuse the comprehensive and precise consistency rules in PurPliance \cite{bui2021consistency}, which are sensitive to data, action, entity, and purpose. We focus on the collection of 32 data objects defined in the privacy Label and six entity-sensitive purposes (e.g., the Advertising purpose is finely divided into  Third-Party Advertising and Developer’s Advertising). Based on PurPliance, we subdivided the inconsistency into three types:  omitted disclosure, incorrect disclosure, and inadequate disclosure. 
}

\ignore{
Previous work purpose several consistency model between the actual behavior of mobile apps and their privacy policies. 
The consistency model is evolving. 

The most simple consistency model only considers whether sensitive data is disclosed in its privacy policy.  Previous work \cite{slavin2016toward,zimmeck2016automated} consider inconsistency by checking whether the return value of sensitive API is described in the policy (called \textit{Omitted disclosure}). 

Further, the ``action'' (e.g.,collect or not collect) is integrated in consistency model. 
PPChecker \cite{yu2016can} introduce the definition of ``Incorrect disclosure'' which means the privacy policy declares that the app will not collect user information but the app does. 

Recent work (e.g., PoliCheck \cite{andow2020actions}) take into account the entities (third-party vs first-party) of the personal data and proposed an entity-sensitive consistency model. They introduce a new set of conflicting entity-sensitive policy statements to be ``Ambiguous Disclosure''. 

PurPliance \cite{bui2021consistency} extend the PoliCheck consistency model  \cite{andow2020actions} by incorporating data usage purposes. They define that A flow is said to be consistent with a privacy policy if there exists a policy policy statement where the data, entity and purpose are semantically subsumptive or equivalent  to those extracted from the flow. 
}

\vspace{-5pt}
\section{New Inconsistency Model for Privacy Label}

  \begin{figure*}[!ht]
        \centering
         \vspace{-10pt}
        \includegraphics[width=\linewidth]{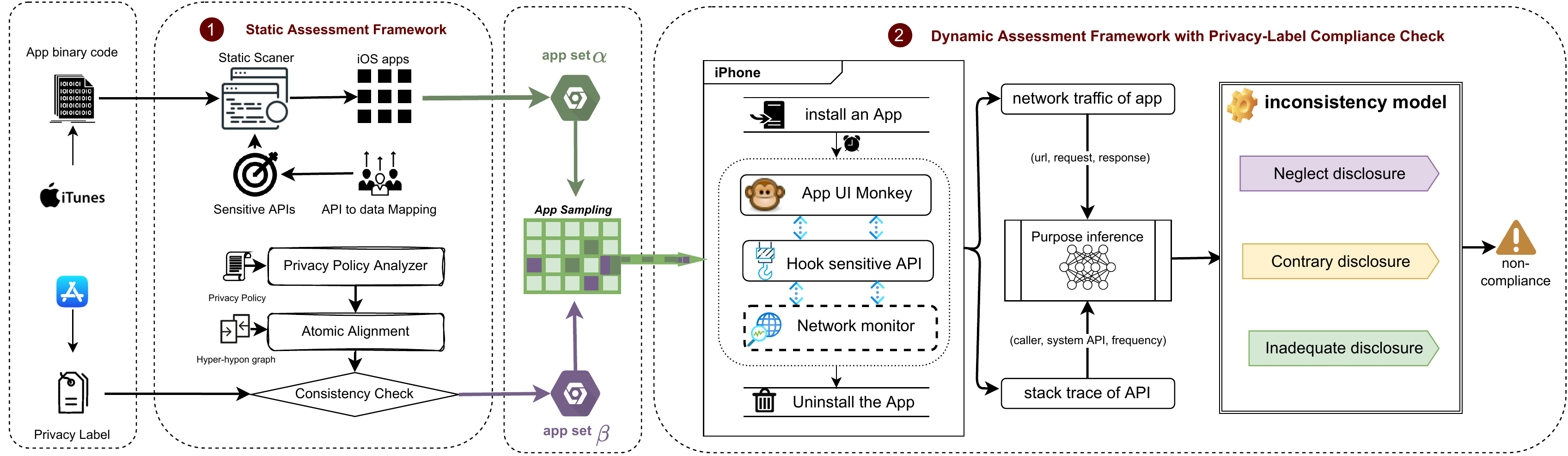}
        \caption{Overview of \toolname{}.}
         \vspace{-10pt}
        \label{f:overview}
    \end{figure*}
    
\label{sec:model}
%We denoted $D_{t}$ to be 32 data objects (e.g., \textit{Email Address, Phone Number etc.}) defined by Apple that should be disclosed in privacy label if collected. Let $P_{t}$ represent the 6 purposes (e.g., \textit{Third-Party Advertising, Analytics, etc.}) also defined by Apple. We model an app’s privacy label as a set of tuple $R(D_{t}, P_{t}) = \{r|r:(d_{r},p_{r}), d_{r} \in D_{t}, p_{r} \subset P_{t})\}$. Each tuple $r$ represent a piece of data collection practice which clearly decribe which data ($d_{r}$) is used for what purposes ($p_{r}$). Note that $p_{r}=\{p_{i}|p_{i}\in P_{t}\}$ can be a set of purposes as a data object can be used for multiple purposes. For example, ``Device ID" can be used for both \textit{Analytics} and \textit{Third-Party Advertising} purposes. 

%\LXcomment{make the terms consistent with Table 1; make the term data object/date item consistent across the paper}
\vspace{-10pt}
In this paper, we propose a new inconsistency model essential for analyzing \textit{privacy label}, which includes three types of inconsistent disclosure based on whether the usage\ignore{collection} purposes of a certain data object are \ignore{inadequately }\textit{omitted}, \textit{contrary}, or \textit{inadequate} in the privacy label. 
%, or completely \qy{change the order to: [completely, contrarily, or inadequately]?} omitted 
%
Compared with previous works~\cite{andow2020actions,bui2021consistency} which defined the consistency model on the predicate sentiment (see Section \ref{sec:background}), our consistency model takes full advantage of the succinct expression of the privacy label to fine-grain the \textit{ommitted disclosure issues} and rationalize them using inconsistent data usage purposes.
More specifically, we aggregate privacy statements in privacy labels and data flows in an iOS app with the correlated data objects (i.e., $\mathbb{S}_f =  \{s:(d_s,q_{s})| d_{f}\sqsubseteq d_{s} \vee d_{f}\sqsupset d_{s}\}$), and identify three types of inconsistencies between the data usage portfolios disclosed in privacy labels and actual data flows in iOS apps.
%and identify three types of inconsistency disclosures between the data usage portfolios of privacy labels and data flows in an iOS apps.
%Compared to the previous individual-level comparisons, our model enables the verification of whether an app developer correctly discloses \textit{all} the usage purposes of a certain data object in the privacy label, as required by Apple. 
We summarize the comparisons with the previous consistency models in Table~\ref{tb:comparison}.
%Specifically, we denote the 32 \textit{data objects} (e.g., \textit{Email Address, Phone Number etc.}) and 6 \textit{data usage purposes} (e.g., \textit{Third-Party Advertising, Analytics, etc.}) defined by Apple as $D$ and $P$, respectively. As required by Apple, such data objects, if collected by mobile applications, should be disclosed in the privacy label along with the associated usage purposes. 
%
Below we introduce the formal representation of the privacy label, data flows, and the semantic relationship between data objects. 
\vspace{-2pt}
\begin{definition}[Privacy Label Representation]
An app's privacy label is modeled as a set of tuples $S(D, Q) = \{s|s:(d_{s},q_s), d_{s} \in D, q_{s}\in Q\}$. Each tuple $s$ represents a privacy statement for a certain privacy-sensitive data item $d_s$, which discloses the purposes $q_s$ that $d_s$ is supposed to be used for (see Figure \ref{fig:privacyLabel}). Note that one data item can be associated with multiple data usage purposes, which is referred to as the data usage portfolio.
%Here, $\mathcal{P}_s=\{q|q\in P\}$ is the \textit{Purpose Portfolio} of data $d_s$ in $s$, which is a set since a data object can be used for multiple purposes. 
For example, ``Device ID"
can be used for both \textit{Analytics} and \textit{Third-Party Advertising}
purposes.
\end{definition}
\vspace{-5pt}
\begin{definition}[Data Flow Representation]
An individual data flow $f$ in an app is represented as a tuple $f=(d_f,q_{f})$, where $d_f \in D$ is a privacy-sensitive data object and $p_{f}$ is the usage purpose of $d_f$ in flow $f$. In mobile apps' data transfer, a certain data object can be associated with multiple data usage purposes in different flows.
%
%Similar to previous work~\cite{bui2021consistency}, we aggregate individual data flows with the same privacy-sensitive data object and different usage purposes into one flow: $\{f|f:(d, q)\}$ \qy{$f=\{f_i|f_i:(d_f,q_{f,i})\}?$ or just let $\{q_f\}$ denote all usage purposes of $d_f$? ( $\{f\}$ represents all flows in an app).}. %Here, $ \mathcal{P}_f=\{q_{f,i}\}$ is the purpose portfolio of $d_f$ with $i=1,2,\dots,K$, where $K$ is the total number of the usage purposes of $d_f$. 

% portfolio of data $d_f$, w
% flows in mobile apps' data transfer
% privacy label
% purpose portfolio
% A \textit{purpose portfolio} is the set of all purposes of a certain data object, summarized from the privacy label or the privacy practice.
\end{definition}
\vspace{-5pt}
\begin{definition}[Semantic Relationship]
Similar to previous work~\cite{andow2020actions,bui2021consistency}, we also use an ontology of data items to capture the relationship between data items (e.g., \texttt{DeviceInfo} is a hypernym of \texttt{DeviceID}).
Given an ontology $o$ and two terms $u, v$, we denote $u\equiv_{o}v$ if $u, v$ are \textit{synonyms} with the same semantic meaning. 
Otherwise, if $u$ is a general term and $v$ is a specific term whose semantic meaning is included in $u$, we denote $v \sqsubset u$ or $u \sqsupset v$. 
In this case, $u$ is called a \textit{hypernym} of $v$, and $v$ is called a \textit{hyponym} of $u$. $u\sqsubseteq_o v$ is equivalent with $u\sqsubset_o v \vee u\equiv_o v$.
\end{definition}
Given the flow $f$ associated with data item $d_f$, we define the flow related privacy label as below. By comparing the purpose portfolio of the flows (i.e., $\{q_f\}$) and their relevant privacy labels (i.e., $\{q_s|s\in\mathbb{S}_f\}$), we reveal the flow-to-label inconsistencies and categorize them into three types of inconsistency disclosures: \textit{Neglect Disclosure}, \textit{Contrary Disclosure}, and \textit{Inadequate Disclosure}. %Note that the three types of inconsistency disclosures can be considered as instances of Omitted Disclosure in~\cite{bui2021consistency}.
\vspace{-2pt}
\begin{definition}[Flow-relevant Privacy Label]
The privacy disclosure $s=(d_s, q_s)$ for data object $d_s$ are relevant to the flow $f=(d_f, q_f)$ (denoted as $s \simeq f$) if and only if $d_{s}\sqsubseteq_o d_{f} $ or $ d_{s}\sqsupset_o d_{f}$. Let $\mathbb{S}_f =  \{s|s\in S \wedge s\simeq f\}$ denote the set of flow-$f$-relevant privacy label. 
\end{definition}
\vspace{-5pt}
\begin{inconsistency}[${[\nVDash]}$ Neglect Disclosure] The privacy label $S$ is an \textit{neglect disclosure} with regard to a flow $f$ if there exists no  flow-$f$-relevant privacy label in $S$:
\vspace{-5pt}
\[
\mathbb{S}_f=\emptyset \ \Rightarrow \ R \nVDash s.
\]
\end{inconsistency} 
\vspace{-5pt}
\begin{inconsistency}[${[\nvDash]}$ Contrary Disclosure] 
The privacy label $S$ is an \textit{contrary disclosure} with regard to a flow $f$ if the flow-$f$-relevant privacy label $\mathbb{S}_f$ exists while the purpose portfolio in $\mathbb{S}_f$ disaccords the purpose portfolio in the flow:
\[
\mathbb{S}_f\neq\emptyset\wedge \{q_{s}|s\in \mathbb{S}_f\} \nsubseteq \{q_{f}\}\wedge  \{q_{f}\}\nsubseteq \{q_{s}|s\in \mathbb{S}_f\} \  \ \Rightarrow \ R \nvDash s.
% \mathcal{R}(s)\neq \emptyset \ \wedge\ \exists p \in \mathcal{P}_{\mathcal{R}(s)}, p \notin p_s   \wedge  p_s \nsubseteq \mathcal{P}_{\mathcal{R}(s)} \  \ \Rightarrow \ R \nvDash s
\]
Here, \textit{disaccords} means some of the data usage purposes in the flows are not included in the privacy label
%\qyue{the privacy label}%$\mathcal{P}_{\mathbb{S}_f}$ %
%%%%%%%%contains usage purposes that are not included in 
%$\mathcal{P}_f$
%\qyue{the flow}
, and vice versa.
\end{inconsistency}
\vspace{-5pt}
\begin{inconsistency}[${[\nvdash]}$ Inadequate Disclosure] 
The privacy label $S$ is an \textit{inadequate disclosure} with regard to a flow $f$ if the  flow-$f$-relevant privacy label $\mathbb{S}_f$ exists while the purpose portfolio in $\mathbb{S}_f$  is a \textit{proper subset} of  the purpose portfolio in the flow:
\vspace{-5pt}
\[
\mathbb{S}_f\neq\emptyset\wedge \{q_{s}|s\in \mathbb{S}_f\} \subset \{q_{f}\}\  \ \Rightarrow \ R \nvdash s.
% \mathcal{R}(s)\neq \emptyset \ \wedge\ \mathcal{P}_{\mathcal{R}(s)} \subset p_s \  \ \Rightarrow \ R \nvdash s
\]
This indicates the privacy label correctly reflects \textit{partial} rather than all usage purposes of the data objects related with $d_f$. Note that in our application (see \S~\ref{sec:inadequate disclosure}), it is possible that the usage purposes programmatically extracted from the data flows are not complete, while this will not erroneously result in false positives of inadequate disclosure.
\end{inconsistency}

\ignore{

% \subsection{Inconsistency Model between \LX{Privacy Label and Privacy Policy}}

\vspace{2pt}\noindent\textbf{Inconsistent type.} Ideally, for each data collection and purpose $s=(d_{s},p_{s})$ in privacy policy, if $d_{s}$ can be aligned with any data type $d_{r}$ in $D_{t}$, then $(d_{r},p_{s})$ should be disclosed in privacy label. This is a consistent disclosure between privacy policy and privacy label based on Apple's policy \textit{``The developer need to provide information about your app’s privacy practices''}.
%The purpose of the label is to help your customers understand what data is collected from your app and how it is used.

%   
Inconsistency can occur at both data level and purpose level. Here, we defined three inconsistency types. They are \textit{Omitted Disclosure}, \textit{Incorrect Disclosure}, and \textit{Inadequate Disclosure} respectively. 
%
%Given an app's privacy policy, we first extract a set of $S$ where $s = (d_{s},p_{s})$, $s \in S$. Given an app's privacy label, we extract a set of $R$ where $r = (d_{r},p_{r})$, $r \in R$.

------------------

Given a privacy policy statement $s = (d_{s},p_{s})$, we define its \textit{associated privacy label} as
$\mathcal{R}(s) = \{r:(d_r,p_r)|d_r: d_{s}\equiv_{o}d_{r} \vee d_{s}\sqsubset_{o}d_{r} \vee d_{s}\sqsupset_{o}d_{r}\}$, where the data object in the associated privacy label is either a synonym, a hypernym, or a hyponym of the data object  in the privacy policy statement. 
Let $\mathcal{P}_{\mathcal{R}(s)} = \{p_r|(d_r,p_r)\in \mathcal{R}(s)\}$ denote the purposes in the associated privacy label of statement $s$.
According to the results of the atomic alignment on the data objects and the purposes, we define three inconsistency relationship between a privacy policy statement $s$ and the privacy label $R$ as below. 

\begin{inconsistency}[${[\nVDash]}$ Omitted Disclosure] The privacy label $R$ is an \textit{omitted disclosure} with regard to a privacy policy statement $s$ if there exists no associated privacy label for the statement: 
\[
\mathcal{R}(s) = \emptyset \ \Rightarrow \ R \nVDash s
\]
\end{inconsistency} 
\begin{inconsistency}[${[\nvDash]}$ Incorrect Disclosure] 
The privacy label $R$ is an \textit{incorrect disclosure} with regard to a privacy policy statement $s$ if the associated privacy label $\mathcal{R}(s)$ exists while the purposes in $\mathcal{R}(s)$ \textit{disaccord} with those in the statement:
\[
\mathcal{R}(s)\neq \emptyset \ \wedge\ \exists p \in \mathcal{P}_{\mathcal{R}(s)}, p \notin p_s   \wedge  p_s \nsubseteq \mathcal{P}_{\mathcal{R}(s)} \  \ \Rightarrow \ R \nvDash s
\]
\end{inconsistency}

\begin{inconsistency}[${[\nvdash]}$ Inadequate Disclosure] 
The privacy label $R$ is an \textit{inadequate disclosure} with regard to a privacy policy statement $s$ if the associated privacy label $\mathcal{R}(s)$ exists while the purposes in $\mathcal{R}(s)$ is a \textit{proper subset} of those in the statement:
\[
\mathcal{R}(s)\neq \emptyset \ \wedge\ \mathcal{P}_{\mathcal{R}(s)} \subset p_s \  \ \Rightarrow \ R \nvdash s
\]

\end{inconsistency}

\qy{explain inadequate disclosure.}
% 针对哪种情况。更关心privacy label的漏报。对data flow的分析不全面不会导致误报。

}

\vspace{-10pt}
\section{Methodology}\label{sec:method}

    % \begin{figure*}[!ht]
    %     \centering
    %      \vspace{-10pt}
    %      % \setlength{\abovecaptionskip}{0.cm}
    %     \includegraphics[width=.8\linewidth]{figure/overview.pdf}
    %     \caption{Overview of \toolname{}.}
    %      \vspace{-10pt}
    %     \label{f:overview}
    % \end{figure*}

%apply a suite of decodings to the traffic flows

% ad mediation services provided the location data embed- ded within the ad link.

\vspace{-5pt}
\subsection{Overview}
\label{sec_method_overview}

In this section, we elaborate on the design and implementation of \toolname{}, a methodology for discovering privacy-label non-compliance from real-world iOS apps \textit{at scale}. 

\vspace{2pt}\noindent\textbf{Architecture}.
Our approach relies on
the extraction of data-purposes pair from dynamic code behavior and analysis of inconsistency with its corresponding privacy label. In particular, the design
of \toolname{} includes three major components: data collection and preprocessing, Static Assessment Framework (\staticFramework{}, \S~\ref{sec_static_framework}), Dynamic Assessment Framework (\dynamicFramework{}, \S~\ref{sec_dynamic_framework}), as outlined in \autoref{f:overview}. 
First, \toolname{} collects a large set of  366,685 iOS apps\ignore{(based on their app IDs we collected, detailed in \S~\ref{sec_data_collection})} from Apple app store (U.S.) from October 29, 2021 to April 26, 2022, with their privacy labels and privacy policies simultaneously collected.
Second, the \staticFramework{} analyzes the app binaries and their privacy labels and policies, and outputs two app sets: (1) apps that use the iOS system APIs whose return data, if collected by the apps, should be disclosed in iOS privacy labels --- denoted as $\alpha$ including 161,262  apps; (2) apps whose privacy label and privacy policy have inconsistent data practice disclosures (regarding data collection and usage purposes) --- denoted as $\beta$ including 164,056 apps.
Last, based on the apps of interest (i.e., sets $\alpha$ and $\beta$), the \dynamicFramework{} includes a dynamic analysis pipeline which performs end-to-end execution (fully automated app UI execution, dynamic instrumentation, and network monitoring). The pipeline strategically takes a subset of the apps for full execution to maximize the precision of detection for privacy-label non-compliance while maintaining the scalability of the study (detailed in \S~\ref{sec_dynamic_framework}). Based on the context that features the data practices (e.g., API caller, stack traces, server endpoints, Web requests, and responses) collected by the pipeline, \dynamicFramework{} infers the actual data collected and purposes of collection, and aligns them, abstracted as tuples $(data, purpose)$, to the 32 data items and 6 purposes defined by Apple. \dynamicFramework{} finally reports privacy label non-compliance by comparing those tuples with disclosures in the vendors' privacy statements.

\vspace{2pt}\noindent\textbf{Data collection}. In our research, we collected 366,685  iOS apps for privacy label compliance checks. Specifically,  we reuse the more than one million Android app names from prior works~\cite{wang2021understanding}, assuming that the iOS and Android versions of an app might share similar names. This approach turned out to be efficient: we collected 485,024 unique app Bundle IDs from the Apple app store (U.S.). Given those Bundle IDs, we ran an app crawler on 10 iPhones to download and decrypt those apps. In this way, we collected 366,685 iOS apps, which cover 27 app categories. Meanwhile, for each iOS app, we also collect its privacy label and privacy policy. \looseness=-1

\ignore{
    Our approach relies on
    the extraction of data-purposes pair from dynamic code behavior and analysis of inconsistency with its corresponding privacy label. In particular, the design
    of \toolname{} includes three major components: data collection, static privacy label assessment (includes static analysis and comparison between privacy policy and privacy label), dynamic analysis, as outlined in Figure~\ref{f:overview}. 
    First, \toolname{} collects a batch of iOS App bundle IDs, which further used to download iOS apps through iTunes and crawl corresponding privacy label from App store longitudinally from Oct 29, 2021 to Apr 26, 2022.
    Further, the static analyzer takes app binary code and target system APIs as its input, and outputs candidate apps with the target APIs existing in their binary code;  the PP vs. PL comparer also outputs candidate apps whose privacy labels diverge from its privacy policy regarding data collection and usage disclosures. 
    The intersection between those two sets takes candidate apps into three different sets ($\mu$, $\gamma$, $\nu $) in order to describe different levels of data inconsistency in apps as detailed in ~\ref{s:sampling}. Next, \toolname{} perform app sampling from $\mu$, $\gamma$, and $\nu $ respectively.
    Finally, the dynamic analyzer execute each sampled app by using an UI automation tool and meanwhile, collect the callback traces of sensitive system APIs and network traffics in order to infer data collection purposes. By comparing with its privacy label, \toolname{} can output the privacy-label violated apps. 
}

\ignore{
\vspace{3pt}\noindent\textbf{Example}.
We use a real app \textit{Atlanta News from 11Alive}~\cite{overivewExample} to illustrate the three types of inconsistencies (\S~\ref{sec:model}) \toolname{} found between its privacy label and actual data practices, indicating non-compliance:

\xiaojing{update this part based on our discussion}
%Table~\ref{table:overviewExample}.

\vspace{1pt}\noindent$\bullet$\textit{ Neglect Disclosure.} The app transmits a \textit{User ID} along with other diagnostics data (e.g., whether the app runs on a jail-broken iPhone) are transmitted to \textit{google-analytics.com}, indicating the purpose of analytics (based on endpoint functionalities and other key features, see \S~\ref{}). However, \toolname{} found that the \textit{User ID} is neglected to be disclosed in its privacy label.  
In the meantime, \toolname{}'s \staticFramework{} found that the app's privacy label and privacy policy are consistent (regarding the date types/items framed by Apple privacy label), indicating dual non-compliance of the app's privacy disclosure.

%Additionally, PLCC doesn't find related-policy statements generated from its privacy policy, which indicated the incomplete  privacy policy may lead to missing disclosure in privacy label. 

\vspace{1pt}\noindent$\bullet$\textit{ Contrary Disclosure.} Based on traffic endpoint, Web requests and responses (and other features detailed in \S~\ref{}), \toolname{} reports the \textit{Device ID} being transmitted to Google advertising network and attributes its purpose being \textit{Third-Party Advertising}. Contrary to the actual purpose, the app's privacy label discloses the usage purpose of \textit{Device ID} as \textit{App functionality} --- a purpose that is considered easier to convince users to provide personal data~\cite{}. 
Meanwhile, \toolname{}'s \staticFramework{} found that the app's privacy label is inconsistent from privacy policy (in disclosing the purpose), which includes texts \textit{``we sell Device data to Advertising partners to facilitate online advertising...''}. Notably, compared to privacy label, the privacy policy language is often non-structured and opaque, still failing to concretely indicate the data item (\textit{Device ID} versus ``Device data'') when describing the purpose.

%Additionally, compared with related-policy statements generated from its privacy policy, privacy label
 %is also incorrectly disclose \textit{Device data } considering the following sentence \textit{``we sell Device data to Advertising partners to facilitate online advertising...''}. This indicates that the privacy label  failed to summarize data collection practices from privacy policy.  

\vspace{1pt}\noindent$\bullet$\textit{ Inadequate Disclosure.}
The \textit{Precise Location} is transmitted to two receivers: one is send to its own endpoint \textit{api.tegnadigital.com}, which is used to provide local news event based on user location for App functionality purpose; while the other one is sent to amazon-ads system for third-party advertising purpose; However, the app developer only disclosed App functionality purpose for precise location and inadequately disclose third-party advertising purpose. 
Additionally, compared with related-policy statements generated from its privacy policy, \textit{Precise Location} is also inadequately disclose considering the following sentence \textit{``we and third-party partners may automatically log precise geolocation about you if you permit our Services to access it...''}. This indicates that the privacy label also failed to summarize data collection practices from privacy policy.

\textit{Our results shed lights on the dure consistency goals among apps, privacy labels and privacy policies} (detailed in \S~\ref{}).
}
\ignore{
    There is an app, called \textit{Atlanta News from 11Alive}~\cite{overivewExample}, which provides newscasts from local events to users,  have three inconsistency problems at the same time as shown in Table~\ref{table:overviewExample}.
    
    \vspace{1pt}\noindent$\bullet$\textit{Neglect Disclosure.} The \textit{User ID} along with other diagnostics data (e.g.,jailbreak detection) are transmitted to \textit{google-analytics.com} for analytics purpose. However, the \textit{User ID} is omitted disclosed in its privacy label.  
    Additionally, PLCC doesn't find related-policy statements generated from its privacy policy, which indicated the incomplete  privacy policy may lead to missing disclosure in privacy label.

    \vspace{1pt}\noindent$\bullet$\textit{Contrary Disclosure.} The \textit{Device ID} is transmitted to google advertising network for third-party advertising purpose. While in this privacy label, it incorrectly disclose \textit{Device ID} is used to App functionality. 
    Additionally, compared with related-policy statements generated from its privacy policy, privacy label
     is also incorrectly disclose \textit{Device data } considering the following sentence \textit{``we sell Device data to Advertising partners to facilitate online advertising...''}. This indicates that the privacy label  failed to summarize data collection practices from privacy policy.

    \vspace{1pt}\noindent$\bullet$\textit{Inadequate Disclosure.}
    The \textit{Precise Location} is transmitted to two receivers: one is send to its own endpoint \textit{api.tegnadigital.com}, which is used to provide local news event based on user location for App functionality purpose; while the other one is sent to amazon-ads system for third-party advertising purpose; However, the app developer only disclosed App functionality purpose for precise location and inadequately disclose third-party advertising purpose. 
    Additionally, compared with related-policy statements generated from its privacy policy, \textit{Precise Location} is also inadequately disclose considering the following sentence \textit{``we and third-party partners may automatically log precise geolocation about you if you permit our Services to access it...''}. This indicates that the privacy label also failed to summarize data collection practices from privacy policy. 

}

%\vspace{2pt}\noindent\textbf{Ethical discussion}.

\ignore{
\subsection{Data Collection and Preprocessing}
\label{sec_data_collection}

Our dataset consists of XXX iOS apps with their corresponding privacy labels and privacy policies (Table~\ref{}). We elaborate on our dataset and the collection process below. 
    
%Our dataset consists of iOS apps, privacy label corpus and privacy policy corpus, as show in table xxx. Further, We elaborate our data collection. 

\vspace{2pt}\noindent\textbf{iOS app collection.} We collected XXX iOS apps from the official Apple app store (U.S.). %by building a crawler based on the prior tool~\cite{}. 
Using the prior tool~\cite{}, we download iOS apps by providing the app Bundle ID~\cite{bundleID} (identifier of iOS apps similar to package names on Android~\ref{}) to the Apple iTunes API \texttt{XXX} and then decrypt the executable of the iOS apps on jailbroken iOS devices.
Notably, Apple did not release the list of iOS apps, and, thus, to collect iOS apps for our study, we need to first collect a large number of app Bundle IDs. In our study, we gather app Bundle IDs by providing the parameter \texttt{app\_name} to the Apple iTunes Search API~\cite{searchAPI}: \nolinkurl{http[:]//itunes.apple.com/search?term=app\_name&media=software}. The API was designed for searching iOS apps based on names or keywords and will return a set of iOS app bundle IDs whose app names are similar to the provided \texttt{appname}.

To obtain possible \textit{appname}s for iOS at scale, we reuse the more than one million Android app names from prior works~\cite{wang2021understanding}, considering that the iOS and Android versions of an app might share similar names.
    Indeed, for example, when we pass \textit{Radio Pakistan}, an Android app name, to the above API, it will return 36 bundle IDs of iOS apps including \textit{Radio Pakistan, Radio Pakistan Record FM \& AM, Radio Pakistan: Online FM, FM100 Pakistan, Pakistani Radios, Pakistan Radios Pro 1.0, etc.} Our approach turned out to be efficient: we collected 485,024 unique Bundle IDs from the US market using xxx devices within xxx days. Providing the bundle IDs to~\cite{}, we downloaded and successfully decrypted XXX iOS apps (with their executables as .ipa~\cite{} file). %For each App, we will record its app name, app ID, Bundle ID, version and the account which purchase this app. 

%To collect iOS apps, we need to collect iOS app Bundle IDs~\cite{bundleID} (identifier of iOS apps similar to package name on Android~\ref{}), download the apps from iTunes based on the ID, and decrypt the apps by adopting the approach~\cite{iOSNetworkService_Security20}. Notably, Apple did not release the list of iOS apps or their Bundle IDs, and, thus, to collect a large number of iOS apps, we need to

\ignore{
    Each iOS application can be identified by a unique bundle ID. For example, The \textit{Messenger} is identified by the bundle ID (com.facebook.Messenger) and can be accessed from iTunes by using this ID. The bundle ID can be obtained by the the iTunes Search API \cite{iosAPI}, if we can provide the parameter \textit{``appname"}: \url{url = "http://itunes.apple.com/search?term=\%s&media=software"\%(appname).}
    To obtain all the possible \textit{appname} in iOS, we reuse the app names of 1.3M Android apps in \cite{wang2021understanding}. Our assumption is that the same app in iTunes and Google Play usually use the same app name or share at least have some similar terms. 
    For example, if we pass \textit{Radio Pakistan}, an Android app's name,  as a parameter into the above URL, it will return 36 bundle IDs of iOS apps including \textit{Radio Pakistan, Radio Pakistan Record FM \& AM, Radio Pakistan: Online FM, FM100 Pakistan, Pakistani Radios, Pakistan Radios Pro 1.0, etc.} It's a super-efficient way to collect the bundle ID. Currently, we have already collected 485,024 unique Bundle IDs from both US market. 
    
    After obtaining the bundle ID of an App, we can purchase and download a DRM protected .ipa file from iTunes buy using the tool XXX. In total, we use XXX apple accounts successfully download XXX .ipa file. For each App, we will record its app name, app ID, Bundle ID, version and the account which purchase this app. 
}

\vspace{3pt}\noindent\textbf{Privacy label collection}
%\label{dataCollection:privacyLabel}
The privacy label for each iOS app, if provided by the app developers, can be found from the app's App Store page, for example, \url{https://apps.apple.com/us/app/xxx} in which xxx is the app Bundle ID. We implemented a crawler that gathered the privacy labels for all the \appNumber{} iOS apps in October, 2021.
% under the Web tag~\cite{} named App Privacy
Further, in order to conduct a longitudinal study (see \S~\ref{}), we crawled the privacy labels for all the \appNumber{} apps weekly for six months (from Oct 29, 2021 to Apr 26, 2022). In total, we gathered the privacy labels for XXX iOS apps (see our automatic semantic analysis in \S~\ref{} and measurement results in \S~\ref{}).

%of the privacy label of those \appNumber{} apps, our crawler per collected all their privacy labels 

%More specifically, we first characterized the link for each app by adding the app Bundle ID to App Store url: \url{https://apps.apple.com/us/app/id{app ID}}. 

In our implementation, our crawler is based on Python and utilized the Selenium module ~\cite{XXX} to launch browsers and open the app's app store link. After the Web page is loaded, the crawler will locate the keyword ``See Details'' under App Privacy tag~\cite{} of the Web page and click it to get the full privacy label, along with other information including the developer’s privacy policy link. 
%
%Finally the raw page data are parsed to extract privacy label for each app. Altogether, we collected XXX privacy label that contains at lease one highest level (see section ~\ref{background:privacyLabel}), as well as XXX Data Not Collected and XXX Not provide details.

%Privacy label for each app can be found from that app's web page under App Privacy tag in Apple App Store. In order to conduct a longitudinal study of the privacy label of those 485k unique apps, we designed a scraping framework to implement a whole-site crawl weekly for six months from Oct 29, 2021 to Apr 26, 2022.
%

%More specifically, we first characterized the webpage link for each app by adding the unique app ID to App Store url: \url{https://apps.apple.com/us/app/id{app ID}}. We implemented the crawler using Python and utilized the Selenium module ~\cite{XXX} to launch browsers and to send crawling requests. After the app's page is loaded, the crawler will locate the keyword ``See Details'' under App Privacy tag and click it to get the full list of privacy label, along with other information (i.e., Apple's statements, developer’s privacy policy link, etc.). In total, we took XXX snapshots for XXX unique apps.
%
%Finally the raw page data are parsed to extract privacy label for each app. Altogether, we collected XXX privacy label that contains at lease one highest level (see section ~\ref{background:privacyLabel}), as well as XXX Data Not Collected and XXX Not provide details.

\vspace{3pt}\noindent\textbf{Privacy policy collection.}
%\label{dataCollection:privacyPolicy}
The link to an iOS app's privacy policy can be found next to the privacy label on the app's App store page. Every time we crawled a privacy label (see above), our crawler looked for and tried to gather the privacy policy of the same app immediately. In total, we found and gathered the privacy policies for xxx iOS apps (see our automatic semantic analysis in \S~\ref{} and measurement results in \S~\ref{}). We then parsed raw Web pages of privacy policies to extract plain texts using the tool HtmlToPlaintext~\cite{}.

%To gather privacy policy data provided by the app developers, we weekly crawled the developer’s privacy policy of the same apps along with the privacy label synchronously.
%
%The developer's privacy policy link can be retrieved from the parsed result in section ~\ref{dataCollection:privacyLabel}. In total, we took XXX snapshot for XXX unique apps. Note that one privacy policy might be used by multiple apps. For instance, XX privacy policy has been clarified by XXX unique apps.
%We then parsed raw page of privacy policy to plain text using the model HtmlToPlaintext~\cite{}. 

\ignore{
    To gather privacy policy data provided by the app developers, we weekly crawled the developer’s privacy policy of the same apps along with the privacy label synchronously.
    The developer’s privacy policy link can be retrieved from the parsed result in section ~\ref{dataCollection:privacyLabel}. In total, we took XXX snapshot for XXX unique apps. Note that one privacy policy might be used by multiple apps. For instance, XX privacy policy has been clarified by XXX unique apps.
    We then parsed raw page of privacy policy to plain text using the model HtmlToPlaintext~\cite{}. 
}
}
%\subsection{Static Privacy Label Assessment}
\vspace{-5pt}
\subsection{Static Assessment Framework} %for Privacy Label
\label{sec_static_framework}
\vspace{-5pt}
%Although \toolname{} eventually leverages dynamic analysis based approaches (\S~\ref{}) to check inconsistencies between app data practices and privacy labels, and between 

As mentioned earlier (\S~\ref{sec_method_overview}), the \staticFramework{} takes in our data set (app binaries with their privacy labels and privacy policies) and finds out two sets of apps, i.e., $\alpha$ (apps using the iOS system APIs returning \ldataItem{}s) and $\beta$ (apps with inconsistency between privacy label and privacy policy). Those apps will be further validated through dynamic analysis in \S~\ref{sec_dynamic_framework}. To this end, \staticFramework{} includes two sub-components: (1) \textit{Sensitive-API Analyzer} (SAA) that statically analyzes the app binaries and outputs $\alpha$ (\S~\ref{sec:saa}); \textit{Privacy Label-to-Policy Consistency Checker} (\comparer{}) that semantically compares privacy labels with privacy policies and outputs $\beta$ (\S~\ref{sec:plpcc}).

\ignore{
    An essential component of \toolname{} is the static assessment framework (\staticFramework{}) that takes in our data set (iOS app binaries with their privacy labels and privacy policies) and finds out two sets of apps that should be further validated through dynamic analysis (\S~\ref{}): (1) apps that use the iOS system APIs whose return data, if collected by the apps, should be disclosed in iOS privacy labels --- denoted as $\alpha$; (2) apps whose privacy label and privacy policy have inconsistent data practice disclosures (regarding the data items to collect and purposes of collection) --- denoted as $\beta$. To this end, \staticFramework{} includes two sub-components: (1) \textit{Sensitive-API Analyzer} (SAA) that statically analyzes the app binaries and finds out the apps in $\alpha$ (\S~\ref{}); \textit{Privacy Label-to-Policy Consistency Checker} (PLCC) that semantically compares privacy labels with privacy policies and outputs the apps in $\beta$ (\S~\ref{}).
}

%To this end, \staticFramework{} includes two sub-components: (1) Sensitive-API Analyzer (SAA) that statically analyzes the app binaries and finds out the apps that involve the iOS system APIs whose return data, if collected by apps, should be disclosed in iOS privacy labels (see Apple's privacy label requirement in \S~\ref{}); (2) Privacy Label-to-Policy Consistency Checker (PLCC) that takes the privacy label and privacy policy of an app, and reports the inconsistency of data practice disclosure regarding the data items to collect and purposes of collection. We elaborate on \staticTool{} in \S~\ref{} and \comparer{} in \S~\ref{}. 

%Based on semantic analysis and a light-weight code analysis.

\vspace{-5pt}
\subsubsection{Sensitive-API Analyzer}
\label{sec:saa}
\vspace{-5pt}
%\vspace{2pt}\noindent\textbf{Static app analysis}. 
%\vspace{2pt}\noindent\textbf{iOS App static analysis} 

%\xiaojing{summarize the goal and workflow}

%For efficient compliance check, we leverage the Sensitive-API Analyzer (\staticTool{})

Although Apple provides a list of 32 data items (e.g., \textit{Device ID}, \textit{Health}, called \ldataItem{}s) whose collection are expected to be disclosed in privacy labels, such information is insufficient and cannot be directly used \ignore{by policy-makers or analysts }to find out whether an app involves/collects such data. The gap is that we lack a precise and comprehensive mapping between the data items and corresponding iOS system APIs that return the data --- we aim to identify apps that access the data by inspecting their invocation of those APIs. Preparing such a mapping is non-trivial due to (1) the substantial amount of iOS system APIs (e.g. more than 200 k public APIs covering 288 modules on the recent iOS 15.5) whose documents often lack descriptions about the return values or are even completely missing~\cite{appleDocumentation}; (2) the \ldataItem{} being too general without a clear definition of its scope. For example, the data item \textit{Device ID} is general and unique on iOS, including \textit{the identifier for advertisers} (IDFA), \textit{the identifier for vendors} (IDFV), and possibly many others, and it is never apparent for app developers whether accessing, for example, IDFV (or calling specific APIs such as \texttt{identifierForVendor}~\cite{idfv}, the iOS API that returns IDFV), should be disclosed in privacy labels (see measurement results in \S~\ref{sec:measurement}).\footnote{IDFA is a unique identifier assigned by Apple to a user's device that allows an installed iOS app to track user behavior across other companies' apps~\cite{idfa}. The IDFV is more convoluted and essentially consider a privacy-sensitive ID in our research since it enables cross-app user tracking: based on Apple, \textit{``it is useful for analytics across apps from the same content provider and may not be combined with other data to track a user across apps and websites owned by other companies unless the app has been granted permission to track by the user''}~\cite{idfvDefinition}.} \looseness=-1
%Oftentimes, it is unclear whether an iOS API

\ignore{
    We use static analysis to narrow down the scale of apps tested by dynamic analysis. To achieve this, we first manually crafted a target API list whose return value are projected by privacy Label. Further, we utilize a static scanner to select candidate apps with the target APIs existing in their binary code. 
}

\vspace{3pt}\noindent\textbf{\ldataItem{} to iOS-API mapping (l-mapping).} To enable a privacy-label compliance tool like \toolname{}, we propose and release an \ldataItem{} to iOS-API mapping (called \textit{l-mapping}). Specifically, we manually inspected iOS API documents (with development of end-to-end proof-of-concept apps to validate the return values of the APIs) and summarized a comprehensive list of iOS APIs corresponding to five types of \ldataItem{}s: Device ID, Location, Contacts, Health, Performance Data. Notably, to handle \ldataItem{}s that are not returned by iOS system APIs (they may come from third-party libraries or user input, see \S~\ref{sec:plcc}), \toolname{} leverages the dynamic assessment framework (see \S~\ref{sec_dynamic_framework}) being complementary to \staticFramework{}. \looseness=-1

%Notably, not all \ldataItem{}s are returned by iOS system APIs since some of them come from third-party libraries (by calling library APIs) or user input (see \S~\ref{}).

\ignore{
    \vspace{1pt}\noindent$\bullet$\textit{ API to data items mapping.} 
    In order to discover information regarding the relationship between terminology used in privacy label expressed in natural language and API method calls used in the corresponding code, we manually go through the Apple developer documentation ...
    %16_ICSE_Slavin.pdf
}

\vspace{3pt}\noindent\textbf{Static scanner.} 
\ignore{
    We use static analysis to filter out apps that not containing calling sensitive system API in its binary code. For example, if the API \textit{advertisingIdentifier} that returns Device ID doesn't exist in the binary code of the APP, then this app is unable to get the  Device ID. To build this filter, we use XXXX...
}
To support a large-scale study, we developed a lightweight, effective static analysis to screen apps that call iOS system APIs recorded in the \textit{l-mapping}. \toolname{} considers that an app does not collect, for example, IDFA (a Device ID), if the corresponding iOS API \texttt{-[ASIdentifierManager advertisingIdentifier]} cannot be found in the app's binary code. More specifically, our static analysis is adapted for the unique function call mechanism of Objective-C~\cite{objc} and Swift~\cite{swift}. In these languages, API names are composed of two parts: a class name (e.g., \textit{ASIdentifierManager}) and a selector name (e.g., \textit{advertisingIdentifier}). The invocation of an API is compiled to an instruction-level call to the iOS-unique \textit{objc\_msgSend} function with the API's selector name (a string) passed in as the second argument~\cite{egele2011pios,deng2015iris,tang2020ios}. Essentially, that is, an API's selector name string will appear in the binary if the API is called by the app. Hence, our static scanner searches the selector names in the apps' binaries (using the \textit{grep} command~\cite{grep}) and filters out apps that do not invoke APIs in the \textit{l-mapping}.

%leverages the \textit{grep} command to check the (non)existence of sensitive system API's selector names in apps' binary code. If the selector name string is not found in a binary, then the app is not calling the API.

%\xiaolong{We use static analysis to filter out apps that not calling sensitive system API in its binary code. For example, if the API \textit{-[ASIdentifierManager advertisingIdentifier]} that returns Device ID is not called in the binary code of the APP, then this app is unable to get the Device ID. Our static analysis is based on the special function call mechanism of Objective-C and Swift languages. In these languages, API names are composed of two parts: class name (e.g., \textit{ASIdentifierManager}) and selector name (e.g.,\textit{advertisingIdentifier}). Once an API is called, the call site is compiled to an instruction-level call to the special \textit{objc\_msgSend} function with the API's selector name string passed in as the second argument. As a result, an API's selector name string will appear in the binary if the API is called by the apps. Our filter relies on this fact and uses \textit{grep} command to check the (non)existence of sensitive system API's selector names in apps' binary code. If the selector name string is not found in a binary, then the app is not calling the API.}

\vspace{-5pt}
\subsubsection{Privacy Label-to-Policy Consistency Checker}
\label{sec:plpcc}
\vspace{-5pt}

%\vspace{3pt}\noindent\textbf{Inconsistency check between privacy policy and privacy label.}
%\vspace{2pt}\noindent\textbf{Privacy policy comparison analysis} 
To identify apps with inconsistent privacy label and privacy policy, in this subtask, we aim at aligning data objects and their usage purposes mentioned in privacy label and privacy policy for inconsistency check.
More specifically, we use PurPliance~\cite{bui2021consistency} to retrieve data items and the purpose of data usage from the privacy policy, and map them into the taxonomy of the privacy label with 32 data items and 6 purposes (see \S~\ref{sec:background}). Below we elaborate on the mapping process.

\ignore{
\vspace{1pt}\noindent$\bullet$\textit{ Retrieve target data from privacy policy}.
Our privacy policy data collection and pre-processing steps are specified in section ~\ref{dataCollection:privacyPolicy}. Next, we coarsely retrieved data from the parsed privacy policy, before mapping them to privacy label. We extracted three categories of data: data items being collected, purposes of data usage, and first/third party entity.

We first seek to identify which sentence tokens represent a data item. For instance, we aim to identify `phone number', `physical address', and `IP address' as data items if any of them appears in a sentence in the privacy policy. To satisfy this goal, we utilized a statistical-based technique of named-entity recognition (NER) to label data items within each sentence. Specifically, we employed the model provided by PolicyLint ~\cite{XXX} which has adapted the en\_core\_web\_lg model to the privacy policy domain and is capable to identify the data items with high precision. 

After we collect the sentences along with data items within them, we aim to identify the purposes of data usage of those sentence. For example, in the sentence: `For the purpose of sending push notifications, we will access your device ID for updating you of our new products', the `device ID' data items is collected for `marketing' purpose. We limited our scope to four purposes: advertising, marketing, analytics, and app functionality. Because XXX.
To distinguish between different purposes, we observed that various purposes sentences will contain specific keywords, due to the distinctive intentions and actions. For example, the sentences with marketing purpose tends to include words like `promotion', `email', `notification', etc. 
In this case, we utilized term frequency–inverse document frequency (TF-IDF) to extract those keyword features and employed the SVM classifier to conduct binary classification for each purpose that we are interested in.
Our training dataset is the combination of OPP-115 Corpus ~\cite{} and the classification result from Polisis model ~\cite{}. In total, we gathered 990 sentences for advertising, 919 sentences for marketing, 789 sentences for analytics, 693 sentences for app functionality, and 4,211 for others. The performance of our model achieves 97\%, 92\%, 96\% and 81\% precision for each purpose above. 

For the sentences which are labelled as marketing and/or advertising purposes, we also determined the actor/entity (belongs to first party and/or third party) that initiate the action or has access to the data. 
For instance, in sentence: `If you have allowed us to collect data about your precise location in the settings of your device, we [actor, first party] may share it with our ad network partners [entity, third party] in order to tailor the ads that are displayed within the App'.
For this purpose, we constructed a dependency tree using the Spacy Dependency Parser ~\cite{} to parse the whole sentence. Next we extracted subjects whose verbs are within a set of pre-defined verb list, as well as objects of preposition that fit into specific verb-with-preposition combinations (i.e, sell to, share with, and provide to), after removing the sentences with negative statements. We then retrieved the reference of first party itself from the privacy policy (for instance, At the Atlantic Monthly Group, Inc. (``The Atlantic")... where ``The Atlantic" denotes the first party), and from the domain name of the privacy policy (for example, \url{https://netflix.com/privacy} where netflix refers to the first party itself). Finally, we classified the actor/entity involved in the sentence as first party as long as the subject or objects of preposition contains we/us/our or first party reference, and as third party if any other words appeared.

%
%Specifically, for the top layer in privacy label which contains four categories: Data Used to Track You, Data Linked to You, Data Not Linked to You, and Data Not Collected, we only considered the mapping to Data Not Collected category because all the rest three categories are not explicitly mentioned in privacy policies. When no data items can be extracted from the privacy policy using the method described in section ~\ref{comparision:retrieveTargetData}, we labelled it as Data Not Collected.
}

\vspace{1pt}\noindent$\bullet$\textit{ Data object alignment.} 
Apple defined 32 specific data items as the taxonomy of privacy label (Section~\ref{sec:background}). However, given an app's privacy policy, the extracted data object $d_p$ can not be simply matched to the data item $d_l$ of privacy label using keyword matching.
This is because the semantic levels of $d_p$ and $d_l$ are sometimes different. For example, considering the sentence extracted from privacy policy \textit{``we will collect Android ID, IMEI, IDFA ...''}, here the sensitive data objects are \textit{Android ID, IMEI, IDFA}. However, those data objects are not matched to the 32 data items in privacy label, but they are semantically subsumed to \textit{``Device ID''} defined by Apple which represents any device-level ID.

To fill this gap, we enhanced the data object ontology in \cite{policyLint} to determine the semantic relationship (\textit{Synonym}, \textit{Hypernym}, \textit{Hyponym}) between $d_{p}$ and $d_{l}$ and align data objects in different granularity. 
More specifically, 
to recover their subsumption relationships, we use data object ontology in \cite{policyLint}, which is a graph-based data structure where each node represent a data object and each edge represents a relationship among hyponym and hypernym (e.g., ``personal information'' subsumes ``your email address'').
However, the ontology in \cite{policyLint} is built on the corpus of the privacy policy, there miss some nodes and edges defined by the privacy label taxonomy. For example, Apple defined \textit{Emails or Text Messages} includes \textit{subject line}, \textit{sender}, \textit{recipients} and etc.  Those data objects and their subsumption edges are not in the current ontology. Hence, we manually enlarged the ontology to cover such new edges (e.g., emails or text messages $\Rightarrow$ subject line). In total, we added 75 nodes and 193 edges to the original ontology. The updated ontology will be released on publish. 
%We release the updated ontology at XXX.
%
Given the ontology, we align $d_{p}$ and $d_{l}$ if $d_{l}$ is a synonym, hypernym, or hyponym of $d_{p}$. \looseness=-1

\vspace{1pt}\noindent$\bullet$\textit{ Purpose alignment.} 
It's non-trivial to fill the gap between the data usage purpose mentioned in the privacy policy and those in the taxonomy of the privacy label.
For example, the \textit{Advertising} purpose should be differential from the entity (i.e., first-party or third-party) in the privacy label: first-party advertising is aligned with \textit{``Developer’s Advertising or Marketing''} while third-party advertising is corresponding to \textit{``Third-Party Advertising''} based on Apple's definition of purposes. Hence, we crafted a mapping table considering both purpose and entity in the privacy policy, as shown in Table ~\ref{table:purpose_alignment}.

%For the second layer which mainly specified six purposes of data usage, as well as their actor/entity, we chose four categories: Developer’s Advertising or Marketing, Third-Party Advertising, Analytics, and App Functionality. Because XXX. 

\begin{table}[t]
\centering
  \setlength{\abovecaptionskip}{0.cm}
\footnotesize
\caption{Purpose Alignment}
\begin{tabular}{l|c||c}
\toprule

\multicolumn{2}{c||}{\textbf{Privacy Policy}} & \textbf{Privacy Label} \\ \hline
\textit{Purpose}  & \textit{Entity} & \textit{Purpose}\\ \hline
Functionality & First Party &  \makecell{App Functionality\\Product Personalization }    \\ \hline

\multirow{2}{*}{Advertising} & First Party & \makecell{Developer’s Advertising or Marketing} \\ \cline{2-3} 
                        & third Party  & \makecell{Third-Party Advertising} \\ \hline

Analytics & - &  Analytics \\ \hline

Marketing & First party &  Developer’s Advertising or Marketing \\ \hline

Other & - &  Other Purposes \\ \bottomrule

\end{tabular}

\label{table:purpose_alignment}
\vspace{-18pt}
\end{table}

\ignore{
In order to check inconsistency between privacy label and privacy policy, we need to infer the semantic relationship between each term in privacy policy ($d_{s}$) and data object ($d_{r}$) defined in privacy label from the enhanced ontology, denoted as $o$.

\qy{remove this part?}
Here, We use the definition in \cite{policyLint} to describe the semantic relationship (\textit{Synonym}, \textit{Hypernym}, \textit{Hyponym}) between $d_{r}$ and $d_{s}$.

\begin{itemize}
  \item \textbf{Synonyms Relationship}. If $d_{s}$ and  $d_{r}$ are terms partially ordered by an ontology $o$. $d_{s}\equiv_{o}d_{r}$ is true if $d_{s}$ and  $d_{r}$ are synonyms, defined with respect to an ontology $o$. 
 
  \item \textbf{Hyponym Relationship}. If $d_{s}$ is subsumed under the term $d_{r}$ such that $d_{s}\not\equiv_{o}d_{r}$, then $d_{s}$ is the hyponym of $d_{r}$, represented as $d_{s}\sqsubset_{o} d_{r}$. 
  
    \item \textbf{Hypernym Relationship}. If $d_{s}$ consists the term $d_{r}$ such that $d_{s}\not\equiv_{o}d_{r}$, then $d_{s}$ is the hypernym of $d_{r}$, represented as $d_{s}\sqsupset_{o}d_{r}$. 
  
\end{itemize}}

%\vspace{1pt}\noindent$\bullet$\textit{ Discussion}
%\xiaojing{add evaluation results here}

%For the third and fourth layer which defined data types (i.e., User Content  and Contact Info) and specific data items (i.e., Device ID, Precise Location and Contacts) separately, we utilized the data terms in data ontology graph ~\cite{XXX} to perform word match and manually extended it with extracted data items that were unique to our tasks.
%
%More specifically, the ontology graph is a graph-based data structure that capture relationships among hyponym and hypernym. For instance, ``personal information'' subsumes ``your email address''. Here we selected the leaves node as hyponym and their parents as hypernym, to map to the specifc data items and data types separately.

\vspace{-5pt}
\subsection{Dynamic Assessment Framework with Privacy-Label Compliance Check}
\label{sec_dynamic_framework}
%\subsection{Privacy-Label Non-compliance Check}

Based on the apps of interest found by \staticFramework{} (i.e., $\alpha$ and $\beta$), the \dynamicFramework{} performs automatic end-to-end execution on a strategically sampled subset of the apps to maximize the precision of the privacy-label non-compliance detection while maintaining the scalability of the study.
%precisely check the compliance of their privacy labels. 
\dynamicFramework{} includes two key components: (1) a dynamic analysis pipeline that automatically runs the apps with dynamic instrumentation and network monitoring (\S~\ref{sec:dap}) and (2) a compliance checker that reports inconsistencies between the actual data practices and privacy labels (\S~\ref{sec:plcc}). \looseness=-1 %We elaborate on their design as follows.

%\subsection{Inappropriate Privacy Label Recognition}

%\LXcomment{Explain why we need to sample apps and not analyze all. If we do not want to analyze all, do we need to gather and statically analyze that many apps? How many?}
% \vspace{-10pt}
\subsubsection{Dynamic Analysis Pipeline} \vspace{-8pt}
\label{sec:dap}
%for Privacy-Label related Data Practices

%\subsubsection{App sampling strategy (\XY{TO DO})}
%\label{s:sampling}

\begin{figure}[t!]
\center
  \setlength{\abovecaptionskip}{0.cm}
\includegraphics[width = 0.3 \textwidth]{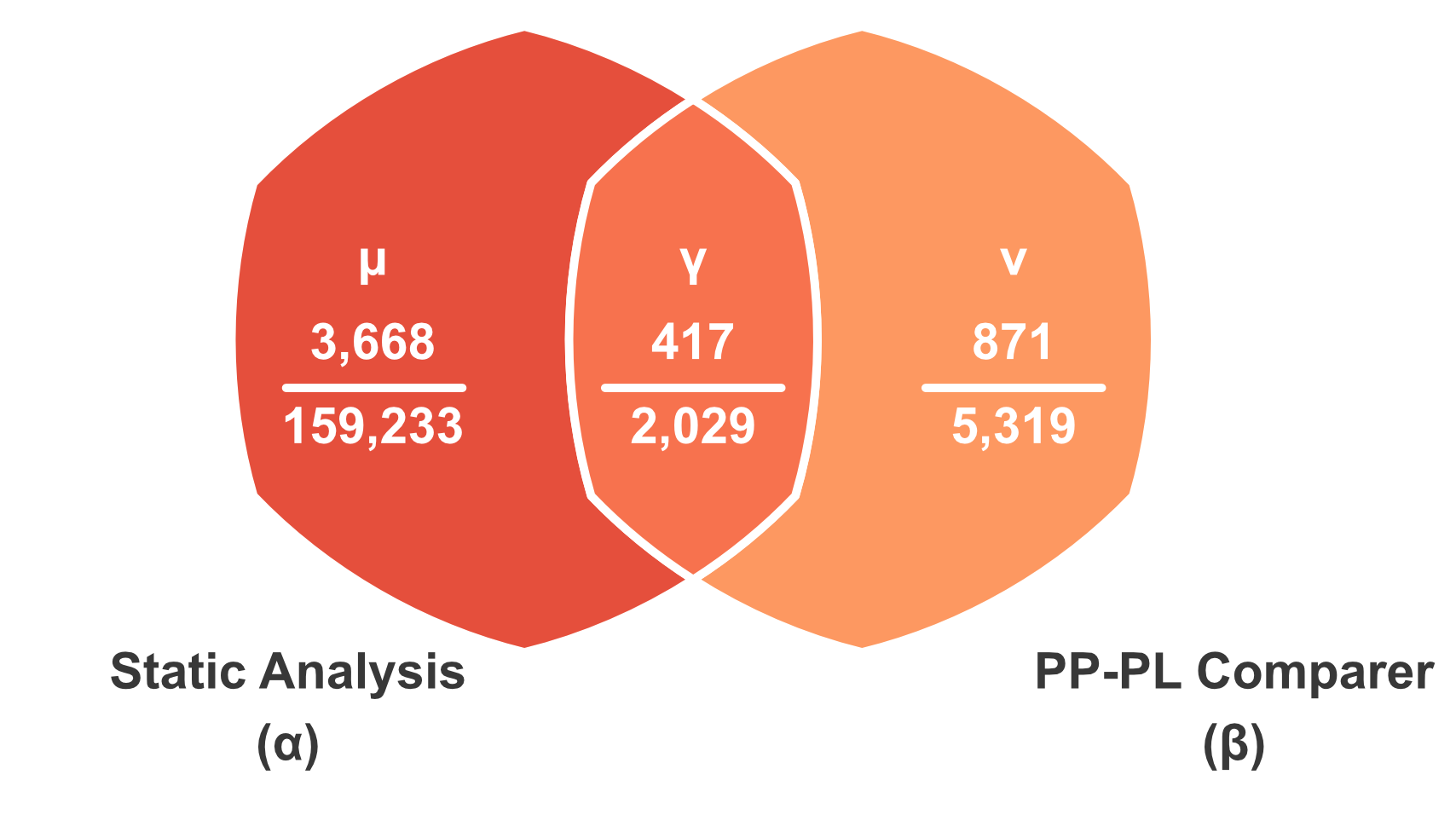}
\caption{App sampling. The denominators on the diagram show the total number of apps in the categories. The numerators represent our sample scopes.}
\label{fig:sampling}
\vspace{-15pt}
\end{figure}

\vspace{3pt}\noindent\textbf{App sampling.}
From our data set of 366,685 apps (\S~\ref{sec_method_overview}), \staticFramework{} yields $\alpha$ of 161,262 apps and $\beta$ of 164,056 apps.
Limited by the scalability of dynamic analysis, we narrow down the test scope while preserving the generalization of results by performing a strategic sampling over the apps. Specifically, we perform an intersection of $\alpha$ and $\beta$, yielding three app sets: $\mu$, $\gamma$, $\nu$ (Figure~\ref{fig:sampling}), which will help characterize different levels of privacy-label noncompliance:

%from static analysis (denoted as $\alpha$) and inconsistent apps from privacy label and privacy policy comparison (denoted as $\beta$). Figure ~\ref{fig:sampling} illustrates that the relationships among this two data sets. We defined three different sets ($\mu$, $\gamma$, $\nu $) in order to describe different levels of data inconsistency in apps.

\ignore{
    \textbf{$\mu$}: 162696 (Device ID: 131635; Precise Location: 145525; Coarse Location: 146293; Contacts: 5919; Health: 1694) \\
    \textbf{$\gamma$}: 2086 (Device ID: 1066; Precise Location: 1000; Coarse Location: 232; Contacts: 10; Health: 10) \\
    \textbf{$\nu$}: 5372 (Device ID: 2217; Precise Location: 2262; Coarse Location: 490; Contacts: 872; Health: 316) \\
}

\ignore{
    Limited by the scalability of dynamic analysis, we narrow down the test scope while preserve the generalization of results by performing the strategic data sampling over apps from static analysis (denoted as $\alpha$) and inconsistent apps from privacy label and privacy policy comparison (denoted as $\beta$). Figure ~\ref{fig:sampling} illustrates that the relationships among this two data sets. We defined three different sets ($\mu$, $\gamma$, $\nu $) in order to describe different levels of data inconsistency in apps.
}

\vspace{1pt}\noindent$\bullet$\textit{ $\mu = \alpha \cap \beta^{c} $}. That is, $\mu$ contains apps in set $\alpha$ that are not in set $\beta$. Intuitively, these are apps that invoke iOS system APIs accessing \ldataItem{} while their privacy labels and privacy policies are consistent (regarding disclosure of \ldataItem{} usage/collection, see \S~\ref{sec:measurement}). Data practices of apps in $\mu$, can be inconsistent with both the privacy labels and privacy policies. In our study, $\mu$ includes 159,233 iOS apps, from which we sampled 3,668 denoted as $\mu_s$ for dynamic analysis (see below). \looseness=-1

\ignore{
    A privacy-label violated app is detected under $\mu$ means a certain inconsistency occurs between code behavior and privacy label, while not in privacy label and privacy policy. Consequently, it indicates the same inconsistency occurs between privacy policy and code behavior. 
}

%a certain inconsistency occurs between label-code while not in label-policy -- indicates the same inconsistency occurs between policy-code -- indicates the process to summarize policy from data behaviour is problematic, while the label correctly reflects the policy; -- the label cannot correctly reflect privacy-sensitive code behavior

\vspace{1pt}\noindent$\bullet$\textit{ $\gamma = \beta \cap  \alpha$}. %$\gamma$ are apps in both $\alpha$ and $\beta$.
$\gamma$ are apps that invoke iOS system APIs accessing \ldataItem{} while their privacy labels and privacy policies are inconsistent. Data practices of apps in $\gamma$, even if inconsistent with privacy labels, can be consistent with privacy policies, or vice versa (consistent/inconsistent with privacy labels/policies, see measurement results in \S~\ref{sec:measurement}). In our study, $\gamma$ includes 2,029 iOS apps, from which we sampled 417 denoted as $\gamma_s$ for dynamic analysis.

%can be inconsistent with both the privacy labels and privacy policies. In our study, $\mu$ includes xxx iOS apps, from which we sampled xxx denoted as $\mu_s$ for dynamic analysis (see below).

%Data practices of apps in $\mu$, can be inconsistent with both the privacy labels and privacy policies. In our study, $\mu$ includes xxx iOS apps, from which we sampled xxx denoted as $\mu_s$ for dynamic analysis (see below).

\ignore{
    A privacy-label violated app is detected under $\gamma$ means a certain inconsistency occurs simultaneously between code behavior vs. privacy label and privacy label vs. privacy policy. It indicates the privacy label does not correctly reflect the code behavior and meanwhile it does not correctly summarized privacy policy. Luckily, in such case, the privacy policy correctly reflects code behaviour.
}

%\LXcomment{the code behavior might be or not be consistent with privacy label. The privacy policy might or might not be consistent with code behaviors.} 
%indicates the process to generate label from policy is problematic, while the policy correctly reflects code behaviour; -- the label cannot correctly reflect privacy-sensitive code behavior

\vspace{1pt}\noindent$\bullet$\textit{ $\nu = \alpha^{c} \cap  \beta$}. % $\nu$ will contain the apps in set $\beta$ that are not in set $\alpha$.
$\nu$ are apps with inconsistencies between privacy labels and privacy policies, while they do not apparently access \ldataItem{}s (i.e., their code is not found to invoke iOS system APIs returning \ldataItem{}s).
Notably, \toolname{} considers that the apps might (1) leverage runtime techniques~\cite{grace2012unsafe,wang2021understanding} (such as reflection) that evade the static screening (\S~\ref{sec:measurement}); (2) access \ldataItem{}s that originate from the app-level code or third-party libraries. For best coverage, our dynamic analysis inspects all possible \ldataItem{}s from the network traffic of the apps (see below). Data practices of apps in $\nu$, can be inconsistent with either privacy labels or privacy policies, or both of them. In our study, $\nu$ includes 5,319 apps, from which we sampled 871 denoted as $\nu_s$ for dynamic analysis. \looseness=-1

%inconsistent with privacy labels, can be consistent with privacy policies, or vice versa (consistent/inconsistent with privacy labels/policies, see measurement results in \S~\ref{}). In our study, $\gamma$ includes xxx iOS apps, from which we sampled xxx denoted as $\gamma_s$ for dynamic analysis.

\ignore{
    Apps under $\nu$ show two properties (1. apps don't collect the targeted data type in static code; and 2. a certain inconsistency occurs between privacy label and privacy policy). However, apps may leverage other run-time tricks (e.g., ~\cite{grace2012unsafe,wang2021understanding}) to avoid static scanner, for example, XXX. In such case, if privacy label doesn't correctly reflect code behavior, then it means privacy policy does correctly; Vice versa, if privacy label correctly reflect code behavior, then it means privacy policy doesn't disclose correctly. Here, we perform dynamic analysis in order to verify which regulations did the bad practices. 
}

%Suppose apps don't leverage run-time contexts~\cite{grace2012unsafe,wang2021understanding} to avoid static scanner, such a certain inconsistency either indicates privacy label over-claims data collection or privacy policy over-claims data collection, which is not harmful to user privacy. In our study, we only focus on data that return by sensitive system API. Therefore, it's nearly impossible for apps to collect data without calling sensitive system API in static code. Hence, we didn't focus on such cases. 

%It will be hard for dynamic analyzer to capture the transmission of targeted data if the corresponding API even not exists in static binary code. 

%label collect, policy not collect => label over claim
%label not collect, policy collect => policy over claim

%If privacy label doesn't correctly reflect code behavior, then it means privacy policy does correctly; Vice versa, if privacy label correctly reflect code behavior, then it means privacy policy doesn't disclose correctly. Here, we perform dynamic analysis in order to verify which regulations did the bad practices. 

%In our study, we didn't focus on such situation. 
%-- indicates both processes to summarize policy from data behaviour and to summarize label from policy are problematic; However, the label can correctly reflect  privacy-sensitive code behavior;

%\XY{TO Discuss}
 
\ignore{To avoid pitfalls of partial datasets and consequent sampling bias, we 

step 0 filter: apps list with sensitive APIs (xiaolong)
step 1 sample\_probality={} (jiale): 
00. Device ID:  0 purpose: xxx%
                1 purpose: xxx%
                2 purpose: xxx%
                ......

01. Precise Location: 0 purpose: xxx%
                      1 purpose: xxx%
                      2 purpose: xxx%
                       ......

02. Coarse Location: 0 purpose: xxx%
                     1 purpose: xxx%
                     2 purpose: xxx%
                       ......

03. Contacts: 0 purpose: xxx%
              1 purpose: xxx%
              2 purpose: xxx%
              ......                       
step 2: 
app total number = 2000
}

\ignore{ % original draft by Yue Xiao
    \subsubsection{App behavior monitoring}
    
    Our analysis pipeline, depicted in Figure XXX, combines the advantages of both static and dynamic analysis techniques to uncover inconsistencies between data flows and privacy labels.  We first utilize static analysis to identify suspected policy-violation apps, and then we use dynamic analysis to confirm such data exfiltrations.
    Specifically, we built tools to automatically download and install apps onto a physical jailbreak mobile iPhone, then interacted with each app individually using a UI automation tool. Meanwhile, we hooked the sensitive system APIs to collect backtraces and used a proxy approach to capture the network traffic generated by this app. This testbed enables us to monitor app run-time behaviors at OS and network layers.
}

%\subsubsection{Dynamic analysis}

%\subsubsection{Pipeline for App Behavior Analysis}
\ignore{
    Our test pipeline schedules each app to run for 3 minutes on jailbreak iPhone. The UI automation tool generates events (e.g., click, swap ...),  the Frida module hooks the sensitive APIs, and the network monitor records traffic during app execution. After each execution, the device goes through a cleaning phase (e.g., stop UI execution, uninstall the app) to run the next app. In the current setup, we can analyze approximately XXX apps/day on XXX phones.
}

\vspace{3pt}\noindent\textbf{App UI execution.}
%\vspace{3pt}\noindent\textbf{App UI execution and dynamic instrumentation.}
Our pipeline automatically installs each app (using the \textit{ideviceInstaller} command~\cite{libimobiledevice}) and schedules it to run on a set of jail-broken iPhones. 
%The UI automation tool generates events (e.g., click, swap ...),  the Frida module hooks the sensitive APIs, and the network monitor records traffic during app execution. After each execution, the device goes through a cleaning phase (e.g., stop UI execution and uninstall the app) to run the next app. In the current setup, we can analyze approximately XXX apps/day on XXX phones.
We leverage an open-source, off-the-shelf UI execution tool, called \textit{nosmoke}~\cite{NoSmoke}, to generate actions and automatically trigger the dynamic execution of an app, such as clicks or swipes, through the user interface (UI).
\ignore{
    Prior works~\cite{lee2013amc,liu2014decaf,reyes2018won,ren2018longitudinal} commonly used UI-automation frameworks such as \textit{monkey}~\cite{Monkey} which often perform actions at random coordinates on the screen and may fail to explore in-depth UI (e.g., those found only after multiple clicks).
}
Notably, \textit{nosmoke} is designed to identify textual information and types of UI elements using the OCR techniques~\cite{smith2007overview} from screenshots. Specifically, it identifies actionable UI elements in the current app window and executes target actions based on a simple configuration (e.g., clicks for a button or edit for a input, see the configuration in Appendix~\ref{table:crawConfiguration} based on depth-first search (DFS) algorithm.
In our pipeline, we set the maximum depth of the UI window stack as 5, and the maximum number of actions on one window as 15; each app is scheduled to run for three minutes and then uninstalled.
With 4 iPhones concurrently running, \toolname{} could execute approximately 600 apps each day.

\ignore{
    We use an open-source UI execution tool, called \textit{nosmoke} \cite{NoSmoke} provided by \textit{Alibaba Inc.} \cite{alibaba}, to generate actions to trigger the dynamic execution of an app, such as clicks, swipes etc., through the user interface (UI).
    Recent research \cite{lee2013amc,liu2014decaf,reyes2018won,ren2018longitudinal} has commonly-used UI-Automation Framework, called \textit{monkey} \cite{Monkey} which performs actions at randomly generated coordinate (x,y) on the current screen. However, without prioritizing actions that have a high possibility of transit from the current page to a new page, the monkey may fail to explore deeper pages. 
    To address this problem, we applied a structural Monkey called \textit{nosmoke} \cite{NoSmoke}, which can strategically identify the a clickable UI element by using the OCR techniques to retrieve textual information directly from screenshots. Specifically, it first identifies clickable elements in the current page and executes the configurable actions (e.g., click, swipe, etc.) based on a DFS crawling algorithm. In our testbed, We set the maximum testing depth of the UI window tree to 5, and the max Action on one page is 15 to prevent staying too long. 
    %droidbot: structural monkey on Android
}

\vspace{3pt}\noindent\textbf{Dynamic instrumentation and network monitoring.}
During the app execution, \toolname{} leverages \textit{Frida}~\cite{FridaAPI}, a dynamic instrumentation tookit, to hook iOS systems APIs returning \ldataItem{}s (see the l-mapping in \S~\ref{sec:saa}). In particular, we inspect the APIs' argument values and return values (using the \textit{onEnter(args)} and \textit{onLeave(retval)} APIs of Frida~\cite{FridaAPI} respectively), and obtain the call-stack traces (using the \textit{Thread.backtrace([context, backtracer])} API~\cite{FridaAPI}). 
\toolname{} matches the APIs' return values with network traffic to confirm the data collection, and \toolname{} adopted a popular network monitoring tool \textit{Fiddler}~\cite{fiddler} to inspect app traffic (capable of TLS decryption and common decoding schemes). \looseness=-1
\ignore{
    Since \toolname{} will match the APIs' return values with network traffic to confirm the data collection (see below), we consider that the app might encrypt an \ldataItem{} before sending to the Internet.  \XY{remove this part??} To account for such a case, we also hook APIs of popular cryptography libraries (see the list in Appendix XXXX): if the input to encryption APIs is an \ldataItem{}, we seek to match the API output with the network traffic.
}

\ignore{
    \vspace{2pt}\noindent\textbf{Sensitive API hooking.}
    We use \textit{frida}, a dynamic instrumentation tookit, to hook sensitive system APIs, spy on crypto APIs and trace the invocation of sensitive API by injecting customized scripts into black box processes. 
    In particular, \textit{``onEnter(args)''} \cite{FridaAPI} is a callback function and can be used to read or write the parameters of hooked functions; \textit{``onLeave(retval)''} \cite{FridaAPI} is also a callback function which contains raw return value. We use \textit{``Thread.backtrace([context, backtracer])''} \cite{FridaAPI} to generate a backtrace for the current thread.
    Hence, we wrote the JS scripts to hook XXX sensitive APIs to inspect the parameters, return value, and backtrace of the API invocation. Then, We attached the Frida session and injected the scripts into the app process during the app execution by the UI automation tool. 
}

%We use a handy network monitoring tool, called \textit{fiddler}~\cite{fiddler} that implements man-in-the-middle interception and uses self-signed certificates to capture all network traffic generated by the app being tested. Specifically, the \textit{fiddler}~\cite{fiddler} installs a root certificate into the iPhone, so it is able to decrypt communications protected by TLS. Besides, it can de-obfuscate and decode network traffic by a simple configuration. What's more, it provides rich interfaces that are written in \textit{JScript.NET} to allow the developer to add customized rules. We built a customized exporter which is able to automatically export network traffic periodically with a timer. 

\ignore{
    We use a handy network monitoring tool, called \textit{fiddler}~\cite{fiddler} that implements man-in-the-middle interception and uses self-signed certificates to capture all network traffic generated by the app being tested. Specifically, the \textit{fiddler}~\cite{fiddler} installs a root certificate into the iPhone, so it is able to decrypt communications protected by TLS. Besides, it can de-obfuscate and decode network traffic by a simple configuration. What's more, it provides rich interfaces that are written in \textit{JScript.NET} to allow the developer to add customized rules. We built a customized exporter which is able to automatically export network traffic periodically with a timer. 
}

\ignore{
    \vspace{3pt}\noindent\textbf{Network traffic monitoring.}
    \toolname{} adopted a popular network monitoring tool \textit{Fiddler}~\cite{fiddler} to inspect TLS-decrypted app traffic. \textit{Fiddler} acts as a man-in-the-middle between the app and remote servers (\textit{Fiddler} allows one to install a root certificate into the iPhone to decrypt TLS traffic). Fiddler can also de-obfuscate and decode network traffic using common schemes~\cite{fiddlerdecode} (e.g., base64, URL decoding, Hex decoding). \toolname{} leverages the convenient programming interfaces of \textit{Fiddler} and automatically export network traffic every ten minutes. 
}

%For data items originate from App-level or user inputs, we infer the data items from keys in network traffic. Based on our observation, the naming of key usually offers meaningful cues to indicate the data items. For example, \textit{device\_identifier} as a key transparently shows the collected data belongs \textit{Deivce ID}. The \textit{blood\_type} as a key can be semantically aligned to the data item \textit{Heath}. Hence, we manually built a key-to-data mapping by inspect XXX network traffic from XXX apps. The examples of mapping is shown in Table~\ref{table:data_scope}.

\ignore{
    As mentioned before, data items typically originate from three sources: Apple system-level API, app-level code including third-party code, or user inputs. To pursue precision instead of recall, we utilize hard-code pattern-based method to extract data items from key-value in network traffic. Specifically, for data items originate from system API, if the return value can be matched in the network traffic, we can examine the data items by using the API-to-Data mapping as detailed in Section XXX. For example, the value \textit{9F0B31E6-980B-4468-9797-0B1F1A8FA56E} existing in network traffic indicated the collected data item is \textit{Device ID}, as this value exactly matched to that returned by the API \textit{advertisingIdentifier}. For data items originate from App-level or user inputs, we infer the data items from keys in network traffic. Based on our observation, the naming of key usually offers meaningful cues to indicate the data items. For example, \textit{device\_identifier} as a key transparently shows the collected data belongs \textit{Deivce ID}. The \textit{blood\_type} as a key can be semantically aligned to the data item \textit{Heath}. Hence, we manually built a key-to-data mapping by inspect XXX network traffic from XXX apps. The examples of mapping is shown in Table~\ref{table:data_scope}.
}

\ignore{
    \vspace{3pt}\noindent\textbf{Data item Inference.} 
    \label{s:data_inference} As mentioned before, data items typically originate from three sources: Apple system-level API, app-level code including third-party code, or user inputs. To pursue precision instead of recall, we utilize hard-code pattern-based method to extract data items from key-value in network traffic. Specifically, for data items originate from system API, if the return value can be matched in the network traffic, we can examine the data items by using the API-to-Data mapping as detailed in Section XXX. For example, the value \textit{9F0B31E6-980B-4468-9797-0B1F1A8FA56E} existing in network traffic indicated the collected data item is \textit{Device ID}, as this value exactly matched to that returned by the API \textit{advertisingIdentifier}. For data items originate from App-level or user inputs, we infer the data items from keys in network traffic. Based on our observation, the naming of key usually offers meaningful cues to indicate the data items. For example, \textit{device\_identifier} as a key transparently shows the collected data belongs \textit{Deivce ID}. The \textit{blood\_type} as a key can be semantically aligned to the data item \textit{Heath}. Hence, we manually built a key-to-data mapping by inspect XXX network traffic from XXX apps. The examples of mapping is shown in Table~\ref{table:data_scope}.
}

\vspace{-10pt}
\subsubsection{Privacy-Label Compliance Check} 
\label{sec:plcc}
\vspace{-10pt}

\vspace{3pt}\noindent\textbf{Overview.}
Based on Apple, an \ldataItem{} is deemed ``collected'' by a vendor only if it is transmitted to the Internet~\cite{iosprivacylabel}. %Notably, Apple's accountability criteria is more strict than many prior works on Android~\cite{gordon2015information,zimmeck2019maps,arzt2014flowdroid} which commonly assumed the data as leaked out once an API returning the data was invoked\ignore{(in those studies, only static inspection of Android API invocation could suffice)}. 
For our compliance check, \toolname{} first infers the \ldataItem{} that is actually collected based on both network traffic and evidence found in dynamic instrumentation. Another key challenge is to infer the vendors' actual purposes for the collection of each \ldataItem{}s (e.g., \textit{App functionality}, \textit{Third party advertising}, falling under categories defined by Apple), which we tackled by adapting a modeling and learning-based approach. 
%Then \toolname{} infers the purposes for the collection of each \ldataItem{}s (e.g., xxx, xxx, see categories defined by Apple \S~\ref{}), and 
\toolname{} finally reports whether the tuple (\ldataItem{}, purpose) representing actual data practices is consistent with the app's privacy labels (see the consistency model in \S~\ref{sec:model}). 
%\subsubsection{Purpose Inference} 

\ignore{
    Apple defines several purposes for developers to specify in the privacy label to help their customers understand how the collected data is used. However, Apple only provides the general definition of each purpose, which seems less clear and sufficient for developers to specify correct purposes. Hence, we seek to provide a purpose classifier for developers to identify the purpose of data transmitted off the device to the network. To achieve this, we first simply the client-side code behavior from the aspect of both network traffic and execution trace generated by dynamic execution. Next, we collect groundtruth apps from both Apple-developed Apps and wide apps. Further, we derive a number of computational features from client-side code behavior. Finally, 
    we apply supervised machine learning to train a purpose classifier based on the proposed features.
}

\vspace{3pt}\noindent\textbf{Inference of \ldataItem{} collection.} 
\label{s:ldatainfer}
%\vspace{3pt}\noindent\textbf{Inference of \ldataItem{} collection.} 
%\label{s:data_inference} 
%, and thus, their studies were often much easier to scale up (i.e., only static inspection of Android API invocation could suffice).
In our study, we adopt an approach that aims to match the network traffic to data values observed in dynamic instrumentation~\cite{appcensus}. Specifically, \toolname{} checks if the traffic matches runtime \ldataItem{} values that are found above by instrumenting iOS system APIs (i.e., those in l-mapping).
%or (2) encrypted \ldataItem{} (found above by instrumenting common encryption APIs).
%
For example, in network traffic of the app \textit{TED}~\cite{ted}, \toolname{} found the value \texttt{9F0B31E6-980B-4468-9797-0B1F1A8FA56E} that matched the return value of the API \textit{advertisingIdentifier} (the iOS API that return IDFA, an Device ID) during the app's execution, \toolname{} infers that an Device ID is collected.

For \ldataItem{}s that are not returned by iOS system APIs (from app-level code or user inputs, see \S~\ref{sec:plcc}), \toolname{} look for those \ldataItem{}s in network traffic based on a crafted set of 197 keywords we adapted for \ldataItem{}s. In our study, we inspected the network traffic of 259 iOS apps (113 from Apple and 146 from high-profile vendors), each running for ten minutes with our automatic UI execution (see above) and 15 minutes with manual operations, to collect a comprehensive list of keywords, shown in Appendix Table~\ref{table:data_scope}). 
    For example, the keyword \textit{device\_identifier} found in network traffic of at least 5 apps assures \toolname{} that the collected data is a \textit{Deivce ID} (the traffic, often being Web API traffic, includes key/value pairs). \toolname{} used the keyword \textit{blood\_type} to find data collection from traffic aligned to the \ldataItem{} \textit{Heath}.

%\LXcomment{collection purpose vs. usage purpose --- make consistent}

\vspace{3pt}\noindent\textbf{Inference of \ldataItem{} collection purposes.}     
%\toolname{} infers the vendors' actual purposes for the collection of an \ldataItem{} (versus claimed purposes in privacy labels). Specifically, leveraging the context information found in dynamic instrumentation (e.g., call stack traces) and semantics in network traffic, \toolname{} abstracts a six-tuple: $(caller, system API, frequency, endpoint, request, response)$ for each \ldataItem{} found to be actually collected. Such an abstraction encodes essential context information for \toolname{} to infer the collection purpose, which can be mapped to those defined by privacy labels. 
\ignore{
    For example, in analyzing the app xxx, \toolname{} found that the third-party ads library amazon-adsystem \textbf{[caller]} invokes \textit{advertisingIdentifier} \textbf{[system API]} multiple times \textbf{[frequency]}, sends the IDFA (a Device ID returned by the API) to \url{https://aax.amazon-adsystem.com/e/msdk/ads?} \textbf{[endpoint]} along with other advertising information in Web requests \textbf{[request]}, and the Web \textbf{[response]} includes advertisement content. 
    \LXcomment{Xiaojing, start editing here:}
    \toolname{} introduced a learning based approach xxx.
    based on the classifier and the tuple, \toolname{} infers that the collection of a Device ID (i.e., an \ldataItem{}) is for the purpose ``Third-Party Advertising'' (one of six purposes framed by Apple, see \S~\ref{}). 
    %\toolname{} abstracts the context of the data flow: the third-party advertiser associate a Device ID with other advertising data to show ads when users use the app xxx. 
    %Hence, the purpose of this tuple can be identified as third-party advertising. 
}
\ignore{
    \vspace{2pt}\noindent\textbf{Client -Side Code Behavior Simplification}.
    Actually, what data is collected for what purpose can be inferred by dynamically monitoring the app client-side code behavior. Specifically, we can extract a six tuple from dynamic execution trace and network traffic: (caller, system API, frequency, endpoint, request, response). For example, the amazon-adsystem \textbf{[caller]} invokes \textit{advertisingIdentifier} \textbf{[system API]} multiple times \textbf{[frequency]} and send to \textit{\url{https://aax.amazon-adsystem.com/e/msdk/ads?}} \textbf{[endpoint]} along with advertising slots information in this request body \textbf{[request]} and return third-party advertising content \textbf{[response]}. From this six tuple, we can describe this piece of code behavior like this: the third-party advertiser associate Device ID (return value of \textit{advertisingIdentifier}) with other advertising data to show ads when users visit apps. Hence, the purpose of this tuple can be identified as third-party advertising. 
}
In our study, we analyze a set of high-profile iOS apps and their privacy labels, and identify their features for purpose prediction.
More specifically, we select features tailored to the purpose categories and definitions from Apple, particularly those considered to be robust,
in the sense that missing these features might disable the classifier to correctly differentiate different purposes. 
To this end, we leverage  13 features, categorized into three groups: external app information, traffic information, and call trace information. Among them, 6 features have never been used in previous research. 
The external app information is extracted from the app download page on the Apple store, including \textit{Bundle ID, App name, Company name}.
We also leverage the context information found in dynamic instrumentation (e.g., call stack traces) and semantics in network traffic to abstract a six-tuple: $(caller, system API, frequency, endpoint, request, response)$ for each \ldataItem{}. Such an abstraction encodes essential context information for \toolname{} to infer the collection purpose, which can be mapped to those defined by privacy labels. 
A description
of these features is provided in Table~\ref{table:feature}. 

\ignore{

    \vspace{2pt}\noindent\textbf{Ground truth collection}. In order to infer the most-likely purpose from client-side code behavior, we collect ground-truth from 113 apps developed by Apple and XXX wide apps in US market using the following criteria: 
    
    \begin{itemize}

      \item \textbf{App score}. We selected  those apps with high score (larger than 4.0) reviewed by its customers from top XXX in each app category (26 categories in total) to be candidate apps. We assumed high-scored apps have better privacy practices. As shown in \cite{appcomments}, the low score apps are more likely violating market policies because the users of low-score apps are more or less complaining that this app contains aggressive advertising behaviors, and the other even reported that this app might be leaking data or even malicious.
    
      \item \textbf{App popularity}. We sorted those candidate apps by its popularity, measured by its number of rating and pick the top XXX from each app category.
    
        \item \textbf{Amenable to dynamic analysis}. In order to dynamically analyze app code behavior, we need to collect both network traffic from a man-in-the-middle proxy and call trace from frida. We exluded those apps that adopt SSL Certificate Pinning \cite{certificPin} to prevent traffic minoring or apply \textit{``Block Frida Toolkits''} \cite{blockfridatoolkits} to prohibit code instrumentation. 
      
    \end{itemize}

}

\ignore{
    \vspace{2pt}\noindent\textbf{Ground truth labeling}. 
    Given each app in ground truth, we manually interact with it for xxx mins. Meanwhile, we automatically hook sensitive API using frida and record network traffic using fiddler. From frida logs, we can extract the caller, the invocation frequency and the return value of sensitive API. From traffic records, we can obtain the endpoint, the request and response body. By searching the return value of sensitive API in the request body, we can align API invocation information with network traffic and  obtain the six-tuple (caller, system API, frequency, endpoint, request, response). Further, we manually assigned (data, purpose) pair extracted from privacy label to each corresponding six-tuple. In total, XXX annotators labeled XXX pieces of traffic which contact XXX distinct domains and achieved  XXX agreement score calculated by Fleiss’s kappa~\cite{fleiss1971measuring} as shown in table XXX.
}

\begin{table}[t]
\centering
\footnotesize
\caption{Feature Selection. Starred Features are new features not mentioned in previous research.}
\begin{tabular}{l|c|c}
\toprule
\textbf{Group} & \textbf{Feature} &  \textbf{Source} \\ \hline

\multirow{3}{*}{External app features} & App name & \makecell{ Apple Store } \\ \cline{2-3}  & Company name$^\star$  & \makecell{Apple Store} \\ \cline{2-3}   & Bundle ID & \makecell{ iTunes API} \\ \hline

\multirow{6}{*}{Traffic features} & Domain name  & \makecell{network traffic} \\ \cline{2-3}  & URL paths  & \makecell{network traffic} \\ \cline{2-3}   & Domain role & \makecell{WHOIS, Crunchbase} \\ \cline{2-3}   &\makecell{ Bundle ID-domain \\ similarity} & \makecell{-} \\ 
\cline{2-3}   & KV pairs & \makecell{network traffic} \\ \hline

\multirow{5}{*}{Execution features} & Caller$^\star$  & \makecell{execution trace} \\ \cline{2-3}  & Caller's  company$^\star$  & \makecell{cocoapods} \\ \cline{2-3}   &  \makecell{Caller's company to \\ domain similarity$^\star$} & \makecell{-} \\ \cline{2-3}   &\makecell{Sensitive API$^\star$} & \makecell{-} \\ 
\cline{2-3}   & API frequency$^\star$ & \makecell{execution trace} \\ \bottomrule

\end{tabular}
\label{table:feature}
%\vspace{-10pt}
\end{table}

\vspace{2pt}\noindent\textbf{Implementation and evaluation}.  
Here we elaborate on the implementation and evaluation of our model.

\vspace{1pt}\noindent$\bullet$\textit{ Dataset.}
We collected and manually validated 259 high-profile iOS apps as groundtruth (113 developed by Apple and 146 from other high-profile vendors). %These are the most popular apps in all 26 categories of Apple app store, each with high user-rated scores (higher than 4.0). 
%
%We collect the based on their privacy labels and data practices observed from dynamic analysis
%
Each app goes through the dynamic analysis pipeline (\S~\ref{sec:dap}) complemented with ten-minute manual app usage, and then we obtained the six-tuple for each \ldataItem{} found in app traffic. These six-tuples are manually aligned with (\ldataItem{}, purpose) pairs extracted from privacy labels to establish ground-truth.
In total, four annotators labeled 2,958 pieces of traffic which contact 184 distinct domains and achieved  0.923 agreement score calculated by Fleiss’s kappa~\cite{fleiss1971measuring}.
%\vspace{2pt}\noindent\textbf{Model training}. 
%We build the supervised machine learning model using PyCaret library \cite{pycaret} which is an open-source machine learning package that over 18 algorithms.

The groundtruth dataset consists of 2,958 samples  in five data usage purpose (1,035 as \textit{Analytics}, 739 as \textit{App Functionality}, 445 as \textit{Developer's Advertising or Marketing}, 141 as \textit{Product Personalization}, and 598 as \textit{Third-Party Advertising}). We use 75\% samples as the training set and 25\% samples as the testing set to train and evaluate our approach. The class labels (i.e., purposes) in the training set and the testing set share the same marginal distribution. 
%The groundtruth dataset consists of 7,335 samples  in five data usage purpose (3,237 as \textit{Analytics}, 1,891 as \textit{App Functionality}, 1,548 as \textit{Developer's Advertising or Marketing}, 362 as \textit{Product Personalization}, and 297 as \textit{Third-Party Advertising}). We use 75\% samples as the training set and 25\% samples as the testing set to train and evaluate 

\vspace{1pt}\noindent$\bullet$\textit{ Implementation}.
 We apply PyCaret \cite{pycaret} with Python 3.6~\cite{python}, an open-source machine learning package deployed with 18 algorithms, to select the best model as well as the hyper-parameters for purpose identification. According to the results of 10-fold cross validation on the training set, we use LightGBM~\cite{ke2017lightgbm} as the classifier with learning rate as 0.1, minimum child samples as 20, minimum child weight as 0.001, number of estimators as 100, number of leaves as 100, and other default settings in~\cite{ke2017lightgbm}. We use LabelEncoder in sklearn~\cite{pedregosa2011scikit} to encode categorical features. Features with multiple values (e.g., domain 
role) are represented as Bag-of-Words~\cite{zhang2010understanding} using CountVectorizer in sklearn. For sequence-like features (e.g., keys in the request body), we aggregate such features as a ``document'' for each sample and apply TF-IDF~\cite{ref1} for the feature representation. Finally we concatenate all features as the input to the classifier.

%\vspace{2pt}\noindent\textbf{Implementation}.

\begin{table}
\centering
\begin{footnotesize}
\setlength{\tabcolsep}{2pt}{% 调整列间距
\caption{Evaluation of Purpose Identifier.}
\label{tb:pi}
%\rowcolors{3}{gray!10}{gray!5}
\begin{tabular}{ lccc} 
\toprule
\textbf{Purpose} &\textbf{Precision} &\textbf{Recall}&\textbf{F1}\\
\midrule

 Analytics&0.97&0.98&0.98\\ 
App Functionality   &0.92&0.97&0.94\\ 
Developer's Advertising or Marketing &0.95&0.77&0.85\\ 
 Product Personalization    &0.86&0.68&0.76\\ 
 Third-Party Advertising  &0.97&0.98&0.98\\
 \midrule
 Macro Avg & 0.94 &0.88&0.90\\
 Weighted Avg & 0.95&0.95&0.95 \\
\bottomrule
\end{tabular}}
\end{footnotesize}
\vspace{-10pt}
\end{table}

\vspace{1pt}\noindent$\bullet$\textit{ Experiment results}.
The inconsistent disclosure model (see \S~\ref{sec:model}) focuses to reveal the data usage purposes in practice which are not reflected by the associated privacy label.
Therefore, the false positives in purpose identification (i.e., the predicted purpose does not occur in practice) will result in false alarms in inconsistent disclosure. 
To avoid false alarms as possible, we tune the hyper-parameters of our model towards high precision, which indicates less false positives in each purpose, during the cross-validation on the training set. Then we evaluate Purpose Identifier on the testing set and the results are shown in Table~\ref{tb:pi}. The proposed model achieves high precision (i.e., 0.97, 0.92, 0.95, 0.86, and 0.97) and tolerable recall (i.e., 0.98, 0.97, 0.77, 0.68, and 0.98) for each purpose.  Further, we run our model on 40,999 records from 5,102 (source of application data), and manually verify the prediction of 1,000 randomly selected samples, which achieves 94.3\% accuracy.

 \ignore{
We use 10-fold cross validation to evaluate the performance of purpose inference and found that \toolname{} can achieve an average accuracy, precision, recall of XXX for data purpose inference. We perform an ablation study to evaluate the effectiveness of the new added features used to predict the purposes. As shown in table xxx, the most effective feature is XXX which improves the F1 score by xxx\%.

1. test set: precision of each class. tune model towards high precision. table.
2. verification on 1000 samples: overall accuracy. (precision)}

\vspace{-5pt}
\subsection{Evaluation}
\label{s:overallEval}
% \vspace{-5pt}
This section reports our evaluation study on \toolname{} to understand its effectiveness and performance, and the challenges in
completely identifying privacy label non-compliance apps from a large number of real-world apps.

\vspace{3pt}\noindent\textbf{Experiment settings}. 
We ran \toolname{} on four iPhones with versions (12.4.1, 13.4, 13.7, 14.8.1) and one Mac mini with Apple M1 chip and 8‑core CPU, one Mac mini with Apple M1 chip and 16‑core CPU, and two Mac book pro with Apple M1 chip and 16‑core CPU, to investigate iOS apps. 

To evaluate the overall effectiveness of \toolname{}, we randomly selected 100 apps. Four students (25 apps/per-person) manually interact with those apps and stop when they believe they have explored all possible UIs. Meanwhile, network traffic and call-stack traces are recorded by \textit{fiddler} and \textit{Frida}. In total, 81 apps out of 100 apps can be successfully tested, generating 939 system API invocations and 1,362 network traffic. Further, four students manually go through the generated network traffic and execution traces to summarize the data and its corresponding purposes, then, output 122 non-compliant $(d, q)$ pairs associated with 64 apps.

\vspace{3pt}\noindent\textbf{Evaluation results}. 
On the ground-truth dataset, \toolname{} generates 407 system API invocations and 831 network traffic and further outputs 75 non-compliant $(d, q)$ pairs (49 apps). \toolname{} shows a precision of 89.33\% and a recall of 60.6\%.
On average, it took 185 seconds (180 seconds for dynamically executing an app and 5 seconds for inconsistency analysis) to investigate one app.

%and took around 5s to output the inconsistency results. \xiaojing{experiment environment here}.

\looseness=-1

\vspace{1pt}\noindent$\bullet$\textit{ Falsely detected inconsistency.}  3 false detections reported comes from legitimate traffics that transmit analytics-related data to Apple. 
For example, when the UI automation tool triggers Siri suggestions or looks up, or types in search when interacting with the app, the diagnostics data including device meta, device ID,  and precise location are sent to \textit{fbs.smoot.apple.com}. However, as defined by Apple, the developers are not responsible for disclosing data collected by Apple~\cite{iosprivacylabel}. Further, to erase such false positives, we remove all the traffic sent to Apple domain when analyzing the inconsistency. 
The rest three false positives come from falsely inferred purposes.  We found it challenging to distinguish product personalization from app functionality as they usually shared some similar features (e.g., sent data to its own domain). We elaborate on the falsely detected cases in Appendix~\ref{app:false_detected}.\looseness=-1

\vspace{1pt}\noindent$\bullet$\textit{ Missed cases.} 
\toolname{} falls short in achieving a high recall mainly due to the natural limitation of dynamic analysis-\textit{code coverage}. Compared with manually executing app, \toolname{} can only invoke 43.34\% API and generate 61.01\% network traffic. 
The main reason is that the UI automation tool is configured with limited search depth and ran in fixed time slots, it can not guarantee to explore all possible UIs. 
%For example, there is an app, called \textit{Clue Period & Cycle Tracker}, which tracks the pregnancy period. After going through seven pages recording the user's pregnancy info, health info, and demographic info (e.g., birthday), the app directs users to the main functionality (start a free trial). Hence, \toolname{} failed to detect 3 inconsistencies in this app. 
%
What's more, a lot of apps require users to log in with a registered username and password (many of them need CAPTCHA verification) before assessing app functionality. 34 missing cases are caused by 27 apps that required two-factor authentication (either through email or phone number). 
%
%In addition, some apps encrypted data before transmission, which cannot be easily reversed.  
\looseness=-1

%deeper ui create an account, information, input interest
%False positive
%1. Propose inference

%\vspace{3pt}\noindent\textbf{Challenge of detecting missed apps}.
%Out of 100 apps in total,  19 apps are not amenable to dynamic analysis due to the detection defend against jailbreak environment, Frida hooking and network proxy. Among 19 missing apps, 12 apps protect them from running in Jailbreaking operating environments. 5 apps that adopt SSL Certificate Pinning \cite{certificPin} to prevent traffic minoring and 2 apply \textit{``Block Frida Toolkits''} \cite{blockfridatoolkits} to prohibit code instrumentation. 

%\input{main/4.overview of privacy label.tex}

\vspace{-5pt}
\section{Inconsistency Results and Analysis}
\label{sec:measurement}
\vspace{-5pt}
Running \toolname{} on 6,332 iOS apps and their privacy labels, we show that privacy label non-compliance is prevalent. % in the Apple App Store. 
Altogether, 5,102 iOS apps have been fully tested, while the other apps crash due to their resistance to run in jail-broken environment, Frida hooking or traffic minoring.
Among the tested apps, \toolname{} reveals that 3,281 of them fail to disclose data and purposes (\S~\ref{sec:neglect disclosure}), 1,628 apps contrarily specify purposes (\S~\ref{sec:contrary disclosure}), and 677 apps inadequately disclose purposes (\S~\ref{sec:inadequate disclosure}).

\vspace{-5pt}
\subsection{Neglect Disclosure}
\vspace{-2pt}
\label{sec:neglect disclosure}
Neglect Disclosure indicates that the app developer collects certain data without disclosure. 
Among 5,102 apps, we found 3,281 apps with neglect disclosure. In total, 11,726 data objects are neglected in the privacy label of these apps. Among these non-compliant apps, 2,346 apps lie in $\mu$, where the data disclosure in the privacy label is consistent with that in the privacy policy, indicating that both the privacy label and privacy policy neglect the data in app code behavior. Also, 238 non-compliant apps 
lie in $\gamma$, where the privacy policy is reliable while the privacy label fails to reflect the code behavior. Besides, 434 non-compliant apps lie in $\nu$, indicating that they harvest data originated from user input, generated by app-level code, or bypassed static screening.

\ignore{
Neglect Disclosure indicates that the app developer collects a specific data, but ignore the disclosure of specific data item governed by Apple Privacy Label requirements. 
%
%For example, a weather app, called \textit{KMI-IRM}, collects user's real time precise location and send to its own server \textit{app.meteo.be} in order to query local weather information for App functionality. However, the app developer neglect disclose\textit{ (Precision Location, App functionality)} in its privacy label. 
%
%To understand the pervasiveness of Neglect Disclosure, we utilize the inconsistency model (\S\ref{sec:model}) of \textit{Inconsistency 1} in \textit{Definition 4} by checking if there existing flow-$f$-relevant privacy label in $S$. 
%
Among 5,102 apps, we found 3,281 apps omitted to disclose 11,726 data objects (on average 3.57 data objects per app). Among those non-compliant apps, we found 2,346 apps from $\mu$, where data disclosure in privacy label are aligned with that in privacy policy, which indicated both privacy label and privacy policy failed to correctly disclose the data collection and usage/purpose occur in app code behavior.
%erroneous privacy policy may misguide the generation of privacy label. 
%both privacy label and privacy policy failed to correctly disclose the data collection and usage/purpose occur in app code behavior; 
%
238 non-complaint apps from $\gamma$ indicated that privacy label failed to reflect the code behavior, while privacy policy could correctly reflect code behavior; 
434 non-complaint apps from $\nu$ indicated that they harvest data originated from user input or generated by app-level code or bypassed static screening.  }
%}
%The above results indicate that most app developers are unable to deliver complete data disclosure in privacy label. 

\begin{figure*}[t]
  \centering
  \subfloat[]{
    \label{fig:omitDisclosureDataPurpose}
    \includegraphics[width=0.25\linewidth]{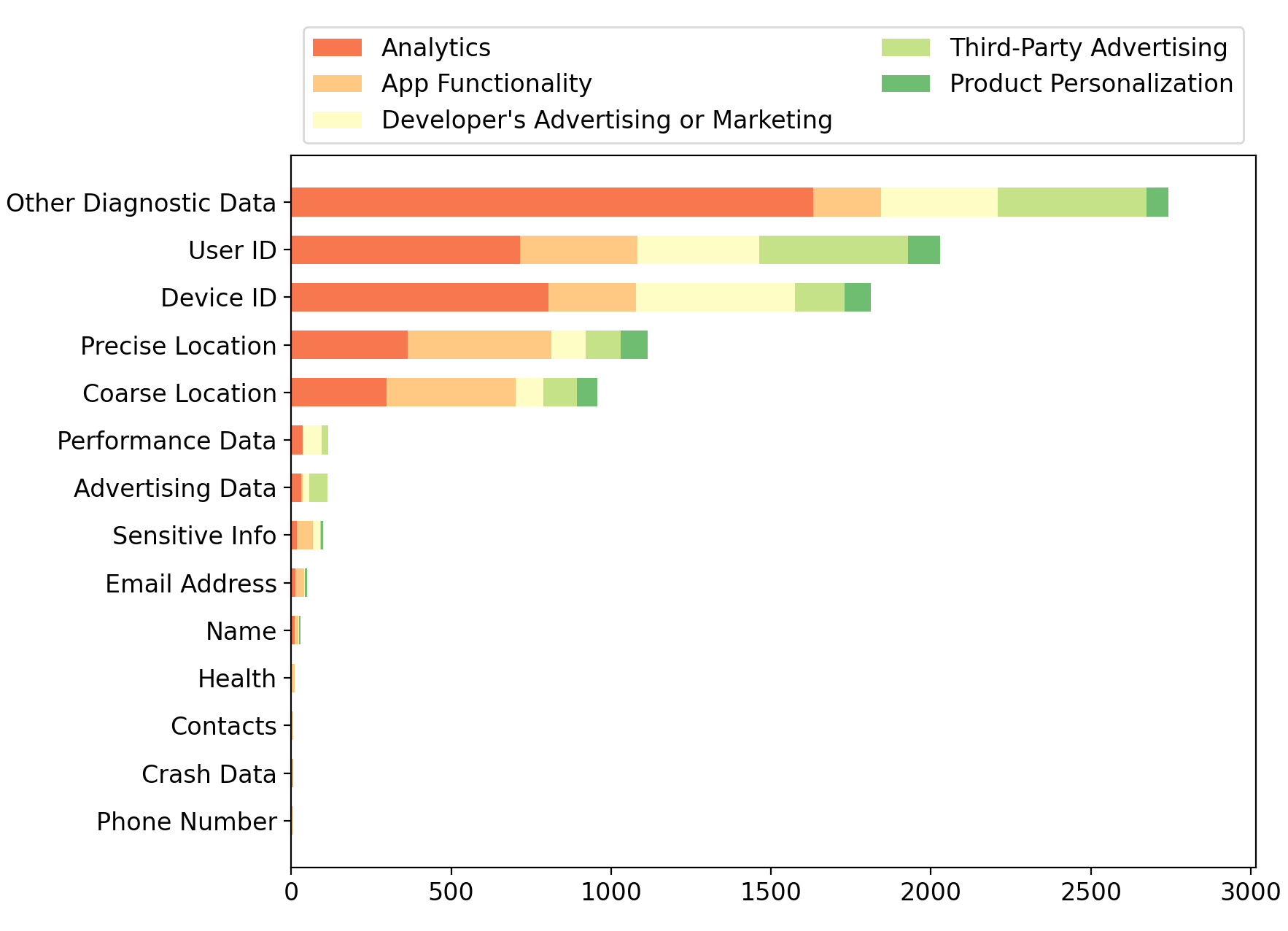}}
  \subfloat[]{
    \label{fig:omitDisclosureDomain}
    \includegraphics[width=0.25\linewidth]{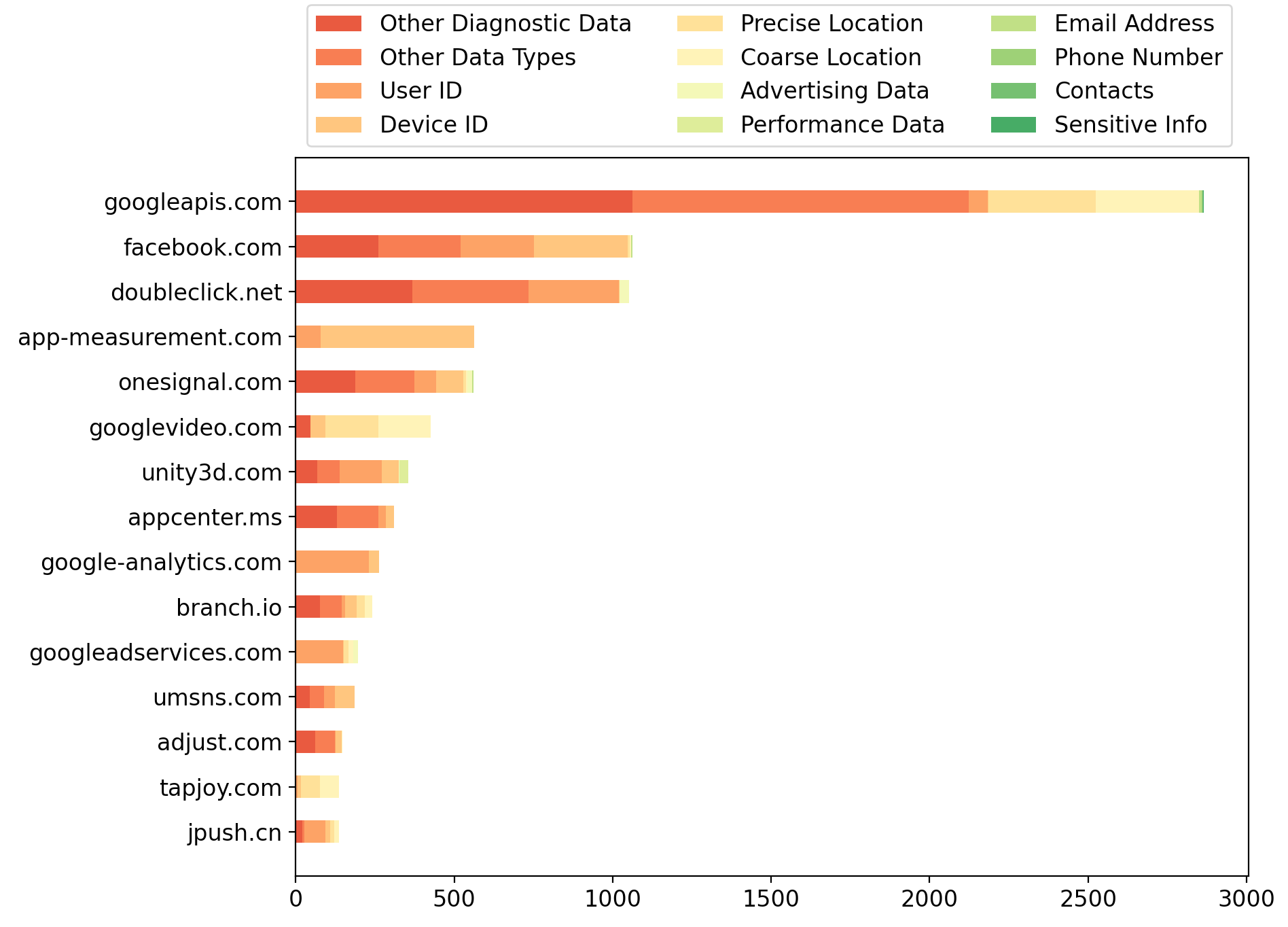}}
  \subfloat[]{
    \label{fig:incorrectStat}
    \includegraphics[width=0.25\linewidth]{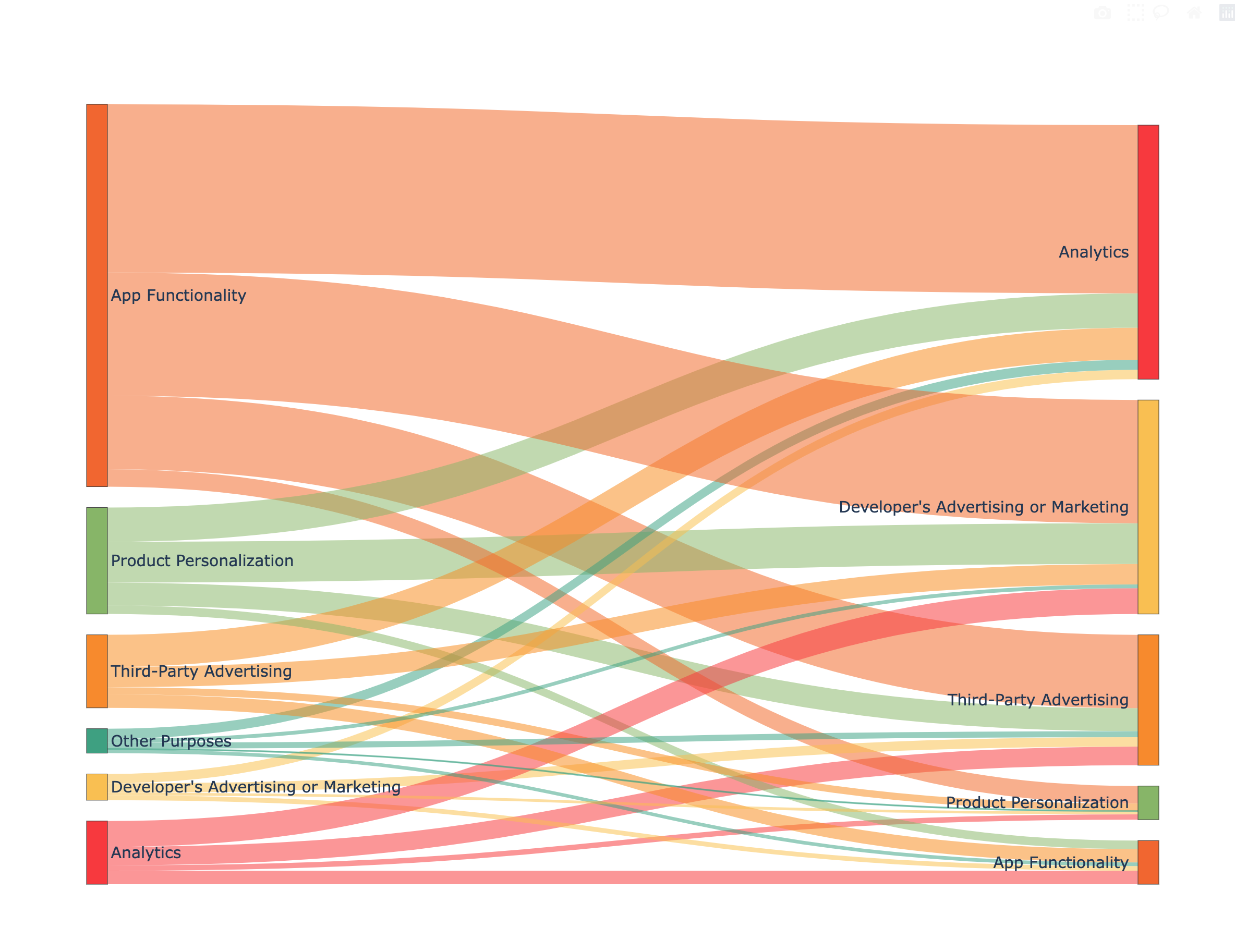}}
  \subfloat[]{
    \label{fig:inadequateStat}
    \includegraphics[width=0.25\linewidth]{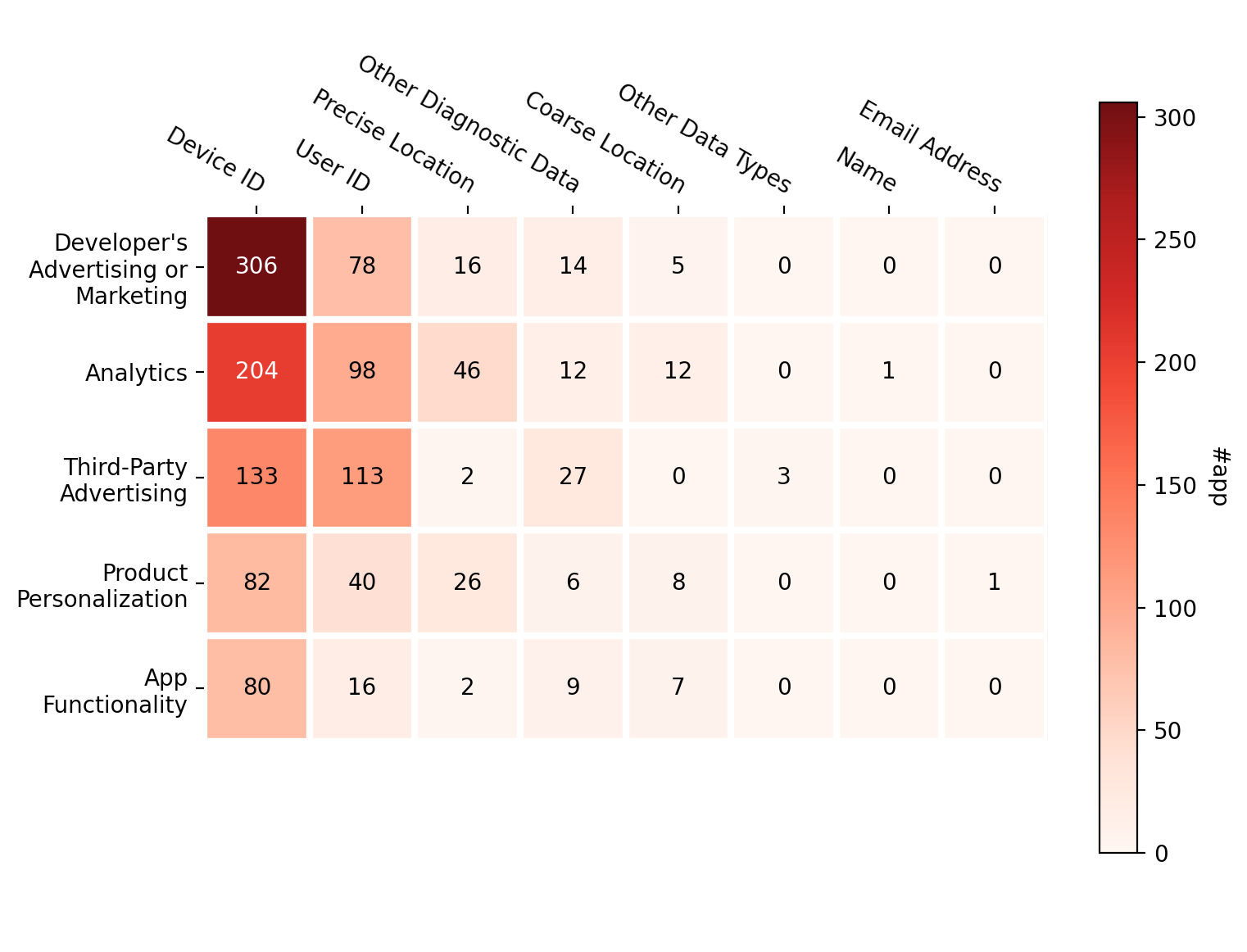}}
    
    \vspace{-5pt}
    \caption{(a) The Distribution of neglected data apps. Each stack represents different data type. The colors on the spectrum show the actual purposes of the apps. (b) The top 15 endpoints that the omitted data are most likely to leak to. Each stack represents different endpoints. The colors on the spectrum show the types of data. (c) The flow between the claimed purposes in the privacy label and actual purposes. The sources on the left denote the purposes in privacy label. The sinks on the right represent the actual purposes of apps. The widths of the flow show the relative portions of apps with contrary disclosure. (d) The distribution of inadequate disclosure in different data types and purposes. Darker color shows the data and purpose more apps fail to adequately disclose in their privacy labels.}
    \vspace{-15pt}
 \end{figure*}

% \begin{figure}[t]
%     \centering
%     \includegraphics[width=\columnwidth]{figure/omitDisclosureDataPurpose.png}
%     \caption{The Distribution of neglected data apps. Each stack represents different data type. The colors on the spectrum show the actual purposes of the apps.}
%     \label{fig:omitDisclosureDataPurpose}
% \end{figure}

% \begin{figure}[t]
%     \centering
%     \includegraphics[width=\columnwidth]{figure/omitDisclosureDomain.png}
%     \caption{The top 15 endpoints that the omitted data are most likely to leak to.
%     Each stack represents different endpoints. The colors on the spectrum show the types of data.}
%     \label{fig:omitDisclosureDomain}
% \end{figure}

\vspace{2pt}\noindent$\bullet$\textit{ Data types and purposes.} 
To understand the disclosure of what kind of data with which purposes are prone to be ignored by app developers, we plot the number of apps neglecting data under each purpose, as shown in Figure~\ref{fig:omitDisclosureDataPurpose}. The results show that \textit{Diagnostic data}, such as device metadata (e.g.,\textit{is\_jailbroken,localized\_Model,GPU\_type, sensors\_singal, screen\_size, os\_version, etc}), is the most likely to be omitted.
Although such data are usually used for technical diagnostics, one can also track a user by the device fingerprinting based on the user’s diagnostic data~\cite{kamara2016not}. 
For example, the traffic sent to a data broker \textit{broker.datazoom.io} consists of 12 device-related data along with user events in the request body, which can be used to profile a user. 
Further, the \textit{User ID}, \textit{Device ID} and \textit{Location} data are also prone to be omitted by developers. Those data are commonly used for third-party advertising, analytics, or marketing. 
%
%The \textit{Performance data} (e.g., battery status, brightness level, etc), which are commonly used to evaluate system effectiveness, are the most likely to be neglected under the analytics purpose. 
%The \textit{Advertising Data} are mainly used for third-party advertising (51.75\%) and developer's advertising and marketing (17.54\%). Besides, 27.19\% of the advertising data are used to perform advertising attribution (e.g., evaluate how an ad account applies credit for clicks, conversions, and sales) for analytics purposes. 
%
In our study, we also observe data that come from user input, like \textit{Email Address, Phone Number} to support the App Functionality purpose, are commonly omitted by app developers.
\ignore{
To understand what kind of data under which purposes are prone to be ignored by app developers, we plot the number of apps of omitted data under each purpose, as shown in Figure~\ref{fig:omitDisclosureDataPurpose}. The results show that the \textit{Other Diagnostic data} including device metadata (e.g.,\textit{is\_jailbroken,localized\_Model,GPU\_type，sensors\_singal, screen\_size, os\_version, etc}) are most prone to be ignored. Although those data usually enable the developers to measure technical diagnostics of the system, unfortunately, it could also be used to track a specific user by device fingerprinting based on the user’s diagnostic data~\cite{kamara2016not}. For example, the traffic sent to data broker \textit{broker.datazoom.io} contains 12 device-related data along with user events in the request body to profile the user. 
Further, the \textit{User ID}, \textit{Device ID} and \textit{Location} data are also prone to be omitted by developers. Those data are commonly used by third-party advertising or analytics or marketing SDKs to track users and fine-tune the ads that show to a user. 
The \textit{Performance data} (e.g., battery status, brightness level, etc), which are commonly used to evaluate system effectiveness, are most omitted under the analytics purpose. 

The \textit{Advertising Data} are mainly used for third-party advertising (51.75\%) and developer's advertising and marketing (17.54\%). The left of them (27.19\%)  are used to perform advertising attribution (e.g., evaluate how an ad account applies credit for clicks, conversions, and sales) for analytics purpose. 
Other omitted data, like \textit{Sensitive Info, Email Address, Phone Number}, which primarily come from user input, are mainly omitted under the App Functionality purpose. }
%
%The \textit{contacts} are only omitted under the app functionality purpose (e.g.,  find/invite your friends) probably due to Apple increased scrutiny of the access to user's contacts after a breach of address book leakage~\cite{}. 

%Specially, xxx apps failed to disclose xxx data objects for the purpose (Third-party adverting, Developer’s Advertising or Marketing, Analytics).

%This data enables the developers to understand real-world field behavior and improve the product based on that information. Unfortunately, it can also reveal information about what DoD users are doing with the systems and what causes them to fail

%Our system creates a unique ID that it associates with your user’s diagnostic data.identify a specific user.

%That is, we only access the address book with an explicit user action,
%But it is also a breach of trust and an invasion of privacy.

\vspace{1pt}\noindent$\bullet$\textit{ Endpoints of omitted data.}
To examine the endpoint that the data are leaked to, we present a stacked histogram of the top 15 endpoints, ranked by the frequencies of collecting each targeted data, as shown in Figure~\ref{fig:omitDisclosureDomain}. The top three endpoints (googleapis.com, doubleclick.net, graph.facebook.com) all belong to third-party libraries. This indicates that the app developers are generally unclear about the data practice of third-party partners (\S\ref{sec:root}). %We elaborate on the cause analysis in \S\ref{sec:root}. 
\ignore{To examine what endpoint the data is most likely being leaked to, we present a stack histogram to show the top 15 endpoints ranking by the times of collecting each targeted data, as shown in Figure~\ref{fig:omitDisclosureDomain}. The top three endpoints are googleapis.com, facebook.com, and doubleclick.net, all of which belong to third-party libraries, which indicated that the app developers are generally unclear about the data practice of third-party partners. We elaborate on this cause analysis in \S\ref{sec:root}. }

%What are those companies that collect data stealthily ?(the distribution of the data endpoint of omitted-disclosure)\\

\vspace{1pt}\noindent$\bullet$\textit{ App categories.}
To examine which app categories are most likely to neglect data disclosure, we investigate the distribution of non-compliant apps according to their app categories. The results show that Game apps are most prone to neglect disclosure. We observe 97.15\% of omitted data are leaked to third-parties in game apps. 
A more surprising app category with neglect disclosure is youth-using education apps, especially with the omitted data types of Device ID, User ID, and Precise Location.
In our study, we observe that 81.41\% of omitted data are leaked to third-parties in education apps.

%is harmful as those apps are youth using.
%
%Defined by children's Online Privacy protection Act~\cite{}, it is not allowed to collect data that used to tracking users for personalize ads in children apps without a clear understandable and complete disclosure~\cite{reyes2018won}. 
%
 %For example, XXX. 

%Some of them sell ads in their apps to make money
%many do because it helps personalize ads.

%https://www.nytimes.com/interactive/2018/09/12/technology/kids-apps-data-privacy-google-twitter.html

\ignore{
\vspace{1pt}\noindent$\bullet$\textit{ Data collected suddenly to \textit{``Data not collect''.}} In section XXX,  we observe 239 apps change  their privacy label from collecting data to \textit{``Data Not collected''} or \textit{`Not provide Details`''}. By performing dynamic analysis of them, we found 69 apps are still collecting data including \textit{Device ID (37.18\%), User ID (29.83\%),Other Diagnostic Data (21.22\%), Precise/Coarse Location (10\%), Email Address (1.05\%) etc}. XXX data are transmitted to third parties. 
}

%what category of apps are most prone to omitted disclosure\\

%\vspace{1pt}\noindent$\bullet$\textit{ The violation percentage in $\mu, \gamma, \nu$}. As shown in section~\ref{s:sampling}, non-complaint apps detected under different sets represent different levels of inconsistency. In our study, we found XXX  non-complaint apps from $\mu$, which indicated both privacy label and privacy policy failed to correctly disclose the data collection and usage/purpose occur in app code behavior. XXX  non-complaintapps from $\gamma$ indicated that privacy label failed to reflect the code behavior, while privacy policy correctly reflect code behavior. 

%1. to what percentage of their privacy policy is omitted-disclosure, static analysis, or both\\

\vspace{-8pt}
\subsection{Contrary Disclosure}
\vspace{-5pt}
\label{sec:contrary disclosure}
Contrary Disclosure indicates that the app developer tries to declare the data usage in the privacy label, but struggles to correctly disclose the purposes. %For example, xxx.  
%
%To understand the pervasiveness of Incorrect Disclosure, we utilize the formalization definition Inconsistency 2 in Def-
%inition 4 by checking if there existing flow-$f$-relevant contrary privacy label. 
%
\toolname{} reports 1,628 apps incorrectly labeling the purposes of 2,935 data objects. Among those non-compliant apps, we found 973, 202, and 314 from $\mu$, $\gamma$, and $\nu$ respectively.

\vspace{1pt}\noindent$\bullet$\textit{ Data types and purposes.}
As shown in Figure~\ref{fig:incorrectStat}, the app developers are most likely to falsely disclose the other four purposes to \textit{App Functionality}. In our study, we observe that 347 app developers distrustfully disclose purpose as \textit{App Functionality}, while the data is actually used for advertising. Such incorrect disclosure may intentionally deceive users to download and use their app by claiming they have a more acceptable reason to collect user data. As reported in \cite{brush2010exploring}, users were willing to share their location data to help cities plan bus routes or to get traffic information (\textit{App Functionality}), but reluctant to share the same data for ads (\textit{Advertising}) or maps showing one’s travel patterns (\textit{Analytics}). 
As to data types, similar to neglect disclosure, we found the data items most commonly to be incorrectly disclosed are Device ID, User ID, and Coarse Location.

%\vspace{1pt}\noindent$\bullet$\textit{The characteristics of incorrect-disclosed purpose.} To examine what characteristics of incorrect-disclosed purposes that likely confused app developers, we inspect the corresponding piece network traffic. \XY{summarized the characteristics: endpoint, request body xxx}

%pair is App Functionality and Analytics. The definition of App Functionality is xxx, and the definition of Analytics is xxx. The boundary between xxx and xxx may be too ambiguous for developer to make correct decision. 

\vspace{1pt}\noindent$\bullet$\textit{ App categories.}
By examining the app category with the most contrary disclosure, we found that lifestyle apps (related to fitness, dating, food, music, travel, etc) always falsely disclose the purpose of Developer's Advertising and Marketing as App functionality. This is probably due to the rich functionalities (e.g., finding the next new song, restaurant, or destination, etc) en-compressed in such apps; also, the app developers may carelessly consider most data collection operation as a part of app functionality. 

%For example XXX. 

% \begin{figure}[t]
%     \centering
%     \includegraphics[width=\columnwidth]{figure/incorrectStat.png}
%     \caption{The flow between the claimed purposes in the privacy label and actual purposes. The sources on the left denote the purposes in privacy label. The sinks on the right represent the actual purposes of apps. The widths of the flow show the relative portions of apps with incorrect disclosure.}
%     \label{fig:incorrectStat}
% \end{figure}
\vspace{5pt }
\subsection{Inadequate Disclosure}

\label{sec:inadequate disclosure}
Inadequate Disclosure means the developer has already declared one or multiple purposes for a specific data item but fails to disclose all of them. %For example, xxxx. 
%
%To understand the pervasiveness of Inadequate Disclosure, we utilize the formalization definition Inconsistency 3 in Definition 4 by checking if there existing flow-$f$-relevant inadequately disclosed in the privacy label. 
%
The scanning results of \toolname{} show that 677 apps inadequately label the purposes of 1347 data objects, which indicated that developers intend to inform users of data usage but lack of capability to disclose data practices in the app comprehensively. 
Among those non-compliant apps, we found 402, 74, and 133 from $\mu$, $\gamma$, and $\nu$ respectively. 
%
%\vspace{1pt}\noindent$\bullet$\textit{ The inadequate disclosure data types and purposes}. 
As shown in Figure~\ref{fig:inadequateStat}, the most prevailing inadequate purposes are \textit{Developer’s Advertising or Marketing},  \textit{Analytics}, followed by \textit{Third-party Advertising}, ranked by numbers of unique apps with inadequate disclosure.  App Functionality and Product Personalization are less prone to be inadequately disclosed. The results indicate that developers are more familiar with data practices in their own code and less clear about data collection from other vendors integrated into the same app. 
As to data types, similar to neglect disclosure and contrary disclosure, we found the most common data items to be inadequately disclosed are \textit{Device ID, User ID, and Precise Location.}

%\vspace{1pt}\noindent$\bullet$\textit{ The characteristics of inadequate-disclosed purpose.}
%To examine what characteristics of inadequate-disclosed purposes that likely neglected by app developers, we inspect neglected corresponding piece network traffic. \XY{summarized the characteristics: endpoint, request body xxx}

\ignore{
\vspace{1pt}\noindent$\bullet$\textit{ The app categories distribution.}
To examine which app categories most likely inadequately disclose purpose, we plot the distribution of apps according to their app categories as shown in Figure~\ref{fig:inadequateCat}. Similarity with omit disclosure, games apps are most likely to inadequately disclose advertising and analytics purpose. For example XXX
}

% \begin{figure}[t]
%     \centering
%     \includegraphics[width=\columnwidth]{figure/inadequateStat.png}
%     \caption{The distribution of inadequate disclosure in different data types and purposes. Darker color shows the data and purpose more apps fail to adequately disclose in its privacy label.}
%     \label{fig:inadequateStat}
% \end{figure}

\section{Root Cause Analysis and Case Study}\label{sec:root}
\vspace{-5pt}
In this section, we analyze the root causes of privacy label non-compliance with a series of empirical studies.
%Based on the detected apps, we further perform a series of empirical studies to understand the root causes of those privacy label non-compliance.

\vspace{-5pt}
\subsection{Misleading Third-Party Disclosure Guides}
\label{s:erroreousLabelGuidance}
\vspace{-5pt}
To help app developers fill out privacy label correctly, some third-party SDKs \ignore{release their data collection and usage practices, and }released guidelines about what privacy label items the app developers need to disclose if their apps integrate those SDKs. 
Ideally, such guidelines will help app developers generate privacy labels correctly and comprehensively.
However, we find that those guidelines are usually incomplete or even incorrect, which easily leads to non-compliance for the apps. Notably, our findings substantially
complements the recent human subject study~\cite{li2022understanding} which showed that iOS app developers were not aware of the guidelines provided by third-party SDKs.

To understand the quality of third-party SDKs' privacy label guidelines, we select the popular SDKs which are integrated by most iOS apps in our dataset and then evaluate their privacy label guidelines. Specifically, we manually search the privacy label guidelines of the top 30 most popular SDKs and found that 18 SDKs have privacy label guidelines. Then, we investigate network traffic associated with the endpoints of those SDKs to examine its data collection practice and compare it with its statements in privacy label guidelines.
Surprisingly, we found that 14 (out of 18) of them are either inadequately (N=4) or incorrectly (N=9) disclosing data collection practices, or using vague statements (N=1), as elaborated below. The list of SDKs is provided in Appendix Table~\ref{table:guidance}.

%\xiaojing{this table was not explained. should be moved to appendix}.
%\XY{\textit{``N''} in original guidance means the data is claimed ``not collected'' in privacy label guidance, but corresponding data object have been observed in the network  to its endpoints from the network traffics extracted from non-compliance apps. \textit{  ``M''} means the data is missing disclosed in original guidance, but we observe it in network traffic. \textit{``V''} means }

\vspace{1pt}\noindent$\bullet$\textit{ Incomplete guidelines.} The incomplete guidelines mean the SDK providers didn't disclose all the data collected by the SDKs. For example, in our study, we observe that \textit{Flurry Analytics SDK} collected precise location data from 76 apps. However, in its privacy-label guidance, it didn't suggest app developers fill out the privacy label to declare the collection of user location data~\cite{FlurrySDKguidance} (but only ``Device ID'' , ``Product Interaction'', and ``Other Usage Data''). 

%declared that app developers need to check the boxes for ``Device ID'' , ``Product Interaction'', and ``Other Usage Data'' and select ``analytics'' purpose when submit privacy label, but 

\vspace{1pt}\noindent$\bullet$\textit{ Incorrect guidelines.} In our study, we also found that the SDK providers falsely claimed their data collection practices. For example, in the guideline of Appsflyer's~\cite{appsflyerSDKguidance}, it claimed only to collect coarse location but not to collect precise location. However, in our study, we observed the user's precise GPS coordinates 
%(e.g., ``\textit{lat=39.173126220703125, long=-86.52308469189246}''\XY{need mask}) 
were sent to the endpoint \textit{wa.appsflyer.com}. Defined by Apple, coarse location is represented in coordinates with lower resolution (less than three decimal); while precise locations are with higher resolution (same or more than three decimal). Appsflyer could misunderstand the terminology of the data items, leading to incorrect privacy label guidance. 

\vspace{1pt}\noindent$\bullet$\textit{ Vague statements.} The vague guidelines describe the data collection practices in an uncertain tone by using an auxiliary verb (e.g., may, could, might). For example, all the data ``collect'' actions from Facebook SDK guidance~\cite{FacebookSDKguidance} are modified by auxiliary verb \textit{may}, e.g., \textit{``We may receive and process certain contact, location, identifier, and device information associated with Facebook users and their use of your application."} In our study, we indeed found that 694 iOS apps (out of 908), which integrated Facebook SDK, did not disclose the data collection practices of Facebook SDK.
%\xiaojing{what are the suggested privacy labels for those SDKs?} \XY{The privacy label guidance should explicitly disclose what data is collected by default without condition, what data is optional based on what configuration under which version.}

\subsection{Abusing Apple's Requirement}

%\vspace{1pt}\noindent$\bullet$\textit{ Data collection in open web.} 
For apps with web views~\cite{ioswebview}, Apple required that 
\textit{``Data collected via web traffic must be declared, unless you are enabling the user to navigate the open web''}~\cite{iosprivacylabel}. That is, data collection inside app web-views must be disclosed, while if the app redirects users to open browsers for data collections, app developers are not required to disclose it. 
In our study, we observed that some apps exploit this disclosure requirement, to collect private user data while likely evading the privacy-label disclosure requirement. 
For example, the privacy label of the financial app \textit{Banco ABC Brasil Personal}~\cite{brasilapp} only discloses collection of \textit{Identifiers} and \textit{Diagnostics}. However, once users open this app, it redirects users to the browser along with the users' precise geo-coordinates appended to the URL, sending to its server \textit{abrasuaconta.abcbrasil.com.br}. 
More seriously, the webpage further sent the precise location data to 50 unique URLs belonging to 9 different domains including \textit{px.ads.linkedin.com, trc.taboola.com, google-analytics.com}.

To understand how pervasive such abusive behaviors are, we use regex \texttt{(iPhone).*AppleWebKit}
\texttt{(?!.*Version).*Safari} to match the \textit{``userAgent''} in HTTP headers of app traffic, to find out traffic redirecting users to browsers (e.g., Safari). Note that the user agent strings of mobile Safari include the word ``Version'', whereas the user-agent strings of iOS web view (\textit{uiWebView}) do not. Hence, the above regex will not include traffic in web views. 
In our study, we found 169 apps passed 1,148 privacy-sensitive data items (e.g., \textit{Deivce ID, Location data, Diagnostics data}) through browsers to 124 unique endpoints, leading to plausible bypassing of privacy-label disclosure requirement based on Apple's current definition. \looseness=-1

\ignore{
    For apps with web views~\cite{}, Apple required that 
    \textit{``Data collected via web traffic must be declared, unless you are enabling the user to navigate the open web''}~\cite{}. Although such a requirement is not absolutely clear, a typical understanding is, data collection inside web-views of the app must be disclosed, while if the app redirects users to open browsers for data collection practices, app developers are not required to disclose it. 
    In our study, we do observe that some apps abuse this disclosure requirement to hide their data collection practices. 
    For example, an app, called \textit{Banco ABC Brasil Personal}, provides users with financial services (e.g., personal income investments, asset management) with the privacy label only disclosing \textit{Identity} and \textit{Diagnose} collection. However, once users open this app, it redirects users to a webpage along with their precise geo-coordinates sent to its own server \textit{abrasuaconta.abcbrasil.com.br}. 
    More seriously, the webpage further sent precise geo-coordinates data to 50 unique URLs belongs to 9 different domains including \textit{px.ads.linkedin.com, trc.taboola.com, google-analytics.com}.
    To understand how pervasive of such abuse activity, we use regex \textit{\url{ (iPhone).*AppleWebKit(?!.*Version).*Safari}} to map the \textit{``userAgent''} in the traffic header, to obtain those traffic redirecting users to browser (e.g., Safari). Note that standalone mobile Safari user agent strings contain the word ``Version'', whereas \textit{uiWebView} user agent strings do not. Hence, the above regex will not include those traffic in web-views. 
    In our study, we found 169 apps passed 1,148 privacy-sensitive data objects (e.g., \textit{Deivce ID, Location data, Diagnostics data}) through browsers to 124 unique endpoints. 

}
\vspace{-5pt}
\subsection{Noncompliant and Dishonest App Makers}
\label{s:root_cause_app_maker}
\vspace{-5pt}

According to Apple's privacy label requirements \textit{``...(Apps) need to identify all of the data you or your third-party partners collect.."}~\cite{iosprivacylabel}, apps are responsible for the disclosure of data collection by in-app third-party libraries. 
%\vspace{1pt}\noindent$\bullet$\textit{ App maker.}
However, many app owners or developers delegate the app development to app makers, who operate an app builder business to automatically and instantly create apps or landing pages. %Some app developers outsource the development of the app to such third-party app makers. 
In such cases, the actual owners of these apps are never in the position to specify privacy labels, while the compliance of the privacy label relies largely on the awareness and the honesty of app makers.

In our study, we observed 9 app makers that distributed 133 non-compliant apps on behalf of the app developers/owners. %As shown in Table~\ref{table:appMaker}, the top app makers associated with most non-complaint apps are com.goodbarber, com.greencopper, Appy Pie.
Specifically, we observed that all 15 non-compliant apps generated by Appy Pie, a top self-service app maker ranked by ~\cite{appmakerRank}, stated \textit{Data not Collected} in privacy labels (meaning they do not collect data).
However, the privacy policies and app descriptions of these apps disclose certain data collection.
%
%\XY{By check apps made by other three top app makers, we found 85 apps generated by com.goodbarber shared 30 unique privacy labels; 18 apps generated by com.greencopper share 5 unique privacy labels; 5 apps generated by com.yinzcam share 3 unique privacy labels.} \yue{Question : add  other app makers' statistic here or not? as I didn't observe interesting trend here}
%
For instance, the app \textit{Centre Laser Pro}~\cite{Laser20Pro}, an app for beauty treatments, was built by Appy Pie with the privacy label stating \textit{Data Not Collected}. 
However, we observed that coarse location obtained from the IP address was sent to \textit{chatbottest.appypie.com}; also, the precise geographical precise location data along with the user identifier were sent to the endpoint \textit{api.appexecutable.com} which is owned by Appy Pie to perform in-app notification for developer's marketing and advertising purpose. It also sent usage information (e.g., log time, disk space) to \textit{analytics.appypie.com} for analytics purpose. Even worse, Appy Pie also embeds advertising networks such as Google Ads~\cite{googleads} and inMobi~\cite{Inmobi} into apps. Those ad networks also collect and transmit device information (e.g., device ID, screen size, etc) along with other advertising data.

\ignore{
    %\vspace{1pt}\noindent$\bullet$\textit{ App maker.}
    App maker operates an app builder business that delegates app developer to instantly create apps or landing pages. Some app developers outsource the development of app to such third-party app maker. 
    In our study, we observe 9 app makers contribute to distribute 133 non-compliant apps. %As shown in Table~\ref{table:appMaker}, the top app makers associated with most non-complaint apps are com.goodbarber, com.greencopper, Appy Pie.
    Specifically, we observe that all 15 non-compliant apps generated by Appy Pie, a top one self-service app creator ranked by ~\cite{appmakerRank}, claim \textit{Data not Collected}, while privacy policies and app descriptions of those apps are different.
    %
    %\XY{By check apps made by other three top app makers, we found 85 apps generated by com.goodbarber shared 30 unique privacy labels; 18 apps generated by com.greencopper share 5 unique privacy labels; 5 apps generated by com.yinzcam share 3 unique privacy labels.} \yue{Question : add  other app makers' statistic here or not? as I didn't observe interesting trend here}
    %
    For instance, the app \textit{Centre Laser Pro}~\cite{Laser20Pro}, a business app of an beauty treatments company, was built by Appy Pie with the privacy label of \textit{``Data Not Collected''}. 
    However, we observed that coarse location obtained from IP address was sent to \textit{chatbottest.appypie.com}; the precise geographical precise location data along with the user identifier were sent to the endpoint \textit{api.appexecutable.com} which is owned by Appy Pie to perform in-app notification for developer's marketing and advertising purpose. It also sent usage information (e.g., log time, disk space) to \textit{analytics.appypie.com} for analytics purpose. Even worse, Appy Pie also embeds mobile advertising networks such as Google Ads~\cite{googleads} and inMobi~\cite{Inmobi} into applications. Those ad networks also collect and transmit device information (e.g., device ID, screen size etc) along with other advertising data.
}

\subsection{Diverse, Opaque Third-Party Partners}
\label{s:OpaqueDataCollection}

Mobile apps extensively incorporate third-party services (e.g., analytics, advertising, app monetization, or single-sign-on SDK), whose data collection is required to be disclosed in privacy labels by the app developers~\cite{iosprivacylabel}.
The recent human-subject study with 12 iOS app developers~\cite{li2022understanding} showed that developers could have limited awareness of third-party libraries' data use, likely leading to non-compliant privacy labels.
%However, app developers are often unaware of the data collection and usage practices employed by third-party partners used in their apps~\cite{balebako2014privacy,wang2021understanding}.
%
In our study, to understand the non-compliance indeed caused by opaque third-party data collection in the wild,  we characterize the non-compliance with respect to different kinds of third-party service, which can serve as a finer-grained analysis to complement prior works.
%In our study, for understanding the non-compliance indeed caused by opaque third-party data collection in the wild and for a finer-grained analysis complementing the prior work, we characterize the non-compliance with respect to the different kinds of third-party services. 
%Nevertheless, we observe XXX data receivers \xiaojing{why only data receivers not the whole data flow?} are third-party partners from privacy non-compliance  apps. 
%
%To understand the non compliance caused by such opaque data collection, we look into those leaked-data receivers \xiaojing{??}. 
Specifically, for apps we found with non-compliant privacy labels (\S~\ref{sec:measurement}), we re-examine the app traffic and extract domain names associated with data transmission to third-party partners. 
%\XY{Here, to eliminate domains of app developer, we calculate frequency of domain occurrence over apps. If it only appears once, we believe it is high possible belongs to app itself.}
%\xiaojing{how you eliminate the domains of the app developer}. 
%
Next, we look up the domain owners using \ignore{the API of }domain WHOIS~\cite{whois} and Crunchbase~\cite{Crunchbase}.
Further, we use the corporate categories in Crunchbase \cite{Crunchbase} to classify those third-party partners into three categories: \textit{data broker}, \textit{service provider}, and \textit{advertising, analytics, and marketing}. We characterize the nuances of their non-compliance with different third-party partners as follows, which can help platform owners and policy-makers pinpoint issues with third-parties and improve regulations.

\ignore{
    mobile apps extensively incorporate third-party services (e.g., analytics, advertising, app monetization, or single-sign-on SDK). According to Apple's privacy label requirements \textit{``...(App developers) need to identify all of the data you or your third-party partners collect.."} ~\cite{iosprivacylabel}, app developers are responsible for the disclosure of data collection by third-party partners.
    However, app developers are often unaware of the data collection and usage practices employed by third-party partners used in their apps~\cite{balebako2014privacy,wang2021understanding}.
    In our study, to understand the non-compliance caused by such opaque data collection, we look into those third-party services and their data flows. 
    %Nevertheless, we observe XXX data receivers \xiaojing{why only data receivers not the whole data flow?} are third-party partners from privacy non-compliance  apps. 
    %
    %To understand the non compliance caused by such opaque data collection, we look into those leaked-data receivers \xiaojing{??}. 
    Specifically, for apps we found with non-compliant privacy labels (\S~\ref{}), we re-examine the app traffic and extract domain names associated with data \LX{transmission to} third-party partners. 
    %\XY{Here, to eliminate domains of app developer, we calculate frequency of domain occurrence over apps. If it only appears once, we believe it is high possible belongs to app itself.}
    %\xiaojing{how you eliminate the domains of the app developer}. 
    %
    Next, we look up the domain owners using \ignore{the API of }domain WHOIS~\cite{whois} and Crunchbase~\cite{Crunchbase}.
    Further, we use the corporate categories in Crunchbase \cite{Crunchbase} to classify those third-party partners into four categories, \textit{app maker}, \textit{data broker}, \textit{service provider}, and \textit{advertising, analytics, and marketing}. \LX{We characterize the diverse xxx with different third-party partners as follows.}
}

\vspace{1pt}\noindent$\bullet$\textit{ Service provider.} %Service providers are entities which help the app build part of functionalities (e.g., push notification, user-subscription, auditing, \ignore{detecting security incidents, debugging, or performing services on behalf of the business,}etc.)~\cite{CCPAServiceProvider}. 
    \ignore{App developers take the advantage of convenient services provided by service providers, while neglecting the data collection and usage behaviors of them.}
%
%As an instance, the app \textit{Diners Club}~\cite{DinersClub} which provided travel recommendations (e.g.,  hotel, shopping offers, scenic spots) to users.  It utilized weather data query service API \textit{\url{https://api.openweathermap.org/data/3.0/onecall?lat={lat}&lon={lon}&exclude={part}&appid={API key}}}, \xiaojing{XXX API} provided by \textit{Open Weather}~\cite{weatherAPI}, which is a service provider provides weather data forecast.
%It utilized postal address query service API provided by \textit{Saudi Post}~\cite{saudi}, which is a national postal service provider in Saudi Arabia. 
%Specifically, the app sent the precise geographical coordinates along with app registered key to query the weather data used to personalize a list of recommended products to users. However, the app developer did not disclose such data in its privacy label.  
%
In our study, non-compliance of 854 apps is caused by opaque data collection by third-party service providers. Contrary to the previous common understanding which attributes privacy non-compliance to opaque third-party disclosure guides, we find that even when the third-party librareis declare the collection and usage of data, such information sometimes cannot be leveraged by app developers to create compliant privacy label.
%Contrary to prior understanding that commonly attributed privacy non-compliance to opaque third-party data practices, we find that even when third-party libraries offered transparency for their data practices, the app developers still did not leverage them to create compliant privacy labels.
%
%The top service provider associated with most (250) non-compliant apps is OneSignal, which helps apps implement push notifications and in-app messaging.
%Table~\ref{table:serviceProvider} shows the top \qyue{3 service providers} observed in most privacy-violated apps.
%OneSignal, which is used for push notifications and in-App messaging, was integrated by most privacy-violated apps, followed by appcenter, and jpush.
%
The top service provider associated with most (250) non-compliant apps is OneSignal, which helps apps implement push notifications and in-app messaging.
As an instance, an app, called \textit{IUOE 513}, providing general news, training, and course information for their members in the union, utilizes \ignore{the service (e.g., push notifications, in-app messages)}the OneSignal SDK to increase user stickiness. However, some sensitive information (e.g., \textit{social\_security\_number, employee\_ID, job\_site\_location etc}) are sent to \textit{api.onesignal.com} without disclosure in the app's privacy label. We found that OneSignal actually provides configurable APIs \textit{OneSignal.sendTag("key", value: "value")} for app developer to share user information. That is, any information collected by OneSignal is explicitly provided by the app developers (by calling the API of OneSignal SDK), and thus known to the app developers. Despite the transparency, the app's privacy label missed the disclosure related to OneSignal.

%It indicates that such data collection from service providers are aware of by app developers, but developers still ignore the appropriate disclosure.

\ignore{
    As an instance, an app, called \textit{IUOE 513}, providing general news, training and course information for their members in the union, utilizes \ignore{the service (e.g., push notifications, in-app messages)}the OneSignal SDK to increase user stickiness. However, some sensitive information (e.g.,\textit{social\_security\_number, employee\_ID, job\_site\_location etc}) are sent to \textit{api.onesignal.com} without disclosure in the app's privacy label. We found that OneSignal actually provides configurable APIs \textit{OneSignal.sendTag("key", value: "value")} for app developer to share user information. It indicates that such data collection from service providers are aware of by app developers, but developers still ignore the appropriate disclosure.
}

%All other event and user properties can be set using Data Tags. Setting this data is required for more complex segmentation and message personalization.

%#appcenter

%\input{table/topAppMaker}
%bendingspoonsapps (app maker)

%https://chatbottest.appypie.com/getAdress?ip=73.158.37.244
%{
  %"ip": "73.158.37.244",
  %"CountryCode": "US",
  %"CountryName": "United States",
  %"CityName": "San Mateo",
  %"subdivisions": "California",
  %"symbol": "$",
  %"name": "US Dollar",
  %"continent": "North America"
%}
%https://www.google-analytics.com/collect

%https://ipv4.icanhazip.com/

\vspace{1pt}\noindent$\bullet$\textit{ Advertising, analytics, Marketing and their affiliates.}
Based on the definition in CCPA \cite{ccpa}, the advertising, analytics, and marketing SDKs are entities that do not qualify as either data collectors or service providers that can obtain the personal information of a consumer from a business. The app developers usually integrate those SDKs for app monetization, user behavior measurement, scaling marketing campaigns, etc. 
%Previous study~\cite{schindler2022privacy,moonsamy2014android,short2014android}, focused on the Android platform, has demonstrated that third-party libraries leaked privacy data. Our results are aligned with those on Android. 
In our study, we observed 2,086 non-compliant behaviors from those advertising and analytics SDKs in iOS apps. For example, the app \textit{``CloudMall''}, providing coupons and price comparison\ignore{ from e-commerce websites}, declared only to collect \textit{Diagnostics data} for analytics purpose in its privacy label. However, we observed that it sent various user data to different advertising and analytics SDKs (e.g., IDFA/IDFV to Adjust SDK, User ID, and Email Address to Google Analytics SDK).

%\qyue{The top 5 advertising, analytics and marketing parties are Google Analytics~\cite{firebase}, Facebook~\cite{facebookGraph}, 
%Google Ads~\cite{googleAdvertising}, 
%Branch.io~\cite{branchio}, and Adjust~\cite{adjust};} The details are shown in Table~\ref{table:topAdvertiser}.
%\input{table/topAdvertiser}
% Google’s advertising service DoubleClick. Every app transmitted data to various parts of the Google system, and all of them had integrated various Google SDKs, including Google Ads, Google Crashlytics, and Google Firebase. \url{https://nakedsecurity.sophos.com/2020/01/16/apps-are-sharing-more-of-your-data-with-ad-industry-than-you-may-think/}

%Participants also relied on third-party tools for other uses, such as analytics or various other features. Participants seemed generally unaware of the privacy and security practices em- ployed by third-party utilities used in the development of their apps. Many developers had not personally read the terms of service, were unsure if their lawyers or legal departments had done so, and may have even forgotten the names of the ad networks or web traffic analysis companies they had used. ----balebako2014privacy

%\input{table/cloudmall}

\vspace{1pt}\noindent$\bullet$\textit{ Data broker.}
Data brokers are businesses that knowingly collect and sell the personal information of consumers with whom they do not have a direct relationship. A data broker, required by California law, needs to register with the Attorney General. A list of 469 registered data brokers~\cite{databrokers} is provided by the State of California Department of Justice. To check whether the endpoint is owned by a data broker, we first obtain the company name by using Crunchbase API and then look up the above list to determine if the data receiver is a data broker. In our study, we found that 273 non-compliant apps leak sensitive data to 22 data brokers. The top four data brokers are \textit{unity3d},
\textit{adcolony}, \textit{kochava}, and \textit{taboola} (see details in Appendix Table~\ref{table:dataBroker}). %As an example, an app, called \textit{Jigsaw Puzzle HD} sent \textit{Device ID, User ID, performance Data} to \textit{\url{https://control.kochava.com/track/json}} used to track users. 
%https://broker.datazoom.io/broker/v1/logs
%rayjump

\ignore{
\subsection{Confusing Optional Disclosure Requirement}
\label{s:developerMisUnderstanding}
%We also found that app developers are unclear about how to disclose privacy label for its own code behavior.
%
%Cranor’s research group pointed out that app developers had trouble creating accurate labels for their apps because they didn’t understand all the terminology and policy \cite{consumerreports}.
%
In our study, we found app developers tend to misinterpret data collection as optional disclosure due to the failure of considering all criteria of optional disclosure. 
For example, the app \textit{Trace travel} is a travel social software that provides services for individual travel user or small groups of self driving tour.  We observed that it collect user precise location frequently (48 times in three minutes) and sent the data to different paths under its own domain to provide guidance and service such as travel strategy, navigation, voice explanation, travel service and track recording of tourism destination. However, the app didn't disclose precision location in its privacy label. The app developer could misinterpret this data collection as optional disclosure because it has obtain user location permission and sends data to its own server only for providing services for the users. Yet it didn't satisfy the last requirement for optional disclosure which is \textit{``Collection of the data occurs only in infrequent cases that are not part of your app’s primary functionality, and which are optional for the user.''~\cite{iosprivacylabel}}.
From the other side, although in some apps, data collection is only taken place once, the purpose of such data usage is for third-Party Advertising, developer's Advertising or Marketing purposes, or sharing with data brokers, which is also not satisfied with the criteria of optional disclosure. For example, there is a game app, called \textit{Monster Battles: TCG} transmitted \textit{Device ID} to the endpoint \textit{control.kochava.com/track/json} belongs to the data broker \textit{taboola}. Although such data collection only happen once, it is shared with a data broker, which must disclose it to the user based on Apple policy~\cite{iosprivacylabel}.
In our study, we found XXX non-compliance apps satisfied part of criteria of optional disclosure, but not all of them. 
}

%1. Developer are not clear about “optional for disclosure” \\
%2. Developer are not clear what data belongs which category (IDFV, ip address) \\

%The most common location-based services such as locationbased search and navigation rely on relatively infrequent,not-always-on location tracking
\ignore{
\subsection{Compare Privacy Label $\&$ Privacy Policy  }
\label{s:missingUnderstandingDataType}
Serving as a rundown of the privacy policy, the privacy label summarizes the data collection statements in a clear, uniform, and digestible format. 
Therefore, both the errors in the privacy policy itself and the problems in the process of summarization will bring the privacy label into inconsistent disclosure of privacy practices.

More specifically, as for the first cause, if the privacy policy fails to reflect the practical data usage purposes and the privacy label summarizes the problematic privacy policy with full agreement, the flow-to-policy inconsistencies will also appear between the privacy label and the data flows.
In this case, the privacy label is inconsistent disclosure of associated data flows, while it is consistent with the associated privacy policy statements.
Therefore, for the data flows that suffer from flow-to-label inconsistency, we extract their associated privacy policy statements and compare them with the flow-relevant privacy label.
Among 3,281 neglect disclosure, 1,628 contrary disclosure, and 677 inadequate disclosure in flow-to-label inconsistency, there are 2,346, 973, and 402 cases where the privacy label is consistent with the associated privacy policy statements.
In such cases, the inconsistencies between the privacy label and the data flows indeed result from the erroneous privacy policy.
%that the data usage purposes are missing, contrary, or inadequate in the privacy policy.
%
In other words, the privacy label summarizes the unreliable privacy policy where the practical data usage purposes are already missing, contrary, or inadequate, leading to inconsistent disclosure.

In addition, the omissions and errors in refining the privacy label from the privacy policy may also cause the privacy label to fail to properly disclose data usage purposes. 
To study this cause, we extract the statements in the privacy policy and associate them with the privacy label by the relevant data objects.
Further, we analyze the inconsistency between the privacy label and the associated privacy policy statements according to our inconsistency model (see Section~\ref{sec:model}).
Through the analysis, we find the privacy label of XXX apps suffer from XX neglect disclosure, XX contrary disclosure, and XX inadequate disclosure
of the corresponding privacy policy statements.

Besides, for the data objects that cannot be observed in real network traffics (e.g., XXX), we only compare the privacy policy statements and the privacy label associated with them. Among XXX apps, we find XX neglect disclosure, XX contrary disclosure, and XX inadequate disclosure in the associated policy-to-label pairs.
The results indicate that most \qy{to check} app developers fail to properly extract and summarize the privacy policy statements into privacy label.
% privacy label 是privacy policy 的总结
% 因此privacy policy本身的错误 以及label对policy错误的总结都会导致label上的inconsistent disclosure
% specifically, 如果privacy policy 本身有问题， 而label 完全一致地反映policy， 则label 呈现出与policy一致的对于flow的inconsisency
%这种情况对应了label 和flow 有inconsistency， 而label 和flow \XY{policy} 是consistent 的
%因此，我们找到与label-flow inconsistency中 的data相关的policy, 并将其与label比较
%其中， 在xx neglect, xx incorrect, xx inadequate flow-to-label inconsistency 中， 分别有xx, xx, xx cases 的label 和policy是完全一致的。
%这说明这些flow-to-label的inconsistency 实质上是由于 data usage purpose 在policy中就已经被 neglected， incorrect, 或inadequate导致的， 也就是说是因为privacy label是在错误的privacy policy的基础上进行总结，从而错误的disclose data usage purpose.
% 另一方面，在由privacy policy总结label过程中出现的遗漏和错误也会导致privacy label无法正确disclose data usage purpose. 
% 为此， 我们extract了privacy policy中的statement并将其与相关的privacy label通过associated data联系起来， 并分析label与policy之间的inconsistency, according to our inconsistency model (see Section 3). 
% 通过分析我们发现了XX neglect， XX incorrect， XX inadequate disclosure between label and policy. 在这些情况中， privacy label 没能对privacy policy中的statement 进行正确的提取和总结，从而不能全面地反映出app's data transfer 中实际的data usage purpose.

}
\ignore{
\subsection{Inconsistent Privacy Policies.}
\label{s:missingUnderstandingDataType}
\xiaojing{this part should be updated based on our discussion}
The privacy label acts as a rundown of lengthy privacy policy in a clear, uniform, and digestible format. Ideally, the privacy label should appropriately reflect each data collection statement in privacy policy. However, failed to summarize data practices fairly well from privacy policy can also cause that the privacy label can not appropriately reflect code behavior. As defined in \ref{s:sampling},  a non-compliance app detected in the set $\gamma$ indicated the failure of summarization from privacy policy to privacy label. In our study, we detect XXX\% such non-compliance  apps due the obstacle of the correctly mapping privacy label from policy. 
To further examine the reason, we found that such error-prone mapping mainly caused by different data scope and granularity defined in this two regulations. 
As mentioned before, Apple defined a limited set of data items (32) for developers to disclose in privacy label. Differently, required by the law regulators (e.g., GDPR \cite{}, FTC \cite{} etc.), the app developer need to disclose all information relating to an identified or identifiable natural person in privacy policy. Apple required the developer to refer to pre-defined 32 data types and compare them to the data collection practices. Some data item are defined at exact same way in privacy policy (e.g., email address) which is clearly for developer to identify; while some other data items defined at a different semantic granularity (e.g., Device ID defined in privacy label can cover xxxx, xxx, xxx etc in privacy policy) which required developer to know what kind of data is semantically subsumptive to it.  Consequently, given a data object protected in privacy policy, it may be hard for developer to find an appropriate data definition in privacy label. 
}

%2. Redirect data to safari and send to third-party tracking service.
%So, the detection script can be modified to work with the latest version of iOS like so:

\ignore{
\vspace{1pt}\noindent$\bullet$\textit{ Reuse outdated tracking technique.} With increasing strict privacy label disclosure requirements defined by Apple, we observe XXX apps reuse outdated tracking technique (e.g., tracks users’ rough location based on IP address) potentially as a workaround for accessing geolocation without triggering a permission request; For example, an app, called \textit{Nine Squares UAE}, is a restaurant App provides food \& drink services. It queries a geolocation service, which leverages reverse IP lookup to obtain  general location of the user (approximately at the city level). Specifically, it sent a request to \textit{geolocation.onetrust.com} to obtain the response including location-related data 
(\textit{country, state, stateName, zipcode, timezone, latitude, longitude, city, continent}) and further re-sent them to its own server.
}

\ignore{
To facilitate inference of purpose, \toolname{} extracts both code and network
features by dynamically executing the app. 
The code features include the caller, sensitive system API, the frequency of invocation,  which are generated from execution traces. The network features consist of endpoint, request and response which are recorded by a proxy. 
The two types features can concatenate into a six-tuple (caller, system API, frequency, endpoint, request, response) to represent client-side code behavior. In our groundtruth dataset, each app generated XXX tuples on average. 
In the following, we will describe the indication of each element and what kind of combination of them can infer the purpose.  

\vspace{2pt}\noindent\textbf{System API}. We consider an API as privacy sensitive if its return value is governed by the Apple privacy label disclosure policy. We utilize return value by (1) searching it in network traffic to determine whether data is transmitting data off the device; (2) mapping it to 32 data objects defined by Apple to 
determine which data object should be disclosed in privacy label. 

\vspace{2pt}\noindent\textbf{Caller}. The caller of sensitive API can either be app itself or third-party SDKs. The caller can indicate whether a data leakage is passive or active. For example, supposed that there is a traffic which leaked data to the third-party, if caller is app itself, which means the app developer intentionally share data with third-party partners; otherwise, it is third-party SDKs actively call the sensitive API, app developer could be innocent about such data leakage.   
%apply a majority voting heuristic

\vspace{2pt}\noindent\textbf{Frequency}. The frequency shows how many times the sensitive API is being invoked in a certain time slot. This factor help us determine whether data collection meets one of the criteria of optional disclosure define by Apple, which is \textit{``Collection of the data occurs only in infrequent cases that are not part of your app’s primary functionality, and which are optional for the user.''}
%we empirically set a threshold for the code similarity check based on preliminary manual analysis of 1,000 cloaked phishing websites. We consider only two categories to find a threshold: correctly labeled cloaking types and mislabeled cloaking types

\vspace{2pt}\noindent\textbf{Endpoint}. To understand who receive the sensitive data (advertiser or analyzer or service provider or app itself), we collected domain categories provided by Crunchbase \cite{} querying their website. For every domain, Crunchbase returns a list of domain categories.

\subsection{inconsistency check}
Over xxx days, we used xxx ios devices to collect data on xxx apps, each running the monkey for 3 minutes, and intercepted xxx million unique traffic requests. 
}
\vspace{-5pt}
\section{Related Work}
\label{sec:relatedwork}
\vspace{-2pt}
\noindent\textbf{Study on privacy labels.} 
Privacy nutrition labels have been proposed in the literature~\cite{kelley2009nutrition,kelley2010standardizing, kelley2013privacy, emami2020ask,emami2019exploring} and Apple's privacy label is considered the first large-scale adoption.
%The idea of developing privacy or security breakdown labels for a software product is not new. In the early 2010s, Kelley et al.~\cite{kelley2009nutrition,kelley2010standardizing} proposed the simple visualizations called ``privacy nutrition labels''  to inform the user how their personal information is collected, used, and shared by a website, which improves the visual presentation and comprehensibility of privacy policies. 
%
%A few years later, the researchers applied this idea to the mobile app field. Kelley et al.~\cite{kelley2013privacy} designed a short ``Privacy Facts'' display in the Google Play Store, which lists what data the app will collect and use. They found that ``Privacy Facts'' can impact users’ decisions when choosing between comparable apps. 
%
%Recently, Naeini~\cite{emami2020ask,emami2019exploring} proposed a prototype of the IoT privacy and security label to help consumers make more informed IoT-related purchase decisions.
%To address that issue, researchers have tried to improve privacy policies’ readability and usability. (For example, Patrick Kelley and his colleagues developed a nutrition-label style presentation of policies.11) Simpler notices do satisfy an important informational need. 
%
Li et al.~\cite{li2022understanding} perform a human subject study with 12 app developers to understand the challenges to create privacy labels from the developers' perspective. 
%The study reveals that nine out of the 12 participants made errors when creating or re-creating their own apps' privacy label. 
The challenges include knowledge gaps about Apple's privacy concepts, unclear third-party data use, extra workloads, etc. 
Li et al.~\cite{li2022understandinglongitutde} performs a longitude study to understand how promptly developers create and update privacy labels.
Compared with previous works, we perform a systematic study on Apple privacy label compliance. In particular, we first time investigate data-flow to privacy-label inconsistency with new methodology and end-to-end tools, and deliver new insights for improving requirements and regulations for privacy-label design and compliance.% .

%into the privacy label compliance analysis considering code behaviors. 

%a new understanding of underlying causes of erroneous labels (e.g., to what extend erroneous labels are caused by opaque data collection of third-party vendors\S~\ref{s:OpaqueDataCollection} and dishonesty of third-party privacy label guidance\S~\ref{s:erroreousLabelGuidance}, developer's mis-understanding\S~\ref{s:developerMisUnderstanding}, and what data are prone to be mis-interpret from privacy policy\S~\ref{s:missingUnderstandingDataType} etc.)  

\vspace{2pt}\noindent\textbf{Privacy compliance check on mobile apps.} 
%\xiaojing{add compliance check papers here}
In recent years, privacy compliance check~\cite{nan2015uipicker, nan2018finding, zimmeck2019maps,andow2019policylint,slavin2016toward, wang2018guileak, zimmeck2016automated,chen2019demystifying, andow2020actions,yu2016can, breaux2013formal} are evolving from coarse-grained analysis to complex but fine-grained analysis, from the data-level consistency, which checks whether the return value of sensitive API is described in the policy~\cite{slavin2016toward,zimmeck2016automated}, to data-purpose-level consistency~\cite{bui2021consistency}.
PPChecker \cite{yu2016can} integrated ``action'' (e.g., collect or not collect) into the consistency model and introduced an inconsistent case  where the privacy policy declares that the app will not collect user information, but the app does. 
Recent work (e.g., PoliCheck \cite{andow2020actions}) take into account the entities (third-party vs first-party) of the personal data and proposed an entity-sensitive consistency model. They introduce a new set of entity-sensitive conflicting rules. 
%For example,  such a case is considered inconsistent if the data receiver is a third-party from a mobile app but declared in policy to be first-party.
%
PurPliance \cite{bui2021consistency} propose a new consistency model by extending the PoliCheck  \cite{andow2020actions} through incorporating data usage purposes. They define a flow to be consistent with a privacy policy if there exists a privacy policy statement where the data, entity, and purpose are semantically subsumptive or equivalent to those extracted from the flow. 
On contrary to previous work, we define a new consistency model for privacy label compliance analysis (\S\ref{sec:model}), which enables three types
of inconsistency analysis between the data usage portfolios disclosed in privacy labels and actual data flows in an iOS app.

\ignore{
PiOS~\cite{egele2011pios} performed static analysis on iOS app to study the threat of user private data leakage. Compared with PiOS, Iris~\cite{deng2015iris} improved resolution rate of call targets through the combination of static and dynamic analysis.
PMP~\cite{agarwal2013protectmyprivacy} was a crowdsourcing-based system for iOS users to decide fine grained privacy protection settings by allowing substituting anonymized to be sent instead of privacy sensitive information to protect user privacy. 
Deshotels et al.~\cite{deshotels2016sandscout} analyzed the security properties of iOS sandbox profiles and discovered seven classes of exploitable vulnerabilities. 

Rastogi et al~\cite{rastogi2013appsplayground} proposed a dynamic testing technique based on automatic system event triggering. They found 23.8\% (out of 3,968) applications to be leaking information to Internet, among which,  most of them are sending data to third party ads and/or analytics libraries. However, they determine the data receiver is third party ads and/or analytics by checking whether its domain relate to at least two app creators. 
 
Reyes et al~\cite{reyes2018won} presented a dynamic analysis framework to analyze mobile apps’ compliance with the Children’s Online Privacy Protection Act (COPPA). They found 19\% (out of 5,855) identifiers or PII are collected by SDKs. Although many of SDKs provide configuration options to disable tracking and behavioral advertising, app developer usually ignore them.  
}
%\vspace{2pt}\noindent\textbf{Flow inconsistency analysis.} 

%\input{cfp}

\vspace{-10pt}
\section{Conclusion}
\vspace{-5pt}
We performed a systematic study on Apple privacy label compliance, in particular focusing on the inconsistency between privacy labels and app data flows/practices for the first time. Specifically, we design and implement \toolname{} as an automated privacy label compliance analysis system by adapting and synthesizing a set of innovative techniques including automatic iOS app UI execution, NLP, and dynamic and static binary analysis.
Running on 5,102 iOS apps, \toolname{} detects 3,423 privacy label non-compliance, which brings to light the significant impact of such non-compliance apps: they are indeed prevalent in App Store, even associated with high-profile apps and various kinds of privacy-sensitive data collection practices.
Our research further uncovers a set of root causes of privacy label non-compliance including opaque data collection from diverse third-party partners and misleading privacy label disclosure guidance, etc.
Our findings and artifact bring new insight into the privacy label compliance analysis and contribute to improving the design of privacy label and compliance requirements.

\section*{Acknowledgement}
We thank the following graduate/undergraduate students Feifan Wu, Jack Ruocco, Sultan Aloufi, Keegan Allred, and Zixiao Pan for their efforts of data collection and annotation, cases study and analysis.
%

%\section*{Acknowledgement}
%
  
%

%-------------------------------------------------------------------------------
% \normalem
% {
% \footnotesize
% \setstretch{0.5}
% \setlength{\bibsep}{5pt}
\bibliographystyle{plain}
\bibliography{ref}
% }

%\appendix
\begin{appendices}
\section{}
% arara: pdflatex: {shell: true}
\subsection{Falsely detected inconsistency}\label{app:false_detected}
We found three \textit{Product Personalization} have been falsely labeled to \textit{App functionality}. We found it is challenging to distinguish this class to app functionality as they usually shared some similar feature (e.g., sent data to its own domain). For example, there is a false positive comes from an app, called \textit{Trusted House sitters}, which helps user to look for a kind and caring house and pet sitter. In that traffic, Location data is used for product personalization purpose as it is sent to its own server to query nearby pet caring house/sitter and further four pet sitters info (e.g., profile photo, name, interests etc) are returned from server. However, its request also appends the current user profile (user\_name,verification\_status) which may misguide the classifier considers it as user authentication for app functionality purpose. 

The two \textit{Developer’s Advertising or Marketing purposes} have been falsely labeled to \textit{Analytics}. Sometimes,  marketing strategy, especially for promotion to targeted user, are inseparable from applicable analytics metrics (e.g., user event, product meta data, device settings). In such cases, if the request body contains a large proportion analytics metrics served for marketing purpose, the classifier may label it as analytics. 
For example,  a false positive case of an app, called \textit{GroupCal-Shared Calendar}, shared \textit{Phone Number} to \textit{clevertap}, a marketing vendor, to implement cross-channel messaging to accelerate  marketing performance. However, in its request body, it also collects lots of device meta (e.g., \textit{os\_version, Model, Carrier etc}) and user behavior (e.g., \textit{event\_name, event\_data etc}), which leads it to be falsely labeled as \textit{Analytics}. 

\subsection{Figures and Tables}

\begin{table}[ht]
\centering
\footnotesize
\caption{Top Service Provider}
\begin{tabular}{l|c|c|c}
\hline
\textbf{Company} & \textbf{Leaked data} & \textbf{Endpoint} & \textbf{\#}\\ \hline

\scriptsize OneSignal &  \makecell{\scriptsize Device ID, Sensitive Info,\\\scriptsize Location, Email Address,\\ \scriptsize Other Diagnostic Data etc}
 &  \textit{\makecell{\scriptsize api.onesignal.com\\\scriptsize onesignal.com}} & 250\\ \hline

\scriptsize Appcenter &  \makecell{\scriptsize User ID, \\ \scriptsize Other Diagnostic Data}&  \textit{\makecell{\scriptsize codepush.appcenter.ms \\\scriptsize in.appcenter.ms \\}} & 224\\ \hline

\scriptsize Jpush &  \makecell{\scriptsize User ID, Device ID \\ \scriptsize  Other Diagnostic Data} &  \textit{\makecell{\scriptsize gd-stats.jpush.cn\\\scriptsize user.jpush.cn}} & 58\\ \hline

\end{tabular}
\label{table:serviceProvider}
\end{table}

\begin{table}[t]
\centering
\footnotesize
\caption{Top Data Brokers}
\begin{tabular}{l|c|c|c}
\toprule
\textbf{Company} & \textbf{Leaked data} & \textbf{Endpoint} & \textbf{\#}\\ \hline

\scriptsize unity3d &  \makecell{\scriptsize Device ID, User ID,\\\scriptsize Other Diagnostic Data,\\ \scriptsize Performance Data} &  \textit{\makecell{\scriptsize perf-events.cloud.unity3d.com \\\scriptsize cdp.cloud.unity3d.com\\\scriptsize httpkafka.unityads.unity3d.com}} &170 \\ \hline

\scriptsize adcolony &  \makecell{\scriptsize User ID, Device ID\\ \scriptsize Coarse Location\\ \scriptsize Other Diagnostic Data \\ \scriptsize Performance Data} &  \textit{\makecell{\scriptsize ads30.adcolony.com \\\scriptsize iosads4-1.adcolony.com \\ \scriptsize iosads4-6.adcolony.com \\ \scriptsize adc3-launch.adcolony.com}} & 18\\ \hline

\scriptsize taboola &  \makecell{\scriptsize User ID, Device ID \\ \scriptsize Precise Location} &  \textit{\makecell{\scriptsize trc.taboola.com \\\scriptsize trc-events.taboola.com \\\scriptsize 15.taboola.com}} & 13\\ \hline

\scriptsize kochava &  \makecell{\scriptsize User ID, Device ID \\ \scriptsize Performance Data} &  \textit{\makecell{\scriptsize control.kochava.com}} & 11\\ \bottomrule

\end{tabular}
\label{table:dataBroker}
\vspace{-18pt}
\end{table}

\begin{table}[]
\centering
\footnotesize
\caption{Crawl Configurations}
\label{table:crawConfiguration}

 \begin{lstlisting}[basicstyle=\ttfamily\scriptsize]
    /** Crawling Action Target  */
    function NSCrawlerConfig() {
      this.testingPeriod = 0.05 * 60 * 60;
      this.testingDepth = 5;
      this.takeScreenShot = true;
      this.newCommandTimeout = 2;
      this.launchTimeout = 3;
      this.packages = '.*';
      this.maxActionPerPage = 15;
      this.exclusivePattern = '';
      this.platform = 'ios';
      this.clickTypes = ["Button"];
      this.editTypes = ["TextField"];
      this.tabBarTypes = ["TabBar"];
      this.strategy = 'source';
      this.depth = 100;
      this.duration = 1800;
      this.deviceType = 'iPhone';
    }

     \end{lstlisting}

\end{table}

\begin{table*}[ht]
\centering
\footnotesize
\caption{Data items covered by our study}
\begin{tabular}{|l|c|c|}
\hline

\textbf{Data items} & \textbf{Inference Source} &  \textbf{Example}\\ \hline

\multirow{2}{*}{Device ID} & key pattern & \makecell{\textit{device\_id}, \textit{device\_identifier}, \textit{d\_id}, \textit{device\_token}, \textit{unique\_device\_no}, \textit{randomized\_device\_token}}\\ \cline{2-3} 
                        & return value & \makecell{\textit{[ASIdentifierManager advertisingIdentifier]}} \\ \hline
                        
User ID & key pattern & \makecell{\textit{user\_id}, \textit{appsflyer\_user\_id}, \textit{adjust\_user\_id}, \textit{anon\_user\_id}, \textit{customer\_id}, \textit{udid} }\\ \hline

Name & key pattern & \makecell{\textit{name,nick\_name,avatar,full\_Name,first\_name,last\_name, mid\_name,family\_name,middle\_name} }\\ \hline

Email Address & key pattern & \makecell{\textit{email,contact\_email,email\_address, hashed\_email\_address} }\\ \hline

Phone Number & key pattern & \makecell{\textit{display\_phone\_number,mobile\_phone,encrypted\_phone\_number,telephone\_mobile,raw\_phone\_number} }\\ \hline

\multirow{2}{*}{Health} & key pattern & \makecell{\textit{heart\_beat\_frequency\_ms, blood\_type,body\_weight,body\_height,blood\_pressure, biological\_sex, skin\_type} } \\\cline{2-3} 
                        & return value & \makecell{\textit{[HKHealthStore bloodTypeWithError:], [HKHealthStore dateOfBirthComponentsWithError:]}}\\ \hline

Physical Address & key pattern & \makecell{\textit{home\_address, mailing\_address, postal\_address} }\\ \hline

Advertising Data & key pattern & \makecell{\textit{ad\_token,ad\_type, ad\_unit, ad\_info,ad\_values,ad\_url,ad\_slots, adsafe\_type,ad\_format,ad\_placement\_id}}\\ \hline

\multirow{2}{*}{Contacts} & key pattern & \makecell{\textit{name\_prefix,email\_addresses,social\_profiles,phonetic\_given\_name, birthday, postal\_addresses} } \\\cline{2-3} 
                        & return value & \makecell{\textit{[CNContact namePrefix],
[CNContact postalAddresses],
[CNContact birthday]}}\\ \hline

\multirow{2}{*}{Precise Location}& key pattern & \makecell{\textit{latitude, longitude, altitude, lat, lng, alt, geo\_lat}}\\\cline{2-3}   & return value &\makecell{\textit{[CLLocation coordinate], [CLLocationManager requestLocation]}}\\ \hline

Coarse Location& key pattern & \makecell{\textit{ip\_address, city, neighborhood\_address, postal\_code, isp\_address }}\\ \hline

Sensitive Info& key pattern & \makecell{\textit{credentials\_type, religion\_type, social\_security\_number,gender, birthday,access\_token}}\\ \hline

Product Interaction & key pattern & \makecell{\textit{click\_area,,time\_show, video\_mute,click\_sensor\_events,mouse\_move,click\_event, launch\_time} }\\ \hline

\multirow{2}{*}{Performance Data} & key pattern & \makecell{\textit{disk\_storage,battery\_state, battery\_level, storage\_bytes\_available,device\_battery\_percent,battery\_info} }\\ \cline{2-3}   & return value &\makecell{\textit{[UIDevice  batteryLevel],
[UIDevice systemVersion],
[UIDevice batteryMonitoringEnabled]}}\\ \hline

Other Diagnostic Data & key pattern & \makecell{\textit{os\_version,screen\_size,screen\_height,screen\_width,device\_type,operating\_system\_version} }\\ \hline

\end{tabular}
\label{table:data_scope}
\end{table*}

\begin{table*}[ht]
\renewcommand\arraystretch{1.5}
\centering
\scriptsize
\caption{Inappropriate Privacy Label Guidance. \textit{``N''} in original guidance means the data is claimed ``not collected'' in privacy label guidance, but corresponding data object have been observed in the network  to its endpoints from the network traffics extracted from non-compliance apps. \textit{  ``M''} means the data is missing disclosed in original guidance, but we observe it in network traffic. \textit{``V''} means the data collection statements are vague without explicitly describing data not collected or collect in what configuration under which version, but we observe it in the network traffic.}
\begin{tabular}{c|c|c|c|c|c}
\toprule
\textbf{Company} & \multicolumn{2}{c|}{\textbf{Original Guidance}}&  \textbf{Sent Data Example} &\textbf{Endpoint} &\textbf{\# of App}\\ \hline

Adjust~\cite{adjustGuidance} & \makecell{Precise Location \\}&N  & \makecell{ \{ ``name'':``callback\_params'',\\
``value'':``\{"geo\_longitude":"-\#\#.\#\#\#\#\#",\\"geo\_latitude":"\#\#.\#\#\#\#\#"\}''
\} }&\textit{app.adjust.com} & 56\\ \hline

Appsflyer~\cite{appsflyerSDKguidance} & \makecell{Precise Location }&N  & \makecell{ \{ ``locationHref'':``location'',\\
``value'':``\{"long":"-\#\#.\#\#\#\#\#\#",\\"lat":"\#\#.\#\#\#\#\#\#"\}''
\} }&\textit{wa.appsflyer.com} & 1\\ \hline

\multirow{4}*{Clevertap~\cite{clevertapguidance}  }& \makecell{Name}&\makecell{M} & \multirow{4}*{\makecell{\{ 
``evtData'':``\{``email'':``\#\#\#\#\#@gmail.com'',\\``last\_name'':``\#\#\#\#'',``ifaA'':false,``wifi'':false,``\\Longitude'':-\#\#.\#\#\#\#\#\#\#\#\#\#\#\#\#\#,\\ 
``Latitude'':\#\#.\#\#\#\#\#\#\#\#\#\#\#\#\#\#,\\``timezoom'':``\#\#\#\#\#''\}''\} }}
&\multirow{4}*{\textit{eu1.clevertap-prod.com}} &\multirow{4}*{4} \\ 
\cline{2-3}
&\makecell{ Email Address \\}&\makecell{M}&&&\\
\cline{2-3}
&\makecell{Phone Number }&\makecell{M}&&&\\
\cline{2-3}
&\makecell{Precise Location   \\}&\makecell{M}&&&\\

\hline

Branch.io~\cite{branchSDKguidance} & \makecell{Diagnostic Data }&N  & \makecell {``\{"local\_ip":"\#\#\#.\#\#\#.\#\#\#.\#\#\#"\}''}
&\textit{api2.branch.io} & 96\\ \hline

\multirow{2}*{Facebook~\cite{FacebookSDKguidance}  }& \makecell{Device ID} &\makecell{V}& \multirow{2}*{\makecell {``\{``extinfo'':``\#\#\#\#,``advertiser\_id'': \\``5B30BC06-9017-4FA0-8A77-3FB3FFBE3D7D''\}''}}
&\multirow{2}*{\textit{graph.facebook.com}} &\multirow{2}*{694}\\ 
\cline{2-3}
&Coarse Location&V&&&\\
\hline

One Signal~\cite{onesignalSDKguidance} & \makecell{Diagnostic Data} &N & \makecell {
``\{``rooted'': ``true'', ``device\_model'':``iPhone10,4'',\\ ``device\_os'':``14.8.1'',``net\_type'' : 0,``state'' : ``ping''\}}
&\textit{api.onesignal.com} & 203\\ \hline

\multirow{3}{*}{Sentry~\cite{sentryguidance}} & \makecell{Device ID } &\makecell{N}& \multirow{3}{*}{\makecell{
``\{``deviceID'': ``03D1DF26-D5B9-4714-9A0F-7B16A82F3AAF''\,\\ ``lat'':``\#\#.\#\#\#\#\#\#\#\#\#\#\#\#\#\#'', ``lng'':``-\#\#.\#\#\#\#\#\#\#\#\#\#\#\#\#\#'',``ip'' :\\ \#\#\#.\#\#\#.\#\#\#.\#\#\#"\,``battery\_level'' : ``100.0'',``battery\_status'' : ``Full'',\\``orientation'' : ``portrait'',``memory\_size'' : ``2108686336''\}''}} &\multirow{3}*{\textit{ingest.sentry.io}} &\multirow{3}*{9} \\ 
\cline{2-3}
& Precise Location&M&&&\\
\cline{2-3}
&Performance Data&M&&&\\
\hline
       
\multirow{2}*{Applovin~\cite{applovinGuidance}} & \makecell{Advertising Data}  &\makecell{M}& \multirow{2}*{\makecell {``\{``loc\_lat'': ``\#\#.\#\#\#'', ``loc\_long'':``-\#\#.\#\#\#',
``ad\_info'':\{``ad\_unit\_id'':\\``62c188063c420e5c'',``extra\_parameters'':\{``viewport\_height'':``50'',\\``viewport\_width'':``320''\},``n'':2,``ad\_format'':``BANNER''\}\}''
}}
&\multirow{2}*{\textit{api.onesignal.com} }&\multirow{2}*{203}\\ 
\cline{2-3}
&Coarse Location&M&&&\\
\hline

Flurry ~\cite{FlurrySDKguidance} & \makecell{Precise Location\\} &\makecell{M} & \makecell {``\{``fl.longitude.value'':-\#\#.\#\#\#\#\#\#\#\#\#\#\#\#\#\#,``fl.bearing.value'':\\-1``fl.frame.version'':1,``fl.latitude.value'':\#\#.\#\#\#\#\#\#\#\#\#\#\#\#\#\#,\\``fl.horizontal.accuracy.value``:65,``fl.vertical.accuracy.value'':10\}''}
&\textit{data.flurry.com} & 76\\ \hline

Ironsource ~\cite{ironsourceSDKguidance} & \makecell{User ID\\}&N  & \makecell {``\{``userIdType'' : ``IDFV'', ``userId'':\\ ``17B4FAC1-7E0F-42C7-BECD-D3FA2607A853''\}''}
&\textit{\makecell{outcome-ssp.\\supersonicads.com}} & 41\\ \hline                
 
Singular ~\cite{singularSDKguidance} & \makecell{Coarse Location\\} &N & \makecell {``\{``city'':``\#\#\#\#\#''\}''}
&\textit{sdk-api-v1.singular.net} & 6\\ \hline    
         
Chartboost ~\cite{chartboostSDKguidance} & \makecell{Other Diagnostic Data\\}&N  & \makecell{``\{``rooted\_device'':true,``audio\_output\_type'':``Speaker''\\,
``device\_dimensions'':\{``width'':320,``height'':568\},\\``dark\_mode'':false,``os'':``13.4.1''\}''}
&\textit{live.chartboost.com} &48 \\ \hline          
     
Mob~\cite{mobSDKguidance} & \makecell{Coarse Location\\}&N  & {``\{``city'':``\#\#\#\#\#''\}''}
&\textit{api.verify.mob.com} &4 \\ %\hline  

\bottomrule

\end{tabular}
\label{table:guidance}
\end{table*}

\begin{figure}[t]
    \centering
    \includegraphics[width=\columnwidth]{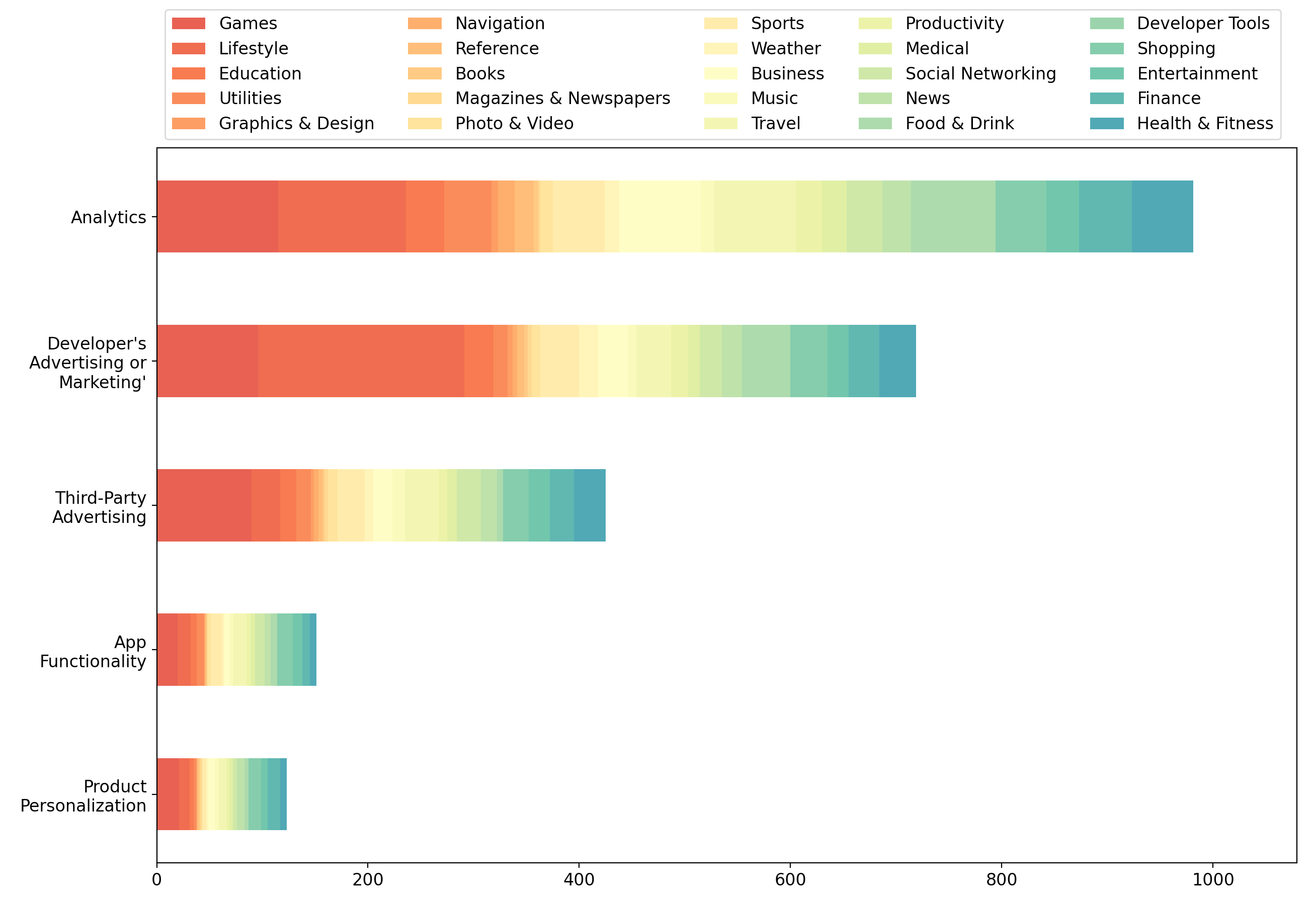}
    \caption{The distribution of iOS apps with incorrect disclosure in their privacy labels. Each stack represents different actual purpose. The colors on the spectrum show the categories of the apps.}
    \label{fig:incorrectCat}
\end{figure}

\begin{figure}[t]
    \centering
    \includegraphics[width=\columnwidth]{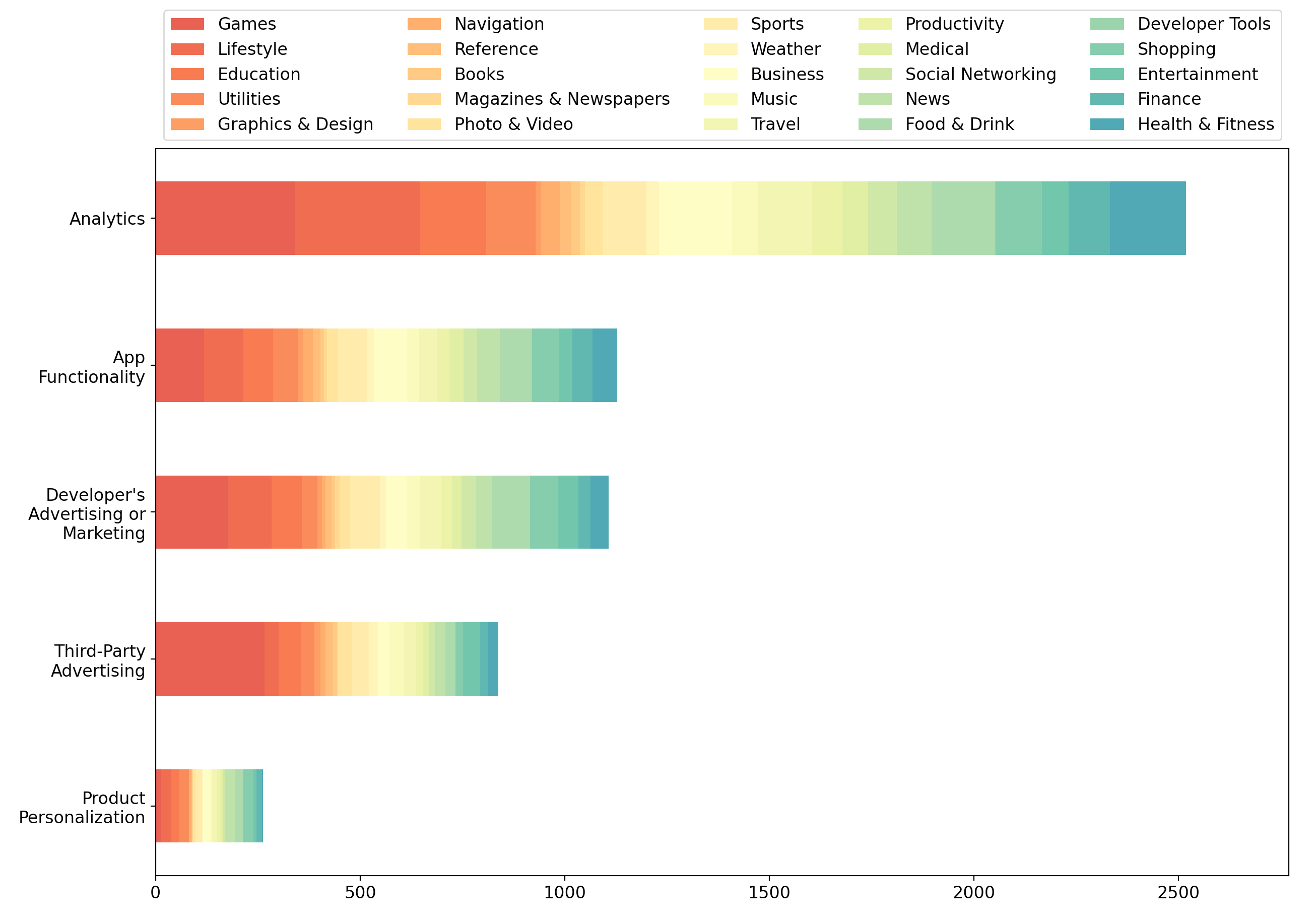}
    \caption{The distribution of iOS apps with neglected disclosure in their privacy labels. Each stack represents different actual purpose. The colors on the spectrum show the categories of the apps.}
    \label{fig:omitDisclosureCategory}
\end{figure}

\begin{figure}[t]
    \centering
    \includegraphics[width=\columnwidth]{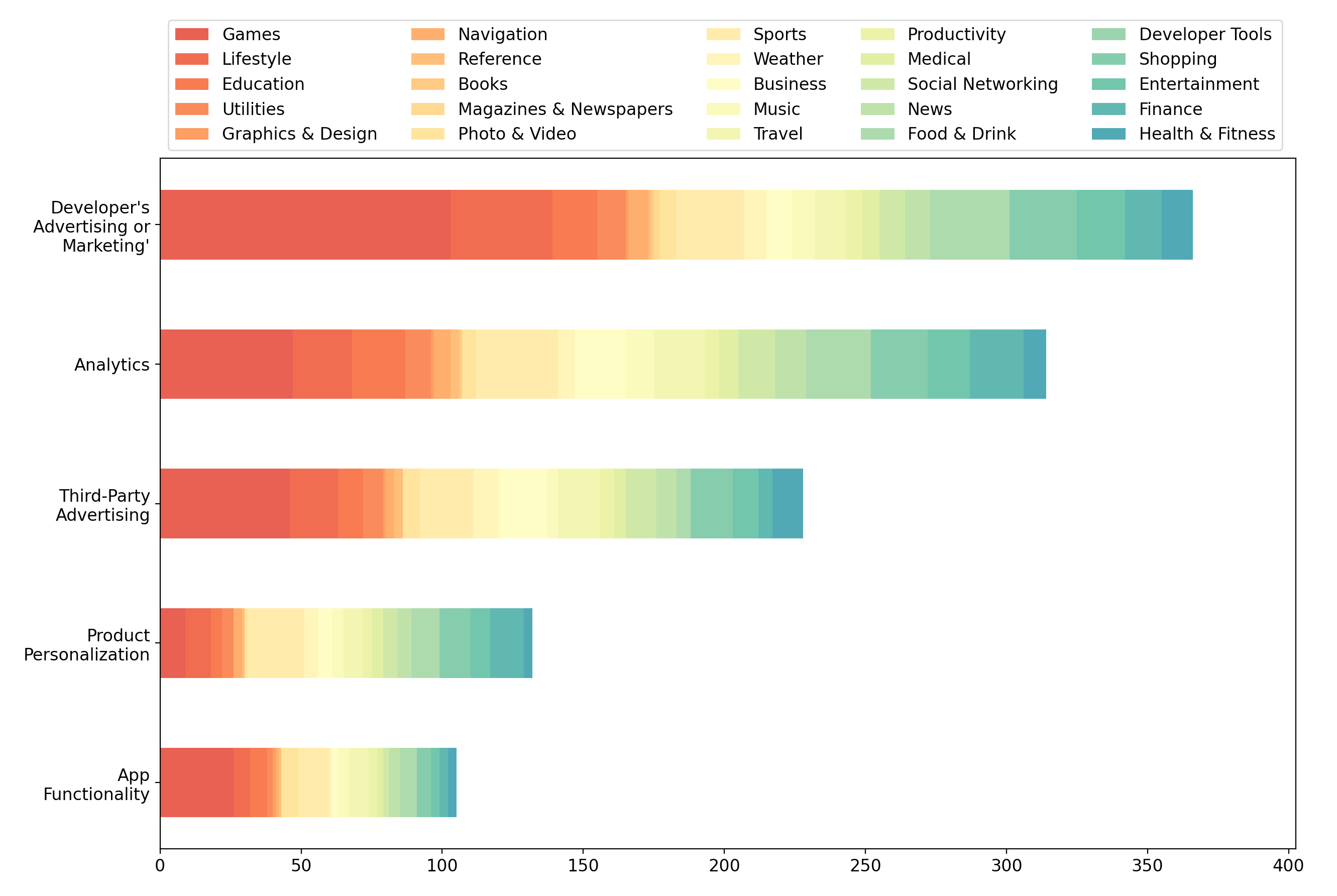}
    \caption{The distribution of iOS apps with inadequate disclosure of their privacy labels. Each stack represents different actual purpose. The colors on the spectrum show the categories of the apps.}
    \label{fig:inadequateCat}
\end{figure}

% {\tt \small
% \begin{verbatim}
% /** Crawling Action Target  */
% function NSCrawlerConfig() {
%   this.testingPeriod = 0.05 * 60 * 60;
%   this.testingDepth = 5;
%   this.takeScreenShot = true;
%   this.newCommandTimeout = 2;
%   this.launchTimeout = 3;
%   this.packages = '.*';
%   this.maxActionPerPage = 15;
%   this.exclusivePattern = '';
%   this.platform = 'ios';
%   this.clickTypes = ["Button"];
%   this.editTypes = ["TextField"];
%   this.tabBarTypes = ["TabBar"];
%   this.strategy = 'source';
%   this.depth = 100;
%   this.duration = 1800;
%   this.deviceType = 'iPhone';
% }
% \end{verbatim}
% }

%\input{table/overviewExample}
\end{appendices}

%%%%%%%%%%%%%%%%%%%%%%%%%%%%%%%%%%%%%%%%%%%%%%%%%%%%%%%%%%%%%%%%%%%%%%%%%%%%%%%%
\end{document}